\documentclass[12pt]{article}
\usepackage[utf8]{inputenc}
\usepackage{setspace}
\usepackage{pdflscape}
\usepackage{longtable}
\usepackage{graphicx}
\usepackage{float}
\usepackage{placeins}
\usepackage{caption}
\usepackage{tikz}
\usepackage[normalem]{ulem}
\usepackage{verbatim}
\usepackage[toc,page]{appendix}
\usepackage{breqn}
\usetikzlibrary{arrows.meta, positioning}
\usepackage[authordate,bibencoding=auto,backend=biber,natbib, doi=false, url = false, isbn = false]{biblatex-chicago}
\AtBeginBibliography{%
  \footnotesize % Adjust the font size (e.g., \small, \footnotesize, \scriptsize)
  \setlength{\itemsep}{0pt plus 0.1ex} % Adjust the spacing between entries
}
\usepackage{booktabs}
\usepackage{siunitx}
\usepackage{multirow}
\usepackage{threeparttable}
\usepackage[margin=1in]{geometry}
\usepackage{adjustbox}
\usepackage{subfiles}
\usepackage{xltabular}
\usepackage{caption} % for '\ContinuedFloat' directive
\usepackage[hidelinks,breaklinks=true]{hyperref}
\hypersetup{
    unicode=false,          % non-Latin characters in Acrobat's bookmarks
    pdftoolbar=true,        % show Acrobat's toolbar?
    pdfmenubar=true,        % show Acrobat's menu?
    pdffitwindow=false,     % window fit to page when opened
    pdfstartview={FitH},    % fits the width of the page to the window
    pdftitle={My title},    % title
    pdfauthor={Author},     % author
    pdfsubject={Subject},   % subject of the document
    pdfcreator={Creator},   % creator of the document
    pdfproducer={Producer}, % producer of the document
    pdfkeywords={keyword1, key2, key3}, % list of keywords
    pdfnewwindow=true,      % links in the new PDF window
    colorlinks=true,       % false: boxed links; true: colored links
    linkcolor=blue,          % color of internal links (change box color with linkbordercolor)
    citecolor=blue,        % color of links to bibliography
    filecolor=magenta,      % color of file links
    urlcolor=blue           % color of external links
}
\newcolumntype{d}{S[
    input-open-uncertainty=,
    input-close-uncertainty=,
    parse-numbers = false,
    table-align-text-pre=false,
    table-align-text-post=false
 ]}
\usepackage{subcaption} % Required for creating subfigures

\usepackage{amsthm}
\usepackage{amsmath}
\usepackage{tabularray}
\usepackage{float}
\usepackage{graphicx}
\usepackage{codehigh}
\usepackage[normalem]{ulem}
\UseTblrLibrary{booktabs}
\UseTblrLibrary{siunitx}

\NewTableCommand{\tinytableDefineColor}[3]{\definecolor{#1}{#2}{#3}}
\AtBeginDocument{%
  \addtolength\abovedisplayskip{-0.5\baselineskip}%
  \addtolength\belowdisplayskip{-0.5\baselineskip}%
}
\usepackage{amsfonts}
\usepackage{enumitem}

\newtheorem{theorem}{Theorem}
\newtheorem{assumption}{Assumption}[section]

\newtheorem{lemma}{Lemma}[section]
\newtheorem{proposition}{Proposition}[section]

\theoremstyle{definition}
\newtheorem{definition}{Definition}[section]

\theoremstyle{remark}
\newtheorem*{remark}{Remark}

\begin{document}
% Test
\title{Can ChatGPT Forecast Stock Price Movements? Return Predictability and Large Language Models
\thanks{Alejandro Lopez-Lira: alejandro.lopez-lira@warrington.ufl.edu, 305 Stuzin Hall, Gainesville, FL 32611; Yuehua Tang:  yuehua.tang@warrington.ufl.edu, 317A Stuzin Hall, Gainesville, FL 32611. We are grateful for the comments and feedback from Dimitris Papanikolaou (the editor), two anonymous referees, Svetlana Bryzgalova, Andrew Chen, Carter Davis, Andrea Eisfeldt, Xiao Han, Ryan Israelsen, Wei Jiang, Ben Lee, Holger von Jouanne-Diedrich, Andy Naranjo, Jay Ritter, Nikolai Roussanov, Jinfei Sheng, Avanidhar (Subra) Subrahmanyam, Baozhong Yang, Jialin Yu, and seminar and conference participants at UCLA, UC Irvine, Florida State University, Bank of Mexico, CEIBS, Peking University HSBC Business School, University of Florida, the SEC, AllianceBernstein, Bloomberg, Qube Research \& Technologies, Santander Bank, UBS Australia, 4th Frontiers of Factor Investing 2024, UBS US Quant Conference, The 36th Australasian Finance and Banking Conference, RSFAS Summer Research Camp, EDHEC Speaker Series, Online Seminars in Finance Series, Artificial Intelligence and the Economy, the 1st New Finance Conference, ITAM Alumni Conference, and the Insightful Minds in Artificial Intelligence Seminar Series. }  
}

\author{
   Alejandro Lopez-Lira and Yuehua Tang \\
   University of Florida
}
\date{\small First Version: April 6, 2023; This Version: \today
}
\maketitle

\vspace{-1cm}
\begin{abstract}
\footnotesize
We document the capability of large language models (LLMs) like ChatGPT to predict stock market reactions from news headlines without direct financial training. Using post-knowledge-cutoff headlines, GPT-4 captures initial market responses, achieving approximately 90\% portfolio-day hit rates for the non-tradable initial reaction. GPT-4 scores also significantly predict the subsequent drift, especially for small stocks and negative news. Forecasting ability generally increases with model size, suggesting that financial reasoning is an emerging capacity of complex LLMs. Strategy returns decline as LLM adoption rises, consistent with improved price efficiency. To rationalize these findings, we develop a theoretical model that incorporates LLM technology, information-processing capacity constraints, underreaction, and limits to arbitrage.

\setstretch{1.3}
\end{abstract}

\footnotesize{
}

\clearpage

% \newpage
\normalsize 

\section{Introduction}

The recent proliferation of generative artificial intelligence and large language models (LLMs) like ChatGPT has significantly impacted multiple domains. Although these models are primarily trained to predict the next word in a sequence, they have exhibited surprising proficiency in complex tasks, such as coding, fueling widespread interest in their emerging capabilities. However, their potential in economic applications remains largely unexplored. A fundamental question in financial markets is how quickly and accurately investors incorporate new information into prices. Understanding this process has proven challenging because it requires measuring both the ``true'' economic implications of news and the market's interpretation of it. To bridge this gap, we investigate whether LLMs can serve as a novel instrument to study market information processing by evaluating their ability to assess the economic content of news headlines and comparing these assessments with how markets actually respond initially and whether there is a subsequent price drift.

Our key finding is that GPT-4 demonstrates a strong ability to predict the immediate market reaction to news headlines, achieving daily portfolio hit rates of approximately 90\% for correctly identifying the direction of the initial price response. This result is notable for two reasons. First, it is not obvious that a general-purpose language model trained without explicit financial supervision should excel at this task. The fact that it does suggests that sophisticated financial reasoning---understanding not just whether news is positive or negative, but also how markets will interpret it---emerges naturally from the model's general language training.\footnote{Unlike prior studies that rely on supervised methods requiring labeled training data and domain-specific fine-tuning \citep{Tetlock2007GivingMarket, Tetlock2008MoreFundamentals, Tetlock2011AllInformation, Garcia2013SentimentRecessions, Calomiris2019HowWorld}, our analysis evaluates LLMs' off-the-shelf ability to interpret news without explicit financial training. This approach requires no labeled data or financial expertise and provides interpretable assessments that any investor can understand.} Second, and more importantly for financial economics, this high accuracy provides a unique window into market information processing. By comparing what GPT-4 identifies as the economic implications of news with how markets actually respond over time, we can quantify instances where markets initially underreact to information that the LLM correctly assesses. This allows us to study \emph{which types of news} create information processing frictions and \emph{why} markets may be slow to incorporate certain information.

Since GPT-4 can accurately understand the economic impact of news, one would expect to observe return predictability patterns consistent with phenomena documented in the literature on delayed information diffusion and limits to arbitrage.\footnote{For example, \citet{bernard1989post}, \citet{chan1996momentum}, \citet{dellavigna2009investor}, \citet{hirshleifer2009driven}, \citet{Jiang2021PervasiveData}, and \citet{Fedyk2023WhenNews}.} Indeed, following the initial reaction to news, stock prices tend to continue drifting in the same direction indicated by GPT-4's assessment over the next one to two trading days. A long-short daily-rebalanced strategy based on GPT-4's assessments of overnight news generates average returns of 34 basis points (bps) per day before transaction costs.\footnote{We obtain similar results using abnormal returns from the CAPM model and the 5-factor model of \citet{Fama2015AModel} as the strategy has low loadings on standard risk factors, including the market factor.} Furthermore, the underreaction is particularly pronounced for smaller stocks and following negative news.  However, exploiting these patterns in practice is only feasible for market participants whose transaction costs are sufficiently low, such as market makers. 
Still, the central contribution of this study lies not only in documenting LLMs' ability to generate return predictability but also in uncovering what LLM assessments reveal about how markets process and react to different types of information. 
%can help understand market underreaction to different types of information.  Size-tercile analysis reveals that while GPT-4 assessments align with initial reactions across all firm sizes, drift occurs only in small caps, demonstrating that large-cap markets process information efficiently while small-cap underreaction persists.

We guide our empirical investigation by developing a theoretical model that incorporates LLM technology, information processing constraints, and limits to arbitrage. Our model allows us to explore the potential impact of LLM information-processing capabilities on market dynamics and generates several testable implications. First, return predictability aligns with theories of delayed information diffusion and limits to arbitrage, with underreaction more pronounced for smaller stocks and short positions. Second, greater LLM model complexity enhances forecasting ability, establishing a critical threshold in AI capabilities above which these technologies can profitably predict returns. Third, as LLMs become more sophisticated and widely adopted, prices should better reflect fundamentals, though return predictability persists in equilibrium.

Empirically, we analyze a comprehensive data set of news headlines relevant to 4,123 U.S. common stocks from major news media and newswires over the period from October 2021 to May 2024. This sample period is selected to avoid look-ahead bias and ensure an out-of-sample evaluation of LLMs like ChatGPT, as the version we evaluate was trained on data up until September 2021.\footnote{\citet{sarkar_lookahead_2024,lopez2025memorization} provide strong evidence that one must use a sample period after an LLM's knowledge cutoff date to avoid lookahead bias and memorization problems.} We prompt ChatGPT to assess the economic implications of each headline for the relevant firm's stock price, categorizing it as positive, negative, or neutral news, and use this signal to study the market's initial reaction and subsequent drift. Consistent with the idea that LLMs are trained to mimic human reasoning, our evidence shows that sophisticated models like GPT-4 can accurately discern the immediate economic implications of a news headline, achieving a daily portfolio hit rate of 93.3\% (88.8\%) for identifying the initial response direction of overnight (intraday) headlines.

We next use LLM assessments to better understand how markets process different types of information. In particular, we develop an interpretability framework %that identifies which categories of news create the most significant information processing frictions. 
using topic modeling to group news into themes and then examine which themes are associated with GPT-4's most accurate predictions of the initial reaction and where the subsequent drift is most pronounced. Our results reveal substantial heterogeneity in how markets process different kinds of information. For certain topics, such as earnings reports, strategic partnership announcements, and clinical trial results, markets react efficiently. GPT-4 scores show strong alignment with initial reactions but minimal drift prediction. In contrast, for topics such as insider stock transactions and specialized conference presentations, we observe underreaction: a strong initial response alignment is accompanied by significant subsequent drift. This suggests that markets are slow to fully incorporate the information in these types of news, given that GPT-4 correctly identifies their implications from the outset.

%\textcolor{red}{This analysis reveals that GPT-4 particularly excels at assessing complex corporate events like insider transactions and healthcare conference presentations, where markets appear to initially underreact. By comparing GPT-4's performance with less sophisticated models like GPT-3, we can also identify which types of information require greater processing capacity to interpret correctly. For instance, GPT-4 shows marked improvements over GPT-3 for complex events like reverse stock splits and industry-specific news, highlighting the types of information where human market participants may face similar processing challenges. }

Consistent with our theory, we find that LLMs' ability to discern the economic implications of news increases with model size. Basic models such as GPT-1, GPT-2, and Llama2-7b exhibit limited off-the-shelf stock assessment capability, with low hit rates in predicting the initial reaction and no significant predictability for subsequent return drift.\footnote{Section A of the Appendix provides details on the full list of 12 LLMs we examine.} More advanced models, such as GPT-3.5 and DistilBART-MNLI, demonstrate stronger assessment ability than the basic models but remain noticeably weaker compared to the state-of-the-art model, GPT-4.  For example, the hit rate and the drift-strategy Sharpe ratios generally increase with model size. A long-short strategy based on next-day drift returns that buys stocks with positive GPT-4 scores and sells those with negative scores delivers an annualized Sharpe ratio of 2.97 over the sample period, compared with 1.66 for GPT-3.5, 1.26 for DistilBART-MNLI, and negative Sharpe ratios for most basic models. Overall, our findings suggest that financial decision-making ability is an emerging capacity of more complex LLMs. %, which aligns well with our conceptual framework. 

%Next, we examine LLMs' forecasting capabilities across headlines with different complexities. We categorize news into low- or high-difficulty using the Flesch-Kincaid Readability Score and study how different LLMs perform with more difficult-to-understand news. Consistent with the notion that a larger number of parameters enhances LLMs' information processing capabilities, only the more advanced LLMs demonstrate strong assessment ability and can predict the return drift in low-readability news. Further, we differentiate between news articles, typically independent of the firms in question, and press releases, which the firms themselves often issue. Basic models, such as GPT-1, GPT-2, and BERT, demonstrate limitations in processing the latter due to their simpler algorithms, which may not adequately interpret data presented with strategic intent. Conversely, more sophisticated models like GPT-4 exhibit sustained predictive accuracy for subsequent return drift, indicating their resilience to potential biases in these communications. 

Next, we examine the performance differences across LLMs within the same family to identify which types of financial information require greater processing capacity for correct interpretation.  Our theory predicts that larger models exhibit the greatest performance advantages where information processing is most challenging, and that is where markets are most likely to display inefficiencies. Comparing GPT-4 to GPT-3.5, the largest advantages appear precisely in categories requiring complex synthesis: insider stock transactions show a 25.2 bps advantage for initial reactions and 13.8 bps for the drift, while healthcare conference participation shows the strongest drift advantage at 20.4 bps. The Llama2 family exhibits even more pronounced differences in predicting initial reactions, with the larger Llama2-70b model outperforming the smaller Llama2-7b model by 316 bps for earnings announcements and 140 bps for stock rating changes. Our evidence suggests that topics requiring deeper analytical reasoning or specialized knowledge benefit most from greater model sophistication.

We further investigate how LLM-based approaches differ from traditional embedding-based methods. Specifically, we directly compare GPT-4's conversational prompting approach with embedding-based supervised learning models trained on actual return outcomes with rolling 6-month windows. The embedding models collapse when training data becomes limited: for intraday news (26,109 observations), performance deteriorates dramatically with OpenAI embeddings achieving a Sharpe ratio of only 1.71 compared to 2.53 for overnight news (105,742 observations). In contrast, GPT-4 maintains consistent performance (Sharpe ratio of 2.63 for intraday), demonstrating that its zero-shot semantic reasoning does not depend on massive training samples. Moreover, unlike traditional methods that require statistical expertise and continuous retraining, LLMs' conversational interface is accessible to any investor. Our model predicts that this accessibility should enhance market efficiency by improving information processing among previously inattentive agents.

%We then study the speed of news assimilation by analyzing the performance of the ChatGPT-based strategies in the week after the news arrival to see how quickly stock prices react to company news. We find that ChatGPT 4 assessment scores accurately capture the immediate reaction to company news and significantly predict returns over the next two trading days but not thereafter. Thus, the market tends to underreact to the news initially and takes about two days to fully incorporate the information contained in the news. 

Finally, considering the widespread adoption of LLMs like ChatGPT and the capabilities they demonstrate, it is imperative to assess their potential impact on market efficiency. Our theoretical framework posits that LLMs can increase investors' information processing capacity and reduce market inefficiencies. As a result, the return predictability should weaken as LLMs' size increases and more investors start using them. While testing this hypothesis represents an empirical challenge, we find some suggestive evidence: a general decline in the performance of the ChatGPT-based strategy during our sample period, over which GPT models' capabilities and adoption skyrocketed. For example, its annualized Sharpe ratio drops from 6.54 in 2021Q4 to 3.68 in 2022, 2.33 in 2023, and to 1.22  over January-May 2024.

Our findings provide several novel insights about financial economics and market functioning. First, and most importantly, we demonstrate that LLMs serve as a powerful instrument for studying what drives information processing frictions in markets. Markets efficiently process transparent, quantifiable information (e.g., earnings reports and clinical trials) but systematically underreact to information requiring complex synthesis (e.g., insider transactions and specialized conference presentations). The pattern of where GPT-4 outperforms GPT-3.5 (complex synthesis tasks) versus where they perform equally (formulaic announcements) reveals which types of information processing create bottlenecks for market participants. This granular mapping of processing frictions, which is impossible to obtain with traditional methods, provides new evidence on the mechanisms underlying delayed price discovery.

Second, the threshold effect in model complexity combined with the limitations of supervised learning approaches indicates that market underreaction may stem from scarcity of genuine semantic reasoning capacity, not merely lack of statistical sophistication. Since LLMs approximate human reasoning patterns, GPT-4's unique success while simpler models and domain-specific approaches fail suggests that the analytical sophistication required to anticipate market responses is relatively rare among investors. This interpretation unifies seemingly disparate findings about underreaction: markets struggle precisely where reasoning capacity is most scarce.

Third, the declining Sharpe ratios as LLMs proliferate provide suggestive evidence that AI adoption is actively reducing information processing frictions in real time. This temporal evolution, together with LLMs' accessibility compared to traditional quantitative methods, suggests that financial markets may be undergoing a structural shift in how efficiently they process complex information. These findings have important implications for understanding how technological change affects market efficiency and the fundamental nature of information processing constraints in financial markets.

\subsection*{Related Literature}

The application of LLMs in economics, particularly ChatGPT, is a relatively unexplored area. In addition to our study, recent research on ChatGPT in economics includes \citet{korinek2023generative}, \citet{Hansen2023CanFedspeak}, \citet{bybee2023ghost}, \citet{Noy2023ExperimentalIntelligence}, and \citet{Manning2024AutomatedSubjects}, each addressing different questions than ours.\footnote{Moreover, \citet{Xie2023TheChallenges} find ChatGPT is no better than simple methods such as linear regression when using numerical data in prediction tasks, and \citet{Ko2023CanPerspective} try to use ChatGPT to help with a portfolio selection problem but find no positive performance. Both results are unsurprising since ChatGPT is better at text-based tasks. In addition, \citet{eisfeldt2023generative} study the impact of ChatGPT on firm value through the channel of labor productivity. Finally, \citet{kogan_et_al_2023} utilize ChatGPT to classify job tasks into routine vs. non-routine categories, and \citet{jiang2022surviving} use ChatGPT to evaluate the complementary vs. substitutive impact of fintech innovations on different occupations.} Furthermore, a contemporaneous work by \citet{chen2023expected} employs a supervised two-step procedure by first embedding news articles in a high-dimensional space using different embedding techniques, including LLM-based embeddings, and then using these embeddings as inputs for a forecasting model to predict stock returns. %While effective, their method requires sophisticated implementation and is better suited for skilled investors. 
Different from their focus, we test the off-the-shelf capabilities of LLMs in extracting economic content from news (accessible to most investors), use them as instruments to study market information processing, and provide a conceptual framework to explain their potential economic impact on financial markets. Additionally, for predicting future returns, the two approaches are complementary: our direct application of conversational AI systems offers accessibility and interpretability, while their two-step procedure provides flexibility in customization and optimization for specific prediction tasks.

We also contribute to the literature that employs machine learning to study finance research questions, including textual analyses of news articles to extract sentiment and predict stock returns.\footnote{See, e.g., \citet{Jegadeesh2013WordAnalysis}, \citet{Rapach2013InternationalStates}, \citet{Hoberg2016Text-BasedDifferentiation}, \citet{Baker2016MeasuringUncertainty}, \citet{Manela2017NewsConcerns}, \citet{Hansen2018TransparencyApproach}, \citet{Gu2020EmpiricalLearning}, \citet{Ke2019PredictingData}, \citet{Ke2019AAnalysis},  \citet{Jiang2019ManagerReturns},  \citet{Cohen2020LazyPrices},  \citet{Freyberger2020DissectingNonparametrically}, \citet{Bybee2023BusinessCycles}, \citet{Lopez-Lira2023RiskReturns}, and \citet{Cong2024TextualInformation}.}  Our unique contribution is being the first to provide a theoretical framework on LLMs and their potential impact on market dynamics and document comprehensive empirical evidence on LLMs' financial decision-making capabilities. Our theoretical model adds to the literature on information incorporation into market prices and introduces LLM technology.\footnote{For example, \citet{VanNieuwerburgh2010InformationUnder-Diversification}, \citet{kyle1985continuous,Kyle1989InformedCompetition}, \citet{Verrecchia1982InformationEconomy}, \citet{davila2018identifying}, and \citet{davila2021trading}.} Our paper also relates to the growing literature on employment exposures and vulnerability to artificial intelligence (AI) related technology by documenting an important task in the financial industry where off-the-shelf LLMs perform well.\footnote{Recent works by \citet{Agrawal2019ArtificialPrediction}, \citet{Webb2019TheMarket}, \citet{Acemoglu2022ArtificialVacancies}, \citet{Acemoglu2022TasksInequality}, \citet{Acemoglu2022HarmsAI}, \citet{Babina2024ArtificialInnovation}, \citet{jiang2022surviving}, \citet{kogan_et_al_2023}, \citet{Noy2023ExperimentalIntelligence}, and \citet{hampole2025artificial} have examined the extent of job exposure and vulnerability to AI-related technology as well as the consequences for employment and productivity.} Finally, our results, along with the interpretability technique we propose to analyze LLMs' performance across topics, contribute to understanding LLMs' potential in processing financial information and forecast stock returns, which can inspire future research on developing LLMs tailored to the financial industry's needs.\footnote{See, for example, \citet{Wu2023BloombergGPT:Finance} on a large language model for finance---BloombergGPT. In addition, see \citet{lerner2024financial} for a recent study on increasingly valuable financial innovations in the U.S.}%[ADD PIXIU AND OTHERS]

\section{Institutional Background}

ChatGPT is a large-scale language model developed by OpenAI based on the GPT (Generative Pre-trained Transformer) architecture. It is one of the most advanced natural language processing (NLP) models developed so far and was trained on a massive corpus of text data to understand the structure and patterns of natural language. The Generative Pre-trained Transformer (GPT) architecture is a deep learning algorithm for natural language processing tasks. It was developed by OpenAI and is based on the Transformer architecture, which was introduced in \citet{Vaswani2017AttentionNeed}. The GPT architecture has achieved state-of-the-art performance in various natural language processing tasks, including language translation, text summarization, question answering, and text completion.

The GPT architecture uses a multi-layer neural network to model the structure and patterns of natural language. It is pre-trained on a large corpus of text data, such as Wikipedia articles or web pages, using unsupervised learning methods. This pre-training process allows the model to develop a deep understanding of language syntax and semantics, which is then fine-tuned for specific language tasks. One of the unique features of the GPT architecture is its use of the transformer block, which enables the model to handle long sequences of text by using self-attention mechanisms to focus on the most relevant parts of the input. This attention mechanism allows the model to understand the input context better and generate more accurate and coherent responses.

ChatGPT has been trained to perform various language tasks such as translation, summarization, question answering, and generating coherent and human-like text. ChatGPT's ability to generate human-like responses has made it useful for creating chatbots and virtual assistants to converse with users. While ChatGPT is designed for general-purpose language-based tasks, it is not explicitly trained to predict stock returns. 

In addition to evaluating ChatGPT, we also assess the capabilities of other prominent natural language processing models such as BERT, BART, DistilBart-MNLI, FinBERT, and Llama2. By considering the more basic models alongside ChatGPT, we can examine the importance of model complexity for stock return prediction based on textual data. Appendix A at the end of the paper provides an overview of the 12 LLMs we study in this paper.

\section{Conceptual Framework: LLMs, Information Processing, and Market Dynamics} \label{sec:framework}

\label{sec:conceptual_framework} 

To understand the economic mechanisms underlying the observed ability of LLMs like ChatGPT to predict stock returns from news headlines and guide our empirical exploration, we develop a conceptual framework grounded in a formal theoretical model (presented in full detail in Section B of the Appendix). This framework not only explains LLMs' return predictability but also explores how their emergence as effective information processors interacts with established market features like bounded human rationality, information frictions, and limits to arbitrage. It aims to provide economic intuition for our empirical findings and to offer new insights into the broader implications of artificial intelligence in financial markets.

\subsection{Key Ingredients}

Our framework considers a market with an asset whose fundamental value is revealed over time. Trading occurs over short horizons following the arrival of public news signals about these fundamentals. Importantly, the market is populated by heterogeneous agents who differ in their capacity to process this news immediately. We include \textit{attentive agents}, who are sophisticated participants capable of quickly processing a significant portion of the information content in news. Alongside them are \textit{inattentive agents}, participants with limited information processing capacity who only grasp a fraction of the news implications initially, fully updating their beliefs with a delay; this element captures aspects of bounded rationality or information acquisition costs. Finally, the framework incorporates \textit{limits to arbitrage} through the presence of noise trading, which represents non-fundamental demand shocks and introduces risk for attentive agents attempting to correct temporary mispricings.
This setup allows for temporary deviations of prices from values implied by fully processed information.

\subsection{LLMs as Information Processors} 

A central element is the explicit modeling of LLMs as information processors whose effectiveness depends on their characteristics. We conceptualize LLM capacity as depending on both \textit{model sophistication} (e.g., measured by model size $k$), where more advanced models can extract information with higher precision, and \textit{news complexity} ($c$), recognizing that even advanced models may struggle with overly complex or nuanced information.
The formal model in the Appendix captures this through a function $\lambda(c,k)$ representing the precision of the LLM's interpretation relative to the total information content in the news. This allows the framework to account for the empirically observed heterogeneity in performance across different LLMs and different types of news.

\subsection{Mechanism of Predictability} 

Within this framework, the potential for LLMs to predict returns arises from a mismatch in information processing speeds and capacities. If an LLM (particularly a sophisticated one) can interpret the implications of news more accurately or quickly than the marginal human trader who sets the current price (often influenced by the slower-reacting inattentive agents), a temporary divergence occurs. The LLM's assessment of the public news ($\mu_{L|s}$ in the formal model) reflects a better estimate of the post-news fundamental value than the current market price ($p_1$). LLM predictability emerges as the price subsequently drifts towards the more fully-processed information level ($p_2 \approx \mu_{A|s}$), aligning with the LLM's initial, more accurate assessment. The presence of noise trader risk ensures that this convergence is not instantaneous or riskless for attentive agents, allowing the predictability to persist in the short run.

\subsection{Key Theoretical Insights}

Our model yields several key economic insights relevant to understanding LLMs in finance.

First, the model establishes the existence of an LLM quality threshold. It posits that there is a critical threshold ($k^*$) in LLM sophistication, below which models cannot generate profitable predictions, and only sufficiently advanced LLMs ($k > k^*$) demonstrate robust predictive power (i.e., Theorem 1). A threshold exists because markets contain inherent frictions, including noise trader risk, transaction costs, and limits to information-processing capacity, that create a hurdle any predictive signal must overcome. Below this quality level, an LLM may extract some information from news, but its signal-to-noise ratio remains too low for reliably profitable trades. The threshold's level depends on market structure: when markets have many attentive participants or high trading volume, prices incorporate information quickly, raising the bar for LLM usefulness. Conversely, noisier markets with high risk aversion or substantial noise trading allow even moderately capable LLMs to identify profitable opportunities. Thus, the threshold reflects both fundamental market frictions and the competitive information environment.

Second, the framework sheds light on the determinants of predictability. Consistent with theories of limits to arbitrage and delayed information diffusion, the model predicts that LLM-based return predictability should be more pronounced where information frictions or arbitrage constraints are higher, such as in smaller, less liquid stocks or following negative news (i.e., Proposition 1). In small-cap or illiquid stocks, fewer sophisticated traders find it worthwhile to pay the fixed costs of active participation, leaving more ``room'' between the initial underreaction and full-information valuation. For negative news, shorting faces higher costs and regulatory restrictions, which limit the ability of attentive arbitrageurs to correct mispricings. In both cases, the LLM exploits gaps that arbitrage forces are slower to close.

Third, our model analyzes the impact on market efficiency from LLM adoption. It predicts that widespread use of capable LLMs can enhance market efficiency by accelerating information incorporation into prices through two channels: empowering less informed (inattentive) agents (i.e., Theorem 2) and augmenting already sophisticated (attentive) agents (i.e., Proposition 3). The mechanism operates through price formation itself. Since the equilibrium price is a precision-weighted average of all agents' assessments, when inattentive agents adopt superior LLMs, their improved beliefs pull the price closer to fundamentals. Similarly, when attentive agents adopt even better LLMs, the most informed component of the price becomes more accurate, further improving price discovery.

Finally, a key corollary concerns the potential erosion of LLMs' predictability. As high-quality LLMs become widely adopted, the very predictability they initially exploited may diminish or disappear (i.e., Proposition 2). This reflects a price impact effect: profitability depends on the gap between the LLM's assessment and the current market price. As more traders act on the same LLM signal, their collective demand drives the initial price closer to the LLM's valuation, shrinking this gap. When adoption becomes widespread and LLM capabilities approach those of the most informed traders, LLM's information advantage vanishes as everyone sees the same signal already reflected in prices. Thus, the LLM's predictive success inherently contains the seeds of its own obsolescence as a trading signal.

This conceptual framework, backed by the formal model in the Appendix, provides a structured way to interpret our empirical results. It helps explain why different LLMs perform differently, why predictability varies across stocks and news types, and offers insights into the potential evolution of market efficiency as AI technologies proliferate. It guides our analysis beyond simply documenting predictability to understanding the underlying economic forces at work.

\section{Data}

We utilize four primary datasets for our analysis: the Center for Research in Security Prices (CRSP) daily returns, news headlines, RavenPack news database, and the NYSE Trade and Quote (TAQ) database. The sample period begins in October 2021 and ends in May 2024. This sample period ensures that our evaluation is out-of-sample as the training data of the ChatGPT models stops in September 2021 (e.g., \citet{sarkar_lookahead_2024,lopez2025memorization}).\footnote{We further address potential lookahead bias concerns by comparing GPT-4 to GPT-3.5 performance over time in Online Appendix C. GPT-4 performs better in the latter part of the sample, inconsistent with the model having knowledge outside of its cutoff date.}

We obtain daily stock returns, open prices, and close prices from the CRSP. Our sample consists of all the stocks listed on the New York Stock Exchange (NYSE), the National Association of Securities Dealers Automated Quotations (NASDAQ), and the American Stock Exchange (AMEX), with at least one news story covered by a major news media or newswire. Following prior studies, we focus our analysis on common stocks with a share code of 10 or 11. Our intraday price and return data come from the TAQ database. We clean the TAQ data and construct minute-by-minute observations of intraday volume-weighted average prices following the cleaning procedures in the literature (e.g., \citet{bollerslev2016roughing}).

We first collect a comprehensive news dataset for all CRSP companies using web scraping based on the company ticker. The resulting dataset comprises news headlines from various sources, such as major news agencies and financial news websites. For each company, we collect all the news during the sample period and then match the headlines with the identifying information from RavenPack. Matching with RavenPack assures that only relevant news will be used for the experiment. They closely monitor the major financial news distribution outlets and have a quality procedure matching news, timestamps, and entity names, which solves any errors that arise from the web scraping procedure.

After the matching procedures, our final sample comprises 159,137 firm-headline-date observations covering 4,123 unique companies from October 2021 to May 2024, representing about 85\% of the 4,875 firms in CRSP during this period.\footnote{Please see Table \ref{tab:news_descriptive_stats} for more details of our news sample. In addition, we have carefully compared our sample of firms with the CRSP universe in Section A of the Online Appendix. We find that firms with matched news are generally larger and more liquid with more volatile returns. Note that the same headline text may appear for multiple firms or dates. For example, a company's monthly dividend announcement appears each month as a separate observation. Of these 159,137 observations, 139 (0.09\%) could not be classified into overnight versus intraday timing due to missing timestamp information, leaving 158,998 observations for our timing-based analyses.} In the sample, about 67.5\% of the headlines correspond to press releases, and the remaining 32.5\% are news articles. Based on the release time, we group news into overnight vs intraday news because the entry times for the two types of news are different when forming trading strategies. About 82\% (or 129,953) of the headlines are classified as overnight news as they are released either before 9 a.m. or after 4 p.m. on a trading day, and the remaining 18\% are classified as intraday news.\footnote{We use 9 a.m. as the cutoff in the morning to classify overnight news so that there is time to process and trade when the market opens. Results are similar if we use alternative cutoffs such as 9:15 or 9:30 a.m.} Being the majority of our sample, we emphasize overnight news in some of the later analyses.  

We employ the ``relevance score" (between 0 and 100) from Ravenpack to indicate how closely the news pertains to a specific company. A 0 (100) score implies that the entity is mentioned passively (predominantly). Our sample requires news stories with a relevance score of 100. We limit it to complete articles and press releases and exclude headlines categorized as `stock-gain' and `stock-loss' as they only indicate the daily stock movement direction. To avoid repeated news, we require the ``event similarity days" to exceed 90, which ensures that only new information about a company is captured.\footnote{Furthermore, we eliminate duplicates and overly similar headlines for the same company on the same day. We gauge headline similarity using the Optimal String Alignment metric (the Restricted Damerau-Levenshtein distance) and remove subsequent headlines with a similarity greater than 0.6 for the same company on the same day.}

\section{ChatGPT Prompt} \label{sec:prompting}

In this section, we discuss how we prompt ChatGPT to extract information from news headlines and illustrate it with an example. 

%\subsection{Prompt}

A prompt is a short text that provides context and instructions for ChatGPT to generate a response. The prompt can be as simple as a single sentence or as complex as a paragraph or more, depending on the nature of the task. Prompts enable ChatGPT to perform a wide range of language tasks, such as language translation, text summarization, question answering, and even generating coherent and human-like text. They allow the model to adapt to specific contexts, generate responses tailored to the user's needs, and perform tasks in different domains.

We use the following prompt in our study and apply it to the publicly available headlines.

\begin{quote}
    \emph{Forget all your previous instructions. Pretend you are a financial expert. You are a financial expert with stock recommendation experience. Answer ``YES" if good news, ``NO" if bad news, or ``UNKNOWN" if uncertain in the first line. Then elaborate with one short and concise sentence on the next line. Is this headline good or bad for the stock price of  \_company\_name\_ in the short term? \\ Headline:  \_headline\_}
\end{quote}

We ask ChatGPT to assume the role of a financial expert with experience in stock recommendations. The terms \_company\_name\_ and  \_headline\_ are substituted by the firm name and the respective headline during the query. The prompt is specifically designed for financial analysis and asks ChatGPT to evaluate a given news headline and its potential impact on a company's stock price in the short term. ChatGPT is requested to answer ``YES" if the news is good for the stock price, ``NO" if it is bad, or ``UNKNOWN" if it is uncertain. ChatGPT is then asked to explain in one sentence to support its answer concisely. The prompt specifies that the news headline is the only source of information provided to ChatGPT. It is not assumed that the headline contains sufficient information to reasonably assess its impact on the stock price since the model can answer it does not know. We set the temperature of GPT models to 0 to maximize the reproducibility of the results.\footnote{Temperature is a parameter of ChatGPT models that governs the randomness and the creativity of the responses. A temperature of 0 essentially means that the model will always select the highest probability word conditional on the text, which will eliminate the effect of randomness in the responses and maximize the reproducibility of the results.}

For example, consider the following headline about Humana in December 2023:

\begin{quote}
    \emph{Cigna Calls Off Humana Pursuit, Plans Big Stock Buyback} 
\end{quote}

And here is ChatGPT 4's response after we replace ``\_company\_name\_" with ``Humana" and ``\_headline\_" with the text above in the prompt:

\begin{quote}
    \emph{NO} 
    
     \emph{The termination of Cigna's pursuit could potentially decrease Humana's stock price as it may be perceived as a loss of a potential acquisition premium.
}
\end{quote}

The news headline states that Cigna Group abandoned its pursuit of a merger with Humana and announced a big stock repurchase plan. This news is value-relevant to two health insurance providers: Cigna and Humana. For Humana, the data vendor's proprietary analytics tool gives a positive sentiment score of 0.65, indicating that the news is perceived as favorable for Humana. However, ChatGPT 4 responds that the information is negative for Humana. It reasons that Humana's stock price could drop due to a loss of an acquisition premium caused by the termination of Cigna's bid. The difference in sentiment scores across the models highlights the importance of understanding the context in prediction tasks. 

With the prompting strategy above, we utilize an API provided by OpenAI to prompt ChatGPT and obtain a recommendation for each headline. We use the ChatGPT model version ``gpt-4-0314" whose training data stops in September 2021 to ensure that our analysis is an out-of-sample evaluation.\footnote{The ``gpt-4-0314" model is a snapshot of GPT-4 from March 14th, 2023. This model will not receive updates, and its training data is up to September 2021.} We transform it into a numerical ``GPT-4 score," where ``YES" is mapped to 1, ``UNKNOWN" to 0, and ``NO" to -1. The average GPT-4 score in our sample is positive (0.31), with the median being zero, and the event sentiment score provided by the data vendor shows a similar pattern. Thus, news headlines in our sample have a positive tilt.\footnote{In our main prompting strategy, we ask ChatGPT to provide an answer first and then the reasoning. We also tried to ask the model to reason first and then answer in a robustness analysis. Based on a sample of 1,000 randomly selected news headlines, we find that ChatGPT 4's recommendations across the two prompting methods are very similar (see Table \ref{table:reasoning_first} of the Online Appendix).} 

In addition to analyzing the performance of ChatGPT 4, we examine the capabilities of more basic models, such as BERT, GPT-1, GPT-2, and Llama2, and compare their performance with that of the more advanced models.\footnote{Llama-2 has a later knowledge cutoff date (September 2022), so their results across model families may be overstated.} We employ a different strategy for the more basic models because those models cannot follow instructions or answer specific questions. For instance, GPT-1 and GPT-2 are auto-complete models. Section C of the Appendix details the prompts we use for these models.

\FloatBarrier

\section{ChatGPT and Market Information Processing} \label{sec:returns}

To evaluate ChatGPT’s ability to forecast stock price movements from news, we begin with what is neither obvious ex-ante nor established in the literature: whether a large language model can correctly predict the market’s immediate reaction to firm-specific news. To the extent that LLMs are trained to mimic human reasoning, sophisticated models may be capable of accurately discerning the immediate economic implications of a news headline. This is the first primary objective of our analysis in this section. 

A natural follow-up question is whether ChatGPT’s ability translates into trading advantages after the announcement. For such interpretive accuracy of news to be practically exploitable by investors who cannot trade on insider information prior to the news release, markets must exhibit some degree of inefficiency, such as underreaction, where prices continue to drift in the direction of the initial impact. Without this subsequent drift, even a perfect reading of the immediate effect would offer no trading opportunity post-announcement. The second objective of this section is to analyze whether ChatGPT scores can predict stock returns after the announcement.

\subsection{Can ChatGPT Forecast Stock Price Movements?} \label{sec:returns_overtime}

To study these two questions, we begin by conducting a portfolio analysis to examine the return patterns around the release of news with positive vs. negative ChatGPT assessment scores. This involves creating long-short portfolios based on ChatGPT's scoring of news headlines (long in the stocks with a positive score and short in the ones with a negative score) and analyzing the performance of these portfolios. Due to the differences in the timing of portfolio formation, we analyze overnight vs. intraday news separately. 

\begin{figure}[htbp]
  \centering
    \caption{Overnight News Returns: Before and After the Release Time}
          \vspace{0.5cm}
  \includegraphics[width=0.8\textwidth]{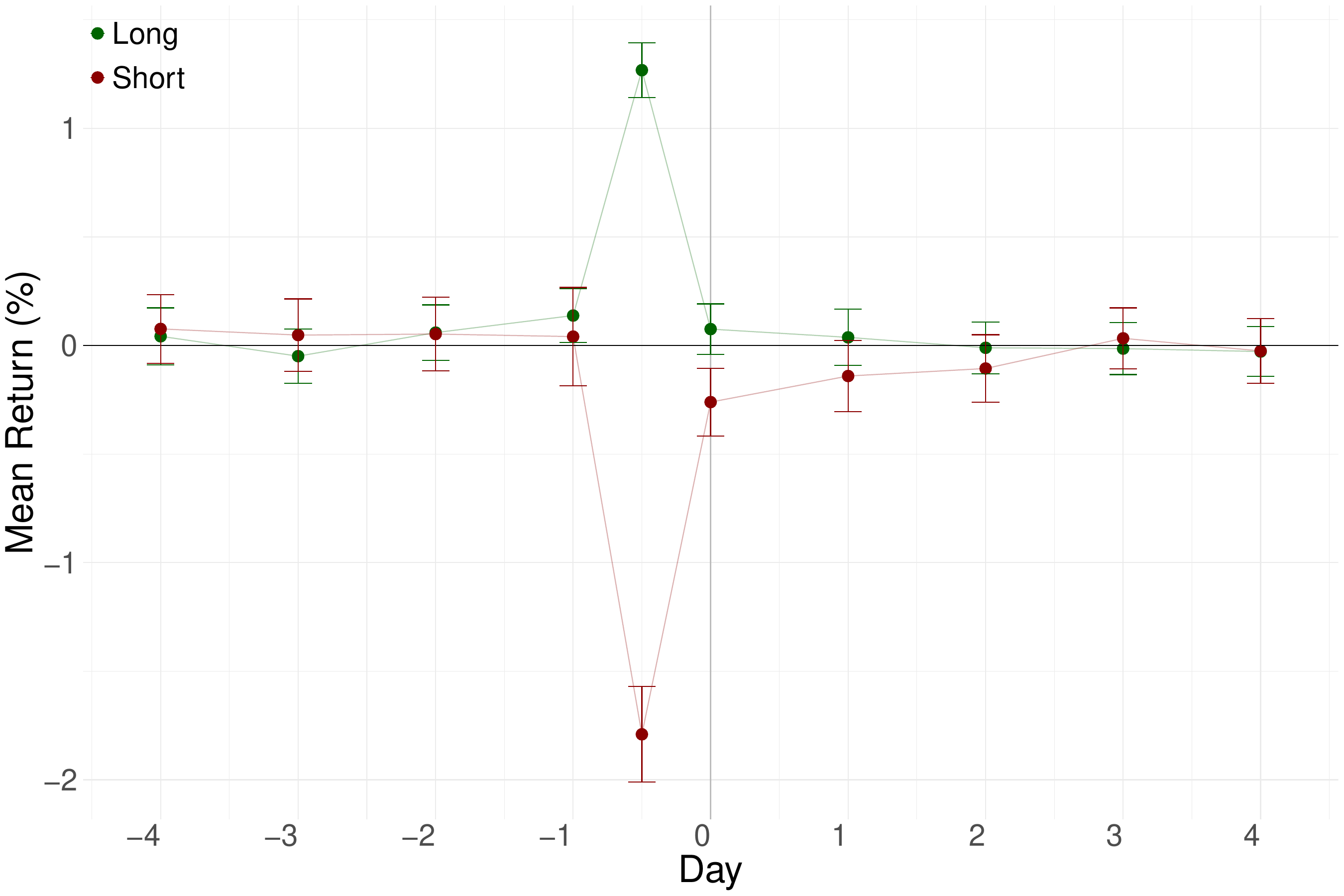}

  \label{fig:daily_ret_overnight_long_short}

  \vspace{0.5cm}
\begin{minipage}{15cm}
\small
This figure presents the average one-day holding period returns over event time for portfolios formed based on GPT-4 assessments of overnight news and their 95\% confidence intervals. 
We form two equal-weighted portfolios based on GPT-4 assessments of overnight news: one for positive scores (Long) and one for negative scores (Short). Overnight news refers to the headlines released after 4 p.m. of trading day $t$-1 and before 9 a.m. on trading day $t$. The plot shows the average returns of each portfolio for day $t$=0 (both the initial reaction and the subsequent drift) and one-day close-to-close returns for the few days before and after day $t$ --- day $t$-4 through $t$-1 and $t$+1 through $t$+4. The dot in between day $t$=-1 and $t$=0 in the plot corresponds to the initial market reaction to news, i.e., the return from the close of day $t$-1 to the market opening of day $t$, and the dot at $t$=0 corresponds to the subsequent drift measured as the return from the market opening to the close of day $t$. The green line corresponds to an equal-weighted portfolio of companies with good news according to ChatGPT 4. The red line corresponds to an equal-weighted portfolio of companies with bad news according to ChatGPT 4. %The initial market reaction is measured as the percentage change from the closing price of day $t$-1 to the opening price of day $t$, and the subsequent drift right after (i.e., the dot at t=0) is measured as the percentage change from the opening price to the closing price of day $t$.

\end{minipage}
\vspace{0.5cm} 
\end{figure}

\begin{figure}[htbp]
  \centering
    \caption{Intraday News Returns: Before and After the Release Time}
          \vspace{0.5cm}
  \includegraphics[width=0.8\textwidth]{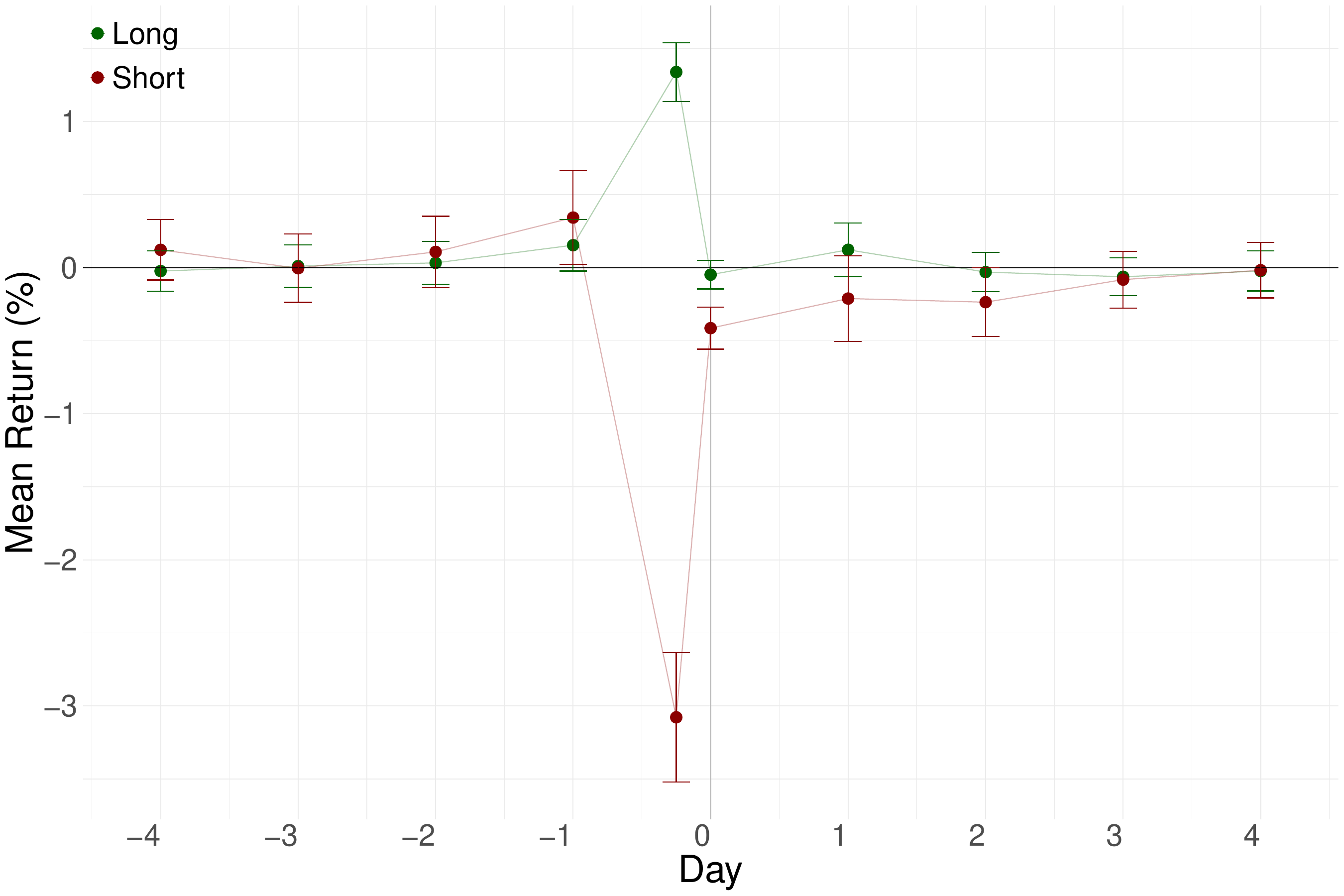}

  \label{fig:daily_ret_intraday_long_short}

  \vspace{0.5cm}
\begin{minipage}{15cm}
\small
This figure presents the average one-day holding period returns over event time for portfolios formed based on GPT-4 assessments of intraday news and their 95\% confidence intervals. We form two equal-weighted portfolios based on GPT-4 assessments of intraday news: one for positive scores (Long) and one for negative scores (Short). Intraday news refers to the headlines released between 9:30 a.m. and 4 p.m. of trading day $t$. %For intraday news released between 9:30 a.m. and 4 p.m. of trading day $t$, we enter the position 15 minutes after the news announcement on day $t$, creating a long-short portfolio that buys companies with good news and short-sells companies with bad news according to ChatGPT 4, and hold it for five full trading sessions. 
The plot shows the average returns of each portfolio for day $t$=0 (both the initial reaction and the subsequent drift) and one-day close-to-close returns for the few days before and after day $t$ --- day $t$-4 through $t$-1 and $t$+1 through $t$+4. The dot in between day $t$=-1 and $t$=0 in the plot corresponds to the initial market reaction to news, i.e., the return from the close of day $t$-1 to 15 minutes after the news announcement on day $t$, and the dot at $t$=0 corresponds to the subsequent drift measured as the return from 15 minutes after the news release to the close of day $t$. The green line corresponds to an equal-weighted portfolio of companies with good news according to ChatGPT 4. The red line corresponds to an equal-weighted portfolio of companies with bad news according to ChatGPT 4. 
\end{minipage}
\vspace{0.5cm} 
\end{figure}

%To study these two questions, we begin by analyzing the return patterns around the release of news with positive vs negative ChatGPT assessment scores. 
Figures \ref{fig:daily_ret_overnight_long_short} and \ref{fig:daily_ret_intraday_long_short} present the results for overnight news and intraday news, respectively. The initial market reaction is measured as the return from the close of day $t-1$ to the opening of day $t$ for overnight news in Figure \ref{fig:daily_ret_overnight_long_short} and from the close of day $t-1$ to 15 minutes after the news release for intraday news in Figure \ref{fig:daily_ret_intraday_long_short}. Both figures demonstrate that ChatGPT's assessment scores align well with the market's \emph{immediate} interpretation of firm-specific news. For instance, the average initial reaction to overnight news with positive ChatGPT scores (the long portfolio) is 1.27\%, while that to overnight news with negative scores (the short portfolio) is –1.79\%; both effects are highly statistically significant. For intraday news, the average initial reaction to those with positive ChatGPT scores is 1.33\%, while that to negative scores is –3.11\%. %This confirms the first condition: ChatGPT can understand the primary economic impact. 
This set of results clearly supports the first hypothesis: ChatGPT can correctly identify the primary economic impact of news headlines.

\centerline{\bfseries [Insert Figures \ref{fig:daily_ret_overnight_long_short} and \ref{fig:daily_ret_intraday_long_short} about here]}

These figures also provide evidence for the second question. Following this initial reaction, stock prices tend to continue drifting in the same direction indicated by ChatGPT's assessment for the next one to two trading days, with the effect being particularly pronounced for negative news. This pattern of underreaction, where the market is slow to fully incorporate the information contained in news, is consistent with the conceptual framework we provide in Section \ref{sec:framework}. Importantly, such market inefficiency provides a window through which an LLM's accurate initial assessment can be translated into a potentially profitable trading strategy.

Table \ref{tab:gpt4_portfolio_performance} provides a comprehensive summary of GPT-4's performance across several key dimensions, including hit rates, mean daily returns, and annualized Sharpe Ratios of the long, short, and long-short portfolios. GPT-4 excels at predicting the initial market reaction, with the long-short portfolio achieving a 93.3\% hit rate (i.e., the percentage of days with positive portfolio returns)\footnote{We use portfolio hit rates rather than headline-level accuracy to weight days equally and measure whether the net daily signal correctly predicts portfolio direction, avoiding cases where individual accuracy misleads about portfolio outcomes.} for overnight news and 88.8\% for intraday news. These initial reaction predictions, if they could be exploited, say by an insider or a trader with early access to information, would translate into strategies with mean returns of 3.06\% for overnight news and 4.44\% for intraday news. 

As expected, GPT-4's ability to predict the subsequent price drift is more modest. It has hit rates of 58\% for overnight news and 55\% for intraday news, and mean returns of 0.34\% and 0.50\%, respectively. These drift strategies achieve annualized pre-transaction-costs Sharpe ratios of 2.97 for overnight news and 2.63 for intraday news. Additionally, the predictability of subsequent drift is significantly stronger for negative news than for positive news. For overnight news, while the long leg has an average daily return of 8 bps and an annualized Sharpe ratio of 0.78, the short leg has an average daily return of 26 bps and an annualized Sharpe ratio of 2.01. These patterns align well with the key ideas of our conceptual framework: GPT-4's primary strength lies in understanding the immediate economic implications of news, while the subsequent drift represents an inefficiency due to limits to arbitrage.

\centerline{\bfseries [Insert Table \ref{tab:gpt4_portfolio_performance} about here]}

%This dual finding, namely that ChatGPT can effectively gauge the immediate economic impact of news and that markets subsequently underreact to this information in a way that creates an exploitable opportunity, provides a foundation for our subsequent analyses. It motivates a deeper exploration into how ChatGPT can predict stock returns post-announcement and the broader implications of its predictive power.  

\subsection{Post-Announcement Drift and Predictive Regressions}
%We begin by conducting a portfolio analysis to evaluate ChatGPT's ability to predict stock price movements. This involves creating long-short trading strategies based on ChatGPT's scoring of news headlines (buying the stocks with a positive score and selling the ones with a negative score) and analyzing the performance of these portfolios. Due to the differences in the entry time of portfolio formation, we analyze overnight vs. intraday news separately. As over four-fifths of the news headlines are released during nontrading hours (i.e., overnight news), we focus on overnight news in much of our baseline analysis. In particular, if a piece of news is released before 9 a.m. on a trading day, we enter the position at the market opening and exit at the close of the same day. If the news is announced after the market closes, we assume we enter the position at the next opening price and exit at the close of the next trading day. All strategies are rebalanced daily.\footnote{ We require at least two firms with positive scores to enter the long leg and similarly with negative scores. Otherwise, we enter just one of the legs. Although we do not use it in the strategy, one could long or short the market portfolio to balance the strategy. Our strategies are not optimized for implementation but rather to show ChatGPT's raw forecasting power.} 

To better illustrate ChatGPT's ability to predict price drift after news release, we analyze the performance of post-announcement trading strategies created based on ChatGPT's assessment scores: buying stocks with a positive score and selling those with a negative score. As more than four-fifths of the news headlines are released during nontrading hours (i.e., overnight news), we focus on overnight news in our baseline analysis. In particular, if a piece of news is released before 9 a.m. on a trading day, we enter the position at the market opening and exit at the close of the same day. If the news is announced after the market closes, we assume we enter the position at the next opening price and exit at the close of the next trading day. All strategies are rebalanced daily.\footnote{We require at least two firms with positive scores to enter the long leg and similarly with negative scores. Otherwise, we enter just one of the legs. Although we do not use it in our strategy, one could consider long or short positions in the market portfolio to balance the strategy. Our strategies are not optimized for implementation but rather to show ChatGPT's raw forecasting power.}

\begin{figure}[htbp]
  \centering
    \caption{Cumulative Returns of Investing \$1 (Without Transaction Costs)}
      \vspace{0.5cm}
  \includegraphics[width=0.8\textwidth]{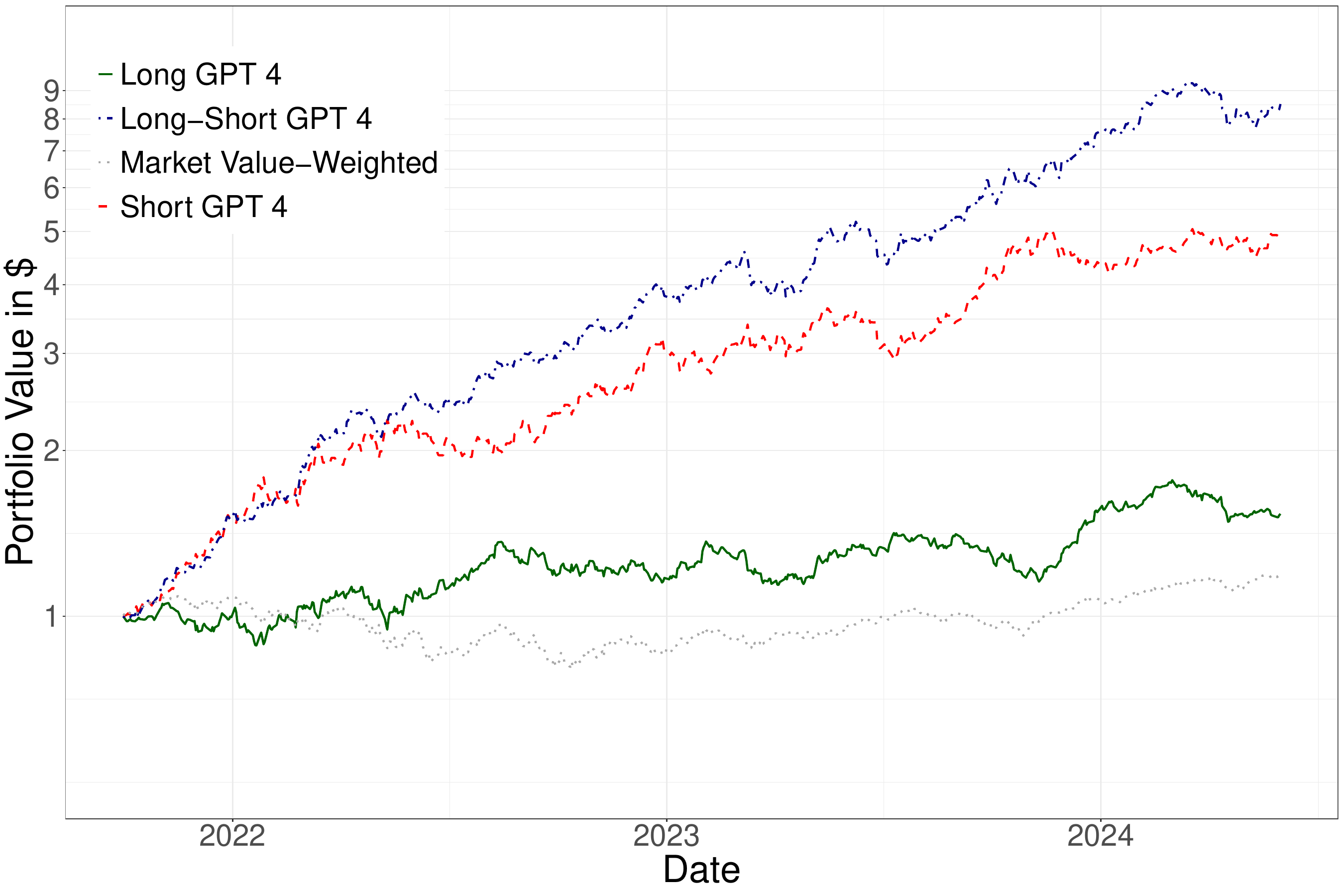}

  \label{fig:cumulative_returns}

  \vspace{0.5cm}
\begin{minipage}{15cm}
\small
This figure shows the performance of trading strategies based on GPT-4's assessment scores of overnight news, without considering transaction costs. If a piece of news is released before 9 a.m. on a trading day, we enter the position at the market opening and exit at the close of the same day. If the news is announced after the market closes, we assume we enter the position at the next opening price and exit at the close of the next trading day. All the strategies are rebalanced daily. The green line corresponds to an equal-weighted portfolio that buys companies with good news, according to ChatGPT 4. The red line corresponds to an equal-weighted portfolio that short-sells companies with bad news, according to ChatGPT 4. The blue line corresponds to an equal-weighted long-short portfolio that buys companies with good news and short-sells companies with bad news, according to ChatGPT 4. The grey line corresponds to a value-weighted market portfolio.
\end{minipage}
\vspace{0.5cm} 
\end{figure}

Figure \ref{fig:cumulative_returns} plots the cumulative returns over our sample period of four different trading strategies (each investing \$1) based on ChatGPT 4 without considering transaction costs: (i) an equal-weighted portfolio that buys companies with good news based on ChatGPT 4, (ii) an equal-weighted portfolio that sells companies with bad news based on ChatGPT 4, (iii) a long-short strategy based on ChatGPT 4, and (iv) a value-weight market portfolio for comparison purposes. This demonstrates the predictive power of ChatGPT scores in forecasting stock returns the next day based on news headlines. For instance, without considering transaction costs, a long-short strategy that buys stocks with a positive ChatGPT 4 score and sells stocks with a negative ChatGPT 4 score earns a cumulative return of around 700\% from 2021m10 to 2024m5, significantly outperforming the market portfolios. While both legs contribute to the predictability, the effect is stronger among stocks with negative news, again consistent with our conceptual framework's predictions regarding underreaction patterns (i.e., Proposition 1).

\centerline{\bfseries [Insert Figure \ref{fig:cumulative_returns} about here]}

\begin{figure}[htbp]
  \centering
    \caption{Cumulative Returns of Investing \$1 With Different Transaction Costs}
          \vspace{0.5cm}
  \includegraphics[width=0.8\textwidth]{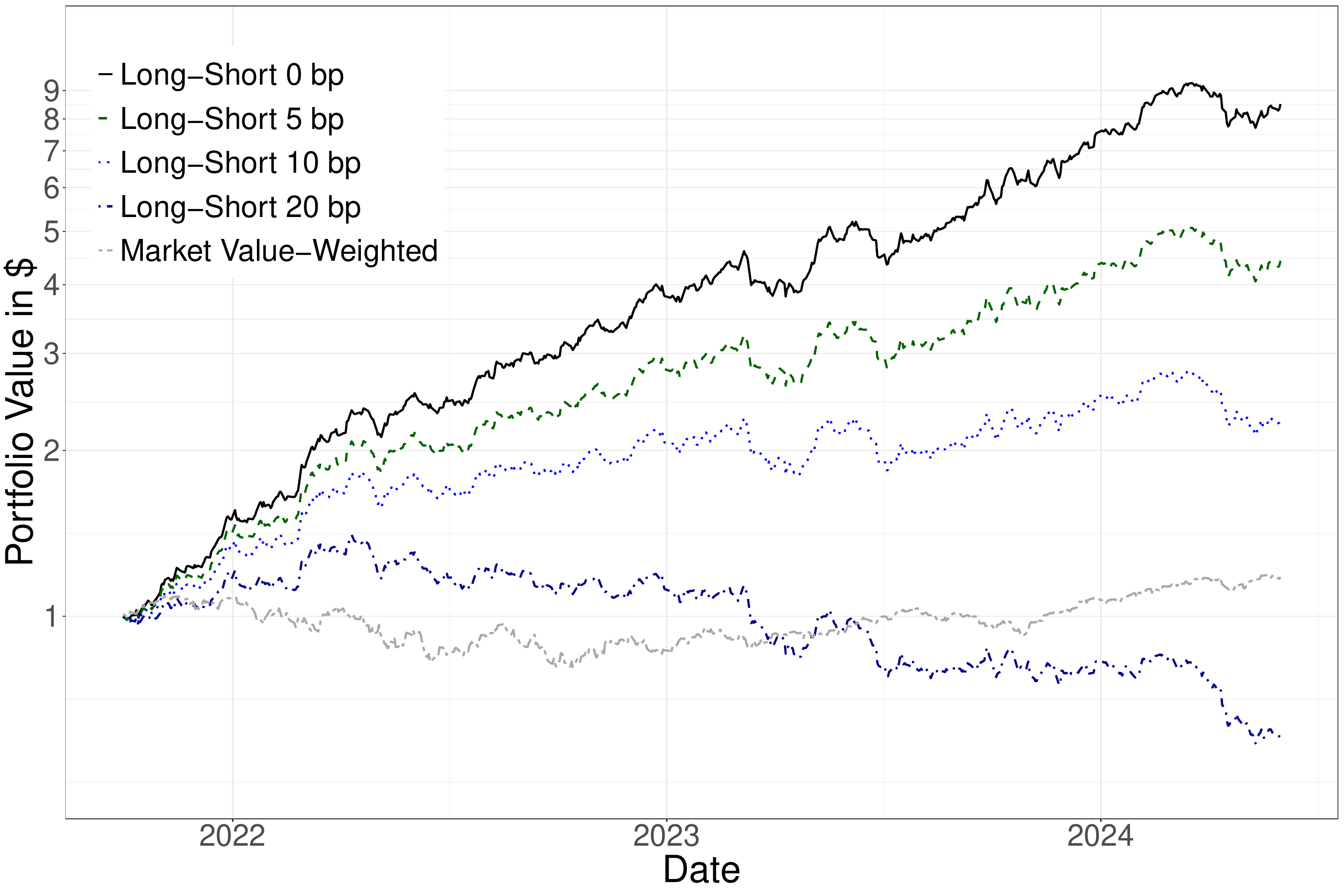}

  \label{fig:cumulative_returns_costs}

  \vspace{0.5cm}
\begin{minipage}{15cm}
\small
This figure shows the performance of trading strategies based on GPT-4's assessment scores of overnight news, considering different transaction costs. If a piece of news is released before 9 a.m. on a trading day, we enter the position at the market opening and exit at the close of the same day. If the news is announced after the market closes, we assume we enter the position at the next opening price and exit at the close of the next trading day. All the strategies are rebalanced daily. The black line corresponds to an equal-weighted long-short portfolio that buys companies with good news and short-sells companies with bad news, according to ChatGPT 4, with zero transaction costs. The dark green line corresponds to the same equal-weighted long-short portfolio with a 5 bps round trip cost. The light blue line corresponds to the same equal-weighted long-short portfolio with a 10 bps round trip cost. The dark blue line corresponds to the same equal-weighted long-short portfolio with a 20 bps round trip cost. The grey line corresponds to a value-weighted market portfolio without transaction costs.
\end{minipage}
\vspace{0.5cm} 
\end{figure}

In Figure \ref{fig:cumulative_returns_costs}, we evaluate the strategy performance under different transaction cost assumptions: 5, 10, and 20 basis points (bps) per round-trip trade. We model implementation frictions as fixed round-trip bps rather than bid–ask spreads because our baseline trades execute at the opening and closing auctions, which are call auctions that establish a single clearing price with no bid-ask spread; so fixed costs parsimoniously capture commissions and price impact. Assuming a transaction cost of 5 bps round-trip, the strategy still earns a cumulative return of over 300\% during our sample period. As we increase the transaction costs to 10 bps round-trip, the cumulative return is still above 100\%. A transaction cost of 20 bps round-trip makes the strategy unprofitable. Hence, deploying this strategy would require minimizing transaction costs, especially price impact, which is challenging for large or unsophisticated investors. 

\centerline{\bfseries [Insert Figure \ref{fig:cumulative_returns_costs} about here]}

To address concerns about high portfolio turnover, we examine partial rebalancing strategies that reduce trading frequency while maintaining signal exposure (detailed results in Online Appendix F). For overnight news with equal-weighted portfolios, the baseline strategy with 100\% daily rebalancing exhibits turnover of approximately 190\% per day, while reducing the rebalancing fraction to 25\% lowers turnover to approximately 46\% per day. At zero transaction costs, both strategies achieve similar Sharpe ratios (2.97 for 100\% rebalancing vs. 2.89 for 25\% rebalancing). However, at 10 basis points per round-trip, the 100\% rebalancing strategy's Sharpe ratio falls to 1.29, whereas the 25\% rebalancing strategy maintains a Sharpe ratio of 1.34, demonstrating greater robustness to implementation costs.

\begin{figure}[htbp]
  \centering
    \caption{Cumulative Returns of Investing \$1 With Different Sample Restrictions}
          \vspace{0.5cm}
  \includegraphics[width=0.8\textwidth]{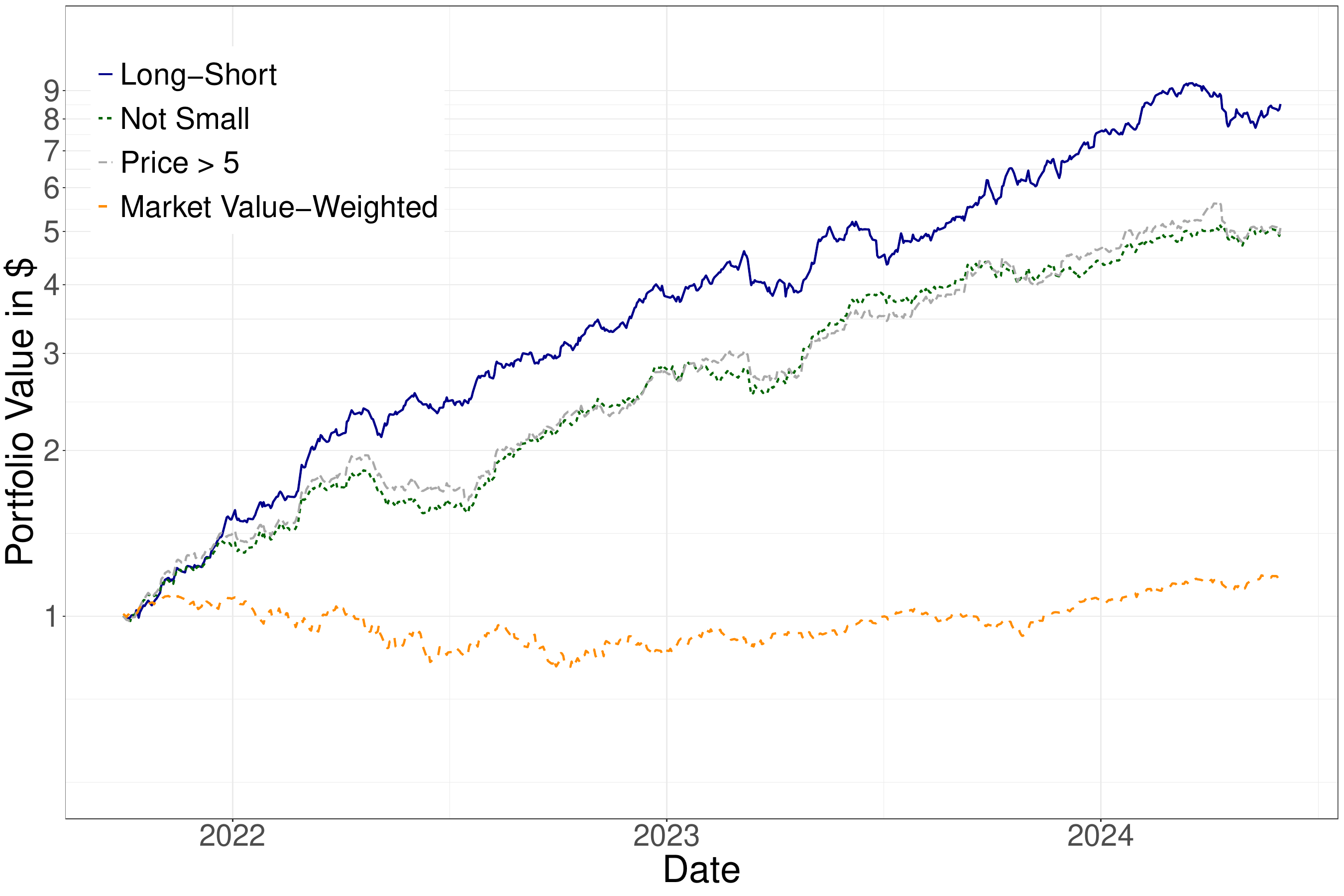}

  \label{fig:cumulative_returns_samples}

  \vspace{0.5cm}
\begin{minipage}{15cm}
\small
This figure shows the performance of trading strategies based on GPT-4's assessment scores of overnight news over different samples, without considering transaction costs. If a piece of news is released before 9 a.m. on a trading day, we enter the position at the market opening and exit at the close of the same day. If the news is announced after the market closes, we assume we enter the position at the next opening price and exit at the close of the next trading day. All the strategies are rebalanced daily. The blue line corresponds to an equal-weighted long-short portfolio that buys companies with good news and short-sells companies with bad news according to ChatGPT 4 for the full sample of U.S. common stocks. The green line corresponds to the same equal-weighted long-short portfolio but over the sample of stocks with a close price greater than \$5 on the prior day $and$ market capitalization above the 20th percentile NYSE market cap breakpoint. The grey line corresponds to a value-weighted market portfolio without transaction costs.
\end{minipage}
\vspace{0.5cm} 
\end{figure}

%The long-short strategy in Figure \ref{fig:cumulative_returns} involves all U.S. common stocks with at least one news headline covering the firm. 
We also analyze the strategy performance by removing small or illiquid stocks from the sample. In particular, as shown in Figure \ref{fig:cumulative_returns_samples}, when we remove from the sample stocks with a close price less or equal to \$5 and stocks with a market capitalization below the 20th percentile NYSE size breakpoint, the cumulative return of the strategy, without considering transaction costs, is still over 300\% during our sample period. It suggests that the predictive power of ChatGPT is not limited to small stocks, but also applies to larger and more liquid stocks.

\centerline{\bfseries [Insert Figure \ref{fig:cumulative_returns_samples} about here]}

%We also illustrate trading strategies based on ChatGPT scores of intraday news, considering three different entry times when forming portfolios: one minute post-release, 15 minutes post-release, and at 4 p.m. on the news day.\footnote{Notice the strategies would need to forecast the total number of news per period, for sizing.} The first strategy enters the position one minute after the news release and exits 15 minutes post-release; the second enters the position 15 minutes post-release and exits at the market close of the news day; and the third enters the position at the close of the news day and exits at the close of the next trading day. Figure \ref{fig:cumulative_returns_entry} of the Appendix shows the cumulative returns of the three trading strategies. The first strategy earns a cumulative return of approximately 100\% over the sample period, while the second and third ones both earn about 300\%. 

In addition to the portfolio analysis above, we also use regressions as the second approach to evaluate ChatGPT's information processing capabilities. Unlike the portfolio analysis in Table \ref{tab:gpt4_portfolio_performance}, which aggregates multiple headlines at the firm-day level, the regression analysis uses individual headlines as observations to preserve the full granularity of news-specific variation in GPT-4 assessments and market responses. Specifically, we run the following linear regressions of the initial reaction or subsequent drift on the ChatGPT score and the sentiment score provided by the data vendor: 
\begin{align} \label{eq:return_gpt} 
r_{i, t}  =   a_i + b_t + \gamma'x_{i,t} + \varepsilon_{i,t+1},
\end{align}

\noindent where the dependent variable, $r_{i, t}$, is stock $i$'s initial market reaction to news or return drift ($r_{t+1}$) over a subsequent trading day after news arrival, $x_{i,t}$ refers to the vector containing the ChatGPT 4 score from assessing stock $i$'s news headlines, and $a_i$ and $b_t$ are firm and date fixed effects, respectively, which account for any observable and unobservable time-invariant firm characteristics and common time-specific factors that could influence stock returns. Standard errors are double-clustered by date and firm. 

Table \ref{tab:regression_analysis} presents the regression results on both initial reactions and subsequent drift, with overnight news in Panel A and intraday news in Panel B. For overnight news, GPT-4's score strongly predicts the initial reaction with a coefficient of 1.326 (t=31.10), indicating that a switch from a negative (-1) to a positive (1) prediction translates to a 2.65 percentage point increase in initial reaction. The drift coefficient is smaller but highly significant at 0.161 (t=6.80), with a negative to positive change in GPT-4 score corresponding to a 32.2 basis point increase in drift returns. For intraday news, the coefficients on GPT-4 score are of similar magnitude: the one for initial reaction is 1.748 (t=13.85) and the one for subsequent drift is 0.160 (t=3.54). Interestingly, when we include both GPT-4 and RavenPack scores to predict the subsequent drift in column (5) of both panels, GPT-4 maintains strong predictive power while RavenPack's coefficient becomes insignificant for drift, indicating that GPT-4 subsumes traditional sentiment analysis methods. 

\centerline{\bfseries [Insert Table \ref{tab:regression_analysis} about here]}

We further examine heterogeneity across firm sizes by including a small-firm indicator (stocks below the 20th percentile of NYSE size breakpoints) and interacting it with GPT-4's score. For overnight news drift, the base GPT-4 coefficient is 0.087 (t=4.09) while the interaction term is 0.404 (t=4.75) in column (6) of Panel A, indicating that predictability is substantially stronger among smaller stocks. This pattern holds for intraday news as well, with an interaction coefficient of 0.683 (t=3.29) for predicting the drift in column (6) of Panel B. Thus, our evidence suggests that the market underreacts to firm-specific news at the daily frequency we examine, consistent with the evidence documented by the extant literature (e.g., \citet{bernard1989post}, \citet{chan1996momentum}, \citet{dellavigna2009investor}, \citet{hirshleifer2009driven}, \citet{Jiang2021PervasiveData}, and \citet{Fedyk2023WhenNews}). Consistent with our conceptual framework, this heterogeneity highlights the significant role of limits to arbitrage in driving the post-announcement return predictability we document.

Next, we conduct an additional regression analysis to gain a deeper understanding of intraday news using different return windows. This analysis uses individual headlines as the unit of observation (the same as in Table \ref{tab:regression_analysis}) rather than the firm-day aggregated data used in Table \ref{tab:gpt4_portfolio_performance}.\footnote{Specifically, the intraday regression sample contains 24,541 individual headline observations representing 21,973 unique firm-day combinations. This differs from the 26,109 firm-day observations in Table \ref{tab:gpt4_portfolio_performance} because: (1) when multiple headlines arrive for the same firm-day, we analyze each headline separately rather than aggregating their scores, and (2) we require complete return data across all three time windows, which results in dropping some observations. This sample is slightly larger than the initial reaction sample in Panel B of Table \ref{tab:regression_analysis} (24,531 observations) because that measure requires previous day's closing data, which is unavailable for some first-trading-day observations that have complete same-day intraday data.} To thoroughly examine the post-release price dynamics, we consider three return windows for each intraday news: (i) from one minute post-release to 15 minutes post-release, (ii) from 15 minutes post-release to the market close of the news day, and (iii) from the market close of the news day to the close of the next trading day. As shown in the first column of Table \ref{tab:intraday_average_by_score}, we find a statistically insignificant coefficient on the ChatGPT 4 score for the return over 1m to 15m post-release, suggesting that the return potential of the ChatGPT strategy is limited in this short window, which is consistent with the pattern in Figure OA1 in the Online Appendix. In contrast, for the return from 15m post-release to the news day's close, the coefficient on ChatGPT 4's score is 0.189 with a $t$-stat of 5.12, while the coefficient is 0.116 with a $t$-stat of 2.31 for the close-to-close return. It suggests that a switch from a neutral (0) to a positive (1) GPT score is associated with a 30 bps return for the strategy that enters the position fifteen minutes post-release and exits it at the close of the next trading day. Thus, it adds significant value to trade on intraday news headlines quickly (within 15 minutes) after the news release to capitalize on the subsequent price drifts during the news day. Finally, we find significant results in the last column of Table \ref{tab:intraday_average_by_score}, which uses the same return window as the drift columns in Panel B of Table \ref{tab:regression_analysis}.

\centerline{\bfseries [Insert Table \ref{tab:intraday_average_by_score} about here]}

In summary, these results demonstrate that GPT-4's advanced language understanding capabilities allow it to accurately assess the economic implications of news headlines. The combination of strong predictive power for initial reactions and significant, yet smaller, drift coefficients reveals a pattern of delayed information incorporation. The subsequent drift occurs because full incorporation of the news requires deeper analytical capacity that is scarce: information is gradually processed by more sophisticated investors, causing prices to continue adjusting days after the announcement. By comparing GPT-4's assessments with actual market responses over time, we identify specific contexts, such as small-stock news and negative announcements, where information processing frictions and limits to arbitrage are most pronounced. Our evidence sheds light on the mechanisms underlying price discovery and market efficiency.

\subsection{Information Processing Across News Types}

While the analysis above demonstrates that market underreaction varies with firm characteristics, a natural question emerges: does the market process different \emph{types} of information at different speeds?

To address this question, we develop a multi-step methodology that examines GPT-4's performance across distinct categories of news content. First, we use clustering on the news embeddings to generate topics for each news and use an LLM to generate the corresponding label. Second, we construct performance scores by multiplying the initial reaction or the subsequent drift by ChatGPT's score. Finally, we run a linear regression of the performance scores on the topics.\footnote{We provide more details of our interpretability methodology in Section E of the Online Appendix.} This approach enables us to decompose market information processing patterns and understand for which types of news ChatGPT-4 is able to accurately comprehend the initial market impact. Furthermore, it allows us to differentiate between topics that suffer from initial underreaction and those that are efficiently incorporated.

Table~\ref{tab:decomposition_analysis} presents this decomposition, separating market responses into three components for each news topic. The initial reaction (\textit{Init}), the subsequent drift (\textit{Drift}), and the total reaction (\textit{Total}) columns show the average market returns in basis points. The right panel displays our key measures: \textit{Init×GPT} quantifies the alignment between GPT-4's assessment and the market's immediate response (GPT score multiplied by initial return), while \textit{Drift×GPT} reveals whether prices continue moving in the direction of GPT-4 prediction (GPT score multiplied by drift return), indicating delayed incorporation of information that GPT-4 initially recognized.

\centerline{\bfseries [Insert Table~\ref{tab:decomposition_analysis} about here]}

Looking at either the initial or total reaction to news headlines, in general, GPT-4 demonstrates the ability to assess the economic content across a wide range of news types. More importantly, the results reveal substantial heterogeneity in how markets process different information types. For certain topics such as earnings and revenue reports, strategic partnership announcements, and clinical trial announcements, markets react efficiently: GPT-4 shows strong initial alignment (\textit{Init×GPT} = 336.7, 92.7, and 308.5 bps, respectively) but minimal drift prediction (\textit{Drift×GPT} = -4.7, 4.3, and -11.4 bps, all statistically insignificant), suggesting that information is incorporated immediately. In contrast, for topics like insider stock transactions, dividend announcements, and healthcare conference presentations, we observe the underreaction pattern: strong initial alignment (\textit{Init×GPT} =71.4, 45.5, and 103.5 bps, respectively) combined with significant drift prediction (\textit{Drift×GPT} = 26.3, 22.3, and 34.2 bps, respectively, all significant at the 5\% level or lower). It suggests that markets are slow to fully process and incorporate the information content for these news types, given that GPT-4 correctly identifies their implications from the outset. 

This heterogeneity provides direct evidence about the nature of information processing frictions in financial markets. Topics exhibiting strong drift patterns indicate news content that is more challenging for the marginal trader to interpret or where limits to arbitrage are more pronounced.  By using GPT-4's assessments as a consistent benchmark across all news types, we can identify which categories of corporate information pose the greatest challenges for real-time market efficiency. The fact that GPT-4 correctly signs both the initial reaction and subsequent drift across diverse news categories demonstrates its ability to extract economically relevant information that markets only gradually incorporate into prices.

\section{Comparing Different Large Language Models}

In this section, we compare different LLMs' abilities to correctly assess the economic implications of news headlines. Since LLMs are trained on human-generated text and learn to approximate human judgments, this comparison serves a dual purpose: it reveals which AI capabilities matter for financial prediction, and more importantly, it provides insights into the information processing challenges that human market participants may face. By examining which models succeed or fail at anticipating market reactions, and which types of information processing tasks separate successful from unsuccessful models, we can better understand the nature of the reasoning required to anticipate market responses. We compare GPT-4 against three alternative approaches: (i) domain-specific models trained on financial text (FinBERT), (ii) supervised learning models trained on actual return outcomes (embeddings), and (iii) LLMs of varying complexity. Our conceptual framework predicts that only LLMs passing a certain threshold in model complexity possess sufficient information processing capacity to correctly interpret news in a way that aligns with how markets ultimately respond, and this threshold may reveal constraints on human information processing in financial markets.

\subsection{Performance Across Models}

To compare the performance of various language models, we conduct a similar portfolio analysis of overnight news as in Table \ref{tab:gpt4_portfolio_performance} using prediction scores from different LLMs. In particular, we consider the following more basic LLMs (listed based on their release time): (i) GPT-1, (ii) BERT, (iii) BERT Large, (iv) GPT-2, (v) BART-Large, (vi) DistilBart-MNLI, (vii) GPT-3.5, (viii) FinBERT, and (ix) Llama2 (7B, 13B, and 70B). All of these LLMs have been released within the past few years (since 2018), highlighting the rapid advancement and proliferation of AI technologies. The more recent LLMs are generally more complex with a larger number of parameters compared to earlier models.The model sizes, measured by the estimated number of parameters, range from 117 million in the initial GPT-1 model to an estimated 1.76 trillion in the most advanced model on our list, GPT-4, as detailed in Appendix \ref{appendix:summaries}.\footnote{While GPT-3's parameter count (175 billion) is officially confirmed by OpenAI, the parameter counts for GPT-3.5 and GPT-4 have not been publicly disclosed. The estimates we use are based on leaked industry reports and are commonly referenced in the literature, but remain unconfirmed by OpenAI.}

Table \ref{tab:average_by_score} reports statistics of long-short portfolios formed based on the assessment scores of news headlines by different models, including the hit rates, annualized Sharpe ratios, average daily returns, Fama-French 5-factor alpha, and average number of stocks in each leg. Our results show a clear pattern: the ability to accurately assess news implications in a way that anticipates how markets will ultimately respond is an emerging capacity of more complex language models. Scores from advanced but not the most complex models, such as DistilBart-MNLI and GPT-3.5, demonstrate some ability to assess news correctly but are noticeably weaker compared to the state-of-the-art model, GPT-4. For instance, the long-short portfolio's hit rates in predicting the initial market reaction are 87\% for DistilBart-MNLI and 93\% for GPT-3.5, compared to 93\% for GPT-4. We observe similar patterns in the hit rate of predicting the subsequent price drift: 56\% for DistilBart-MNLI and 56\% for GPT-3.5, compared to 58\% for GPT-4. The average daily drift returns of the strategies based on DistilBart-MNLI and GPT-3.5 are 0.14\% and 0.29\%, respectively, compared to 0.34\% based on GPT-4.\footnote{We note that the high average daily return for GPT-3.5 is mainly driven by a less diversified short portfolio, which consists of a small number of stocks (on average, six per day). In comparison, the short leg of the GPT-4 strategy has an average of 20 stocks per day. FinBERT, a domain-specific model trained on financial text, shows 90\% accuracy on initial reactions but 48\% on drift (negative Sharpe -0.33). However, FinBERT only makes directional predictions on 25\% of overnight news (assigning neutral to 75\%), limiting the comparability of these results.} The same pattern remains when we evaluate their risk-adjusted performance using the 5-factor model of \citet{Fama2015AModel}. We find that the strategies all have a low loading on risk factors, and their 5-factor alphas are almost of the same magnitude as the corresponding daily average returns. Importantly, the annualized Sharpe ratios of the drift strategies based on DistilBart-MNLI and GPT-3.5 are 1.26 and 1.66, respectively, all of which are lower than 2.97 achieved by GPT-4. 

\centerline{\bfseries [Insert Table \ref{tab:average_by_score} about here]}

Below these three LLMs, the next tier includes BART-large and Llama2-70b, with hit rates of 86\% and 86\% in predicting the initial market reaction, respectively. The average daily drift returns are 0.12\% for BART-large and 0.14\% for Llama2-70b. The annualized Sharpe ratios of the drift strategies are 1.05 and 0.97, respectively. All of these metrics are significantly lower than those of GPT-4. Note that much of the predictability of these LLMs comes from the short leg of the portfolio, even though the short leg has fewer stocks on average than the long leg, a pattern also shared by the most complex model, GPT-4. 

When we use basic models such as GPT-1, GPT-2, and BERT to assess news headlines, their scores are much less accurate in capturing the relevant information of news headlines, as reflected in their substantially lower hit rates (all below 65\%) in predicting the initial market reaction. In addition, their assessment scores are not significantly positively correlated with subsequent return drift. The Sharpe ratios of the drift strategies based on these models are all negative. These findings suggest that simpler models lack the information processing capacity to correctly interpret the economic implications of news headlines, often producing assessments with incorrect directional implications.\footnote{We find very similar performance ranking across the different models when separately examining small vs. nonsmall stocks, with GPT-4 at the top and models like GPT-1, GPT-2, and BERT at the bottom (see Table \ref{tab:average_by_score_small} of the Online Appendix).}

In short, the most complex model we consider, GPT-4, exhibits the greatest accuracy in assessing news implications, with the information processing capacity of LLMs generally increasing as model size grows, which aligns well with our conceptual framework. Therefore, only when AI technologies surpass critical thresholds in model complexity do they possess sufficient sophistication to correctly interpret news in a manner that anticipates market reactions. %This threshold effect reveals that sophisticated financial reasoning, understanding not just sentiment but how markets will interpret news, requires substantial model capacity. This suggests that less sophisticated investors may face similar challenges in processing complex corporate information.
This threshold effect suggests that conventional financial tasks, such as processing and interpreting news, require substantial reasoning capacity and that less sophisticated investors may face similar difficulties when interpreting complex corporate information.

%\subsection{Interpretability and News Types}

%To better understand how ChatGPT interprets the immediate economic impact of news—and why markets might initially underreact to this assessment—we need to examine the specific news characteristics and categories where ChatGPT demonstrates superior understanding compared to both simpler models and potentially human market participants. By identifying these categories, we can gain insights into which types of financial information may be subject to greater processing frictions in the market, making them particularly valuable for LLM-based analysis. This investigation aligns directly with the editor's suggestion to explore whether "these GPT4-based assessments [can] help us unpack and understand how markets process information in real time." The following interpretability analysis serves this purpose by systematically evaluating ChatGPT's decision-making across different news contexts.

% \subfile{Interpretability}

\subsection{News Complexity and News Types}

Having established that GPT-4 outperforms other models overall, we now examine whether this superiority holds across different news characteristics. This analysis serves to distinguish between fundamental market mechanisms (whether underreaction varies with text complexity or source) and model capabilities (whether LLMs struggle more with certain text characteristics). We analyze LLM performance across news complexity levels and source types, which allows us to test whether the topic-level heterogeneity in underreaction documented in the previous subsection is driven by systematic differences in text complexity or information source, or whether it reflects inherent differences in how markets process different types of corporate events.

We categorize headlines based on their Flesch-Kincaid Readability Score, with a daily computed median demarcating them as either low-complexity or high-complexity, and examine performance separately for news articles (journalist-written) versus press releases (firm-issued).\footnote{The Flesch-Kincaid Readability Score measures text complexity based on sentence length and syllables per word. Higher scores indicate text requiring more advanced reading comprehension. The score roughly corresponds to U.S. grade level (e.g., a score of 12 indicates high school senior level). We compute the median separately each day for overnight versus intraday news to ensure balanced comparisons within each trading period.} Table \ref{tab:sharpe_by_complexity} reports hit rates and Sharpe ratios across these categories for 12 different models.

The results reveal that market underreaction operates uniformly across news types. Notably, GPT-4's drift hit rates remain tightly clustered between 55\% and 59\% regardless of complexity or source (high complexity: 55\%, low complexity: 57\%, news articles: 59\%, press releases: 56\%). This narrow range suggests the fundamental market friction, delayed price adjustment to news, is highly consistent. Similarly, GPT-4 achieves exceptional accuracy in predicting initial market reactions across all categories, with hit rates of 89-91\% whether headlines are simple or complex, from journalists or firms. This robustness contrasts sharply with simpler models: GPT-1, GPT-2, and BERT struggle particularly with complex news, achieving initial reaction hit rates below 65\% and negative drift-strategy Sharpe ratios.

While market mechanisms appear uniform, model performance varies. GPT-4's Sharpe ratios range from 1.47 to 2.32 across categories, reflecting differences in classification accuracy rather than differences in market efficiency. Advanced models maintain profitability across all news types, while basic models fail to generate positive returns except in narrow contexts.

The distinction between news sources reveals economically meaningful differences in market information processing. News articles, written by journalists, and press releases, issued directly by firms, may differ in information content and strategic framing. The last six columns of Table \ref{tab:sharpe_by_complexity} compare model performance across these sources. Press releases account for over two-thirds of the sample. GPT-4 achieves consistently high initial reaction hit rates (91\%) for both source types, but demonstrates notably stronger drift prediction for news articles compared to press releases (Sharpe ratio of 2.23 vs. 1.96). The higher drift hit rate for news articles (59\% vs. 56\%) is consistent with journalist-interpreted information being subject to greater initial market underreaction, potentially because third-party analysis contains more nuanced insights that take longer for markets to fully incorporate. When analyzing news articles, several models including BART Large, DistilBart-MNLI, and GPT-3.5 achieve strong performance with drift-strategy Sharpe ratios exceeding 1.5. In contrast, GPT-4's relative advantage is most pronounced for press releases, where its sophisticated language understanding may help navigate corporate strategic disclosure and potentially managed messaging (e.g., \citet{Cao2023HowAI}). %, achieving substantially better performance than simpler models.

\centerline{\bfseries [Insert Table \ref{tab:sharpe_by_complexity} about here]}

\subsection{Which Topics Larger LLMs Understand Better}

In this section, we investigate the comparative advantages of larger models by examining which specific news topics are better understood by more complex models. In particular, we compare the performance of GPT-4 with GPT-3.5 and of Llama2-70B with Llama2-7B across a variety of topics. By comparing models within the same family but of different sizes, our goal is to better understand the financial decision-making capabilities of advanced LLMs across different news contexts.

Table \ref{tab:decomposition_gpt4_vs_gpt3} presents the differential performance across topics, revealing the specific domains where model sophistication provides the greatest advantage. The bottom row shows that GPT-4 outperforms GPT-3.5 overall by 12.2 basis points in predicting the initial reaction and 4.9 basis points in predicting the drift, both highly statistically significant. The improvement is even more pronounced for the two models in the Llama family: Llama2-70B outperforms Llama2-7B by 58.6 basis points in predicting the initial reaction. This suggests that parameter scale provides substantial benefits for financial text interpretation, particularly for tiny models.

The topic-level analysis reveals systematic patterns in where larger models provide the greatest advantages. For GPT-4 versus GPT-3.5, the largest performance gains appear in three categories. First, insider stock transactions show a 25.2 basis point advantage for initial reactions and 13.8 basis points for drift, both highly significant. This suggests that understanding the implications of complex insider trading patterns, which may involve multiple transactions, different types of equity instruments, and varied motivations, benefits substantially from GPT-4's enhanced reasoning capabilities. Second, healthcare conference participation announcements show the strongest drift advantage at 20.4 basis points, indicating that GPT-4 better anticipates how markets will subsequently process information about firms' strategic positioning in the healthcare sector. Third, several topics involving analytical complexity show identical 15.4 basis point initial reaction advantages: earnings and revenue announcements, strategic partnership announcements, and clinical trial announcements. These topics all require an understanding of business implications, competitive positioning, and future cash flow prospects. Interestingly, dividend announcements have virtually no performance difference between GPT-4 and GPT-3.5 (0.2 basis points for the initial reaction and 0.9 for the drift), suggesting that GPT-3.5 is already capable of accurately understanding the implications. There are no statistically significant topics where GPT-3.5 performs better for initial reactions or the drift component.

The comparison between Llama2-70B and Llama2-7B reveals even greater differences, particularly for topics that require quantitative reasoning. In particular, for earnings and revenue announcements, the larger Llama model outperforms its smaller counterpart by 315.7 basis points in the initial reaction. This advantage suggests that interpreting earnings reports is a task where model scale matters greatly for the Llama family. Similarly, large advantages are observed for stock rating adjustments (139.6 basis points), index inclusion announcements (85.0 basis points), and clinical trial announcements (59.7 basis points) in the initial reactions. The magnitude of these differences, substantially larger than those observed in the GPT family comparison, likely reflects the fact that GPT-3.5 was already a highly capable model with extensive training, whereas the gap between Llama2-7B and Llama2-70B represents a tenfold difference in parameters without the same degree of optimization. Across both model families, the pattern is clear: topics requiring deeper analytical reasoning, medical or scientific knowledge, and complex financial interpretation show the largest benefits from increased model sophistication.

\centerline{\bfseries [Insert Table \ref{tab:decomposition_gpt4_vs_gpt3} about here]}

\subsection{ChatGPT vs Embeddings}

In this section, we compare GPT-4's sentiment-based approach with machine learning models based on text embeddings. Supervised embedding models represent a different paradigm: they use supervised learning with rolling-window retraining on actual return outcomes, while GPT-4 operates in a zero-shot manner without any training on our dataset. We evaluate two state-of-the-art embedding approaches: OpenAI embeddings (1,536 dimensions) and MiniLM embeddings (384 dimensions).\footnote{Text embeddings convert text into numerical vectors that encode semantic
  meaning, enabling machine learning models to process textual information.
  Similar texts map to nearby points in the vector
   space. Our analysis employs two widely used embedding models: OpenAI's text-embedding-ada-002 (1,536
  dimensions; \url{https://platform.openai.com/docs/guides/embeddings}) and
  sentence-transformers' all-MiniLM-L6-v2 (384 dimensions;
  \url{https://huggingface.co/sentence-transformers/all-MiniLM-L6-v2}).} Both models use multinomial logistic regression for three-class classification ($\{-1, 0, +1\}$) to match GPT-4's directional prediction methodology. The embedding models are retrained monthly using a rolling 6-month training window, with predictions starting in October 2021 following a 9-month warm-up period.\footnote{We use logistic regression with default sklearn parameters (max\_iter=10, lbfgs solver). Similar to how GPT-4's performance could be improved with better prompting, these embedding results represent a lower bound that could potentially be improved with more extensive hyperparameter optimization or more sophisticated architectures.}

Table \ref{tab:gpt4_embeddings_comparison} reveals substantial performance differences across multiple dimensions. For overnight news, where embedding models have access to 105,742 training samples, GPT-4 achieves a 93.3\% hit rate for initial reactions compared to 84.6\% for OpenAI embeddings and 80.2\% for MiniLM. The performance gap widens considerably for drift prediction, where GPT-4 achieves an annualized Sharpe ratio of 2.97, compared to 2.53 for OpenAI embeddings and only 1.42 for MiniLM. This pattern—where GPT-4's advantage is larger for drift than for initial reactions—is consistent with semantic processing of economic implications, rather than statistical pattern matching, being more important for predicting how prices will continue to adjust after the initial market response.

The intraday results are even more revealing of the fundamental differences between these approaches. While GPT-4 maintains strong performance with a drift-strategy Sharpe ratio of 2.63, the embedding models show substantially weaker performance despite having 1,536 dimensions (OpenAI embeddings) versus GPT-4's zero-shot reasoning. OpenAI embeddings achieve a drift Sharpe ratio of 1.71 for intraday news, representing a 35\% decline from GPT-4's performance. MiniLM, with its 384 dimensions, performs even worse at 0.79, falling 70\% short of GPT-4. This performance degradation occurs despite the embedding models being specifically trained to predict these outcomes through monthly retraining with rolling 6-month windows. We note that both prompting strategies and embedding architectures
could be further optimized, so this comparison represents lower bounds for both approaches.

The key bottleneck appears to be training data availability: the intraday sample of 26,109 firm-day observations is approximately one-quarter the size of the overnight sample (105,742), suggesting that supervised embedding approaches require massive training datasets to capture the subtle patterns in post-announcement drift. In contrast, GPT-4's zero-shot semantic reasoning maintains consistent performance regardless of training data availability, suggesting that GPT-4's language processing capabilities provide a more robust foundation for financial text interpretation than learned statistical associations from limited training data.

\centerline{\bfseries [Insert Table \ref{tab:gpt4_embeddings_comparison} about here]}

The collapse of supervised learning approaches with limited data (embeddings on intraday news) demonstrates that statistical pattern-matching cannot fully substitute for the kind of semantic processing that advanced LLMs provide without sufficient training data. GPT-4's zero-shot reasoning maintains consistent performance regardless of training data availability, revealing a fundamental advantage of language understanding over learned associations. The threshold effect in model complexity, where only the most advanced LLMs succeed, suggests that market underreaction may stem from a scarcity of sophisticated reasoning capacity among investors. 

Since LLMs are trained on human-generated text and approximate human judgments, GPT-4's superiority suggests that the level of analytical sophistication required to correctly anticipate market responses is possessed by relatively few market participants. This interpretation is consistent with markets systematically underreacting to certain types of news, particularly when information is difficult to interpret or limits to arbitrage are binding. As advanced AI tools become more widely adopted, this reasoning capacity becomes more accessible and may improve market efficiency over time, which we examine in the next section.

\section{Speed of Price Response and Market Efficiency}

In a semi-strong efficient market (\citet{Fama1970EfficientWork}), stock prices adjust quickly to new public information. However, as our conceptual framework shows, factors such as information capacity constraints and limits to arbitrage could also matter for the speed of news assimilation, and breakthroughs in AI technologies could significantly change information processing and dissemination in the marketplace and, in turn, impact market efficiency. In this section, we first examine the speed of news assimilation over our sample period and then try to examine the potential impact of LLMs on market efficiency. 

\subsection{Speed of Price Response}
We examine the performance of the ChatGPT-based trading strategies in the week after the news arrival to have a more complete picture of how stock prices react to value-relevant news headlines. By doing so, we aim to shed light on the speed of news assimilation and market efficiency.  

\begin{figure}[htbp]
  \centering
    \caption{Returns of Overnight News Strategy Over Event Time}
          \vspace{0.5cm}
  \includegraphics[width=0.8\textwidth]{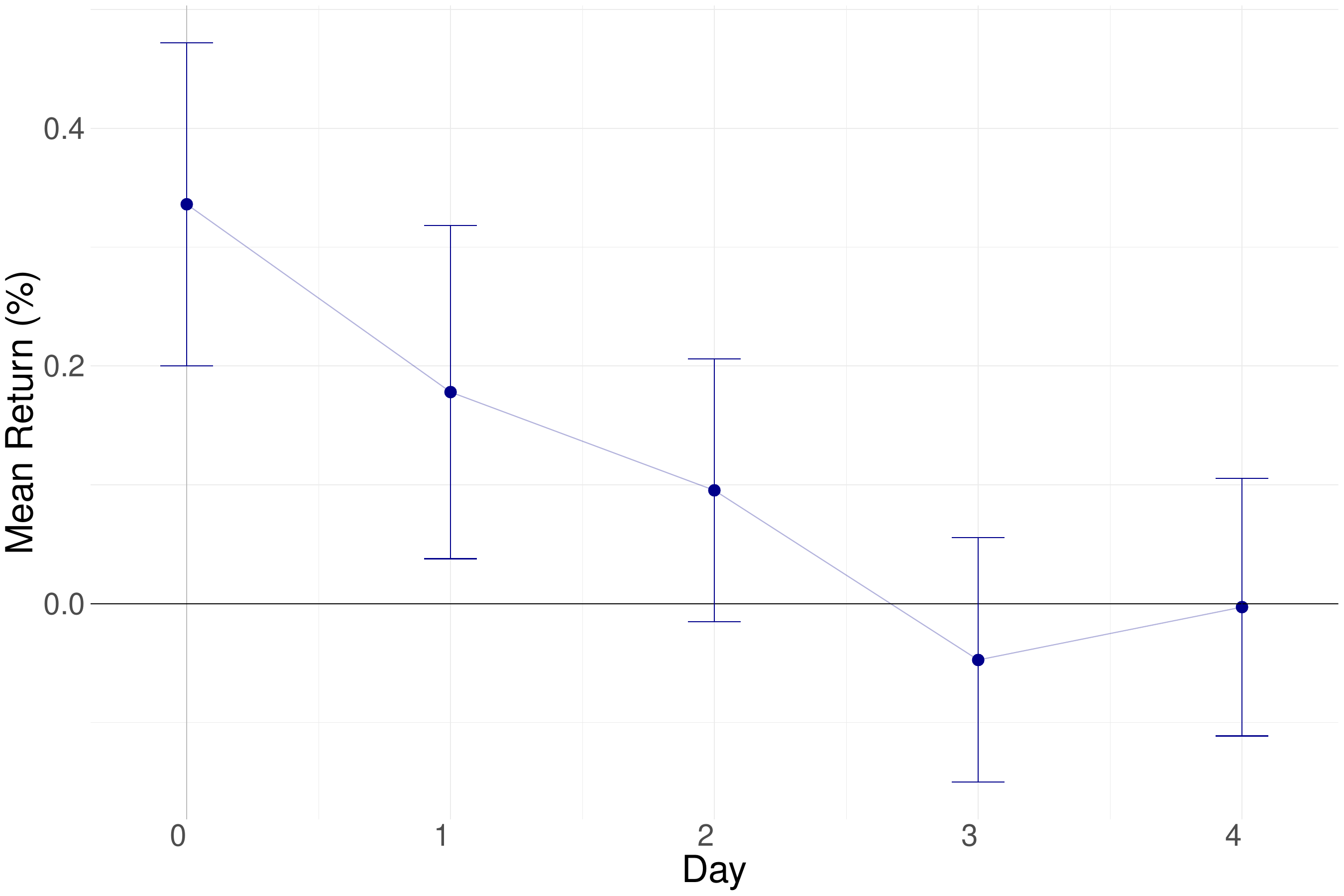}

  \label{fig:daily_returns_overnight}

  \vspace{0.5cm}
\begin{minipage}{15cm}
\small
This figure presents the average one-day holding period returns for the post-announcement drift strategy using overnight news and their 95\% confidence intervals. For overnight news released before 9 a.m. on trading day $t$ but after 4 p.m. of day $t$-1, we enter the position at the market opening of day $t$, creating a long-short portfolio that buys companies with good news and short-sells companies with bad news according to ChatGPT 4, and hold it for five trading days. The plot shows the average returns of the strategy for day $t$=0 (i.e., entering the position at the market opening of day $t$ and exiting at the same day's close) and one-day close-to-close returns for each of the next four days, day $t$+1 through $t$+4. 
\end{minipage}
\vspace{0.5cm} 
\end{figure}

\begin{figure}[htbp]
  \centering
    \caption{Returns of Intraday News Strategy Over Event Time}
          \vspace{0.5cm}
  \includegraphics[width=0.8\textwidth]{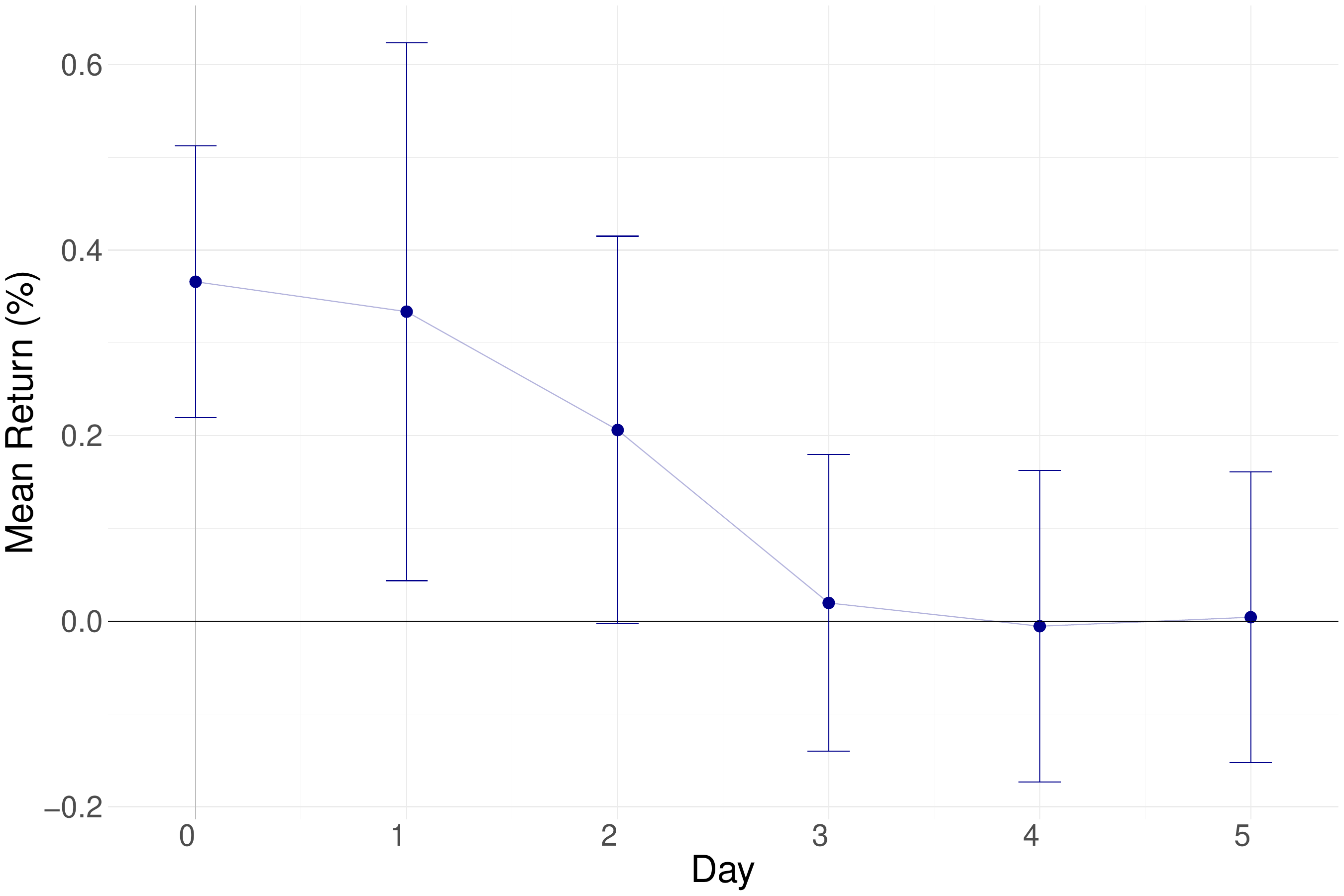}

  \label{fig:daily_returns_intraday}

  \vspace{0.5cm}
\begin{minipage}{15cm}
\small
This figure presents the average one-day holding period returns for the post-announcement drift strategy using intraday news and their 95\% confidence intervals. For intraday news released between 9:30 a.m. and 4 p.m. of trading day $t$, we enter the position 15 minutes after the news announcement on day $t$, creating a long-short portfolio that buys companies with good news and short-sells companies with bad news according to ChatGPT 4, and hold it for five full trading sessions. The plot shows the average returns of the strategy for day $t$=0 (i.e., entering the position 15 minutes after the news announcement on day $t$ and exiting at the same day's close) and one-day close-to-close returns for each of the next five days, day $t$+1 through $t$+5. 
\end{minipage}
\vspace{0.5cm} 
\end{figure}

We start with forming long-short portfolios based on GPT-4 assessment scores the same way as in Section \ref{sec:returns_overtime}. Rather than just holding the long-short strategy for the trading day after the news arrival, we hold it for five full trading days. Figure \ref{fig:daily_returns_overnight} presents the average one-day holding period returns for the strategy for overnight news and their 95\% confidence intervals. Specifically, the plot shows the average returns of the strategy for the news day $t$=0 (i.e., entering the position at the market opening and exiting at the same day's close) and one-day close-to-close returns for each of the next four days. The results show that GPT-4 scores based on news headlines can predict returns over the next two days, but not afterward. The average daily return for the overnight news strategy is 34 bps on the first trading day after the news arrival and 19 bps on the day after. We repeat the same analysis in Figure \ref{fig:daily_returns_intraday} for the strategy formed based on intraday news, entering the position 15 minutes post-release and holding it for five full trading days. Again, we find a significant return for the post-news intraday trading session (32 bps) and the day after (30 bps), but not afterward. Thus, our evidence suggests that information contained in the firm-specific news in our sample is absorbed into market prices in about two days.      

\centerline{\bfseries [Insert Figures \ref{fig:daily_returns_overnight} and \ref{fig:daily_returns_intraday} about here]}

\subsection{Market Efficiency}
Having established that price adjustment occurs within two days, we now examine whether this adjustment speed has changed over time as LLM adoption has grown. GPT models have witnessed substantial growth in their adoption since 2022 with the GPT-3.5 model being released in March 2022, and ChatGPT, fine-tuned from a GPT-3.5 model, being launched in November 2022.\footnote{See media coverage on the widespread adoption: e.g., ``ChatGPT sets record for fastest-growing user base" by Krystal Hu, \emph{Reuters}, February 2, 2023; ``ChatGPT Is The Fastest Growing App In The History Of Web Applications" by Cindy Gordon, \emph{Forbes}, February 2, 2023. Statistics from \emph{Similarweb} show that the number of monthly visits to the ChatGPT platform was approximately 152 million in November 2022 and increased all the way to over 1.5 billion by the end of 2023.} Consistent with the general trend, \citet{sheng2024aihedgefunds} document a sharp increase in 2022 in the adoption of generative AI by asset management firms such as hedge funds. Given the widespread adoption of the recent LLMs and the forecasting capabilities we document, one important question to examine is that of the potential impact of their adoption on market dynamics such as market efficiency. 

Our conceptual framework posits that LLMs can increase investors' information processing capabilities and reduce market inefficiencies. Thus, return predictability using ChatGPT's prediction score of news headlines is expected to weaken as LLMs' model size increases substantially and an increasing number of market participants use them for processing information for investment purposes. While directly testing this hypothesis is empirically challenging, we examine whether there are any changes over time in the performance of the long-short strategies formed based on the ChatGPT 4 scores. 

\begin{figure}[htbp]
  \centering
    \caption{Year-by-Year Performance of Overnight News Strategy: Sharpe Ratios}
      \vspace{0.5cm}
  \includegraphics[width=0.8\textwidth]{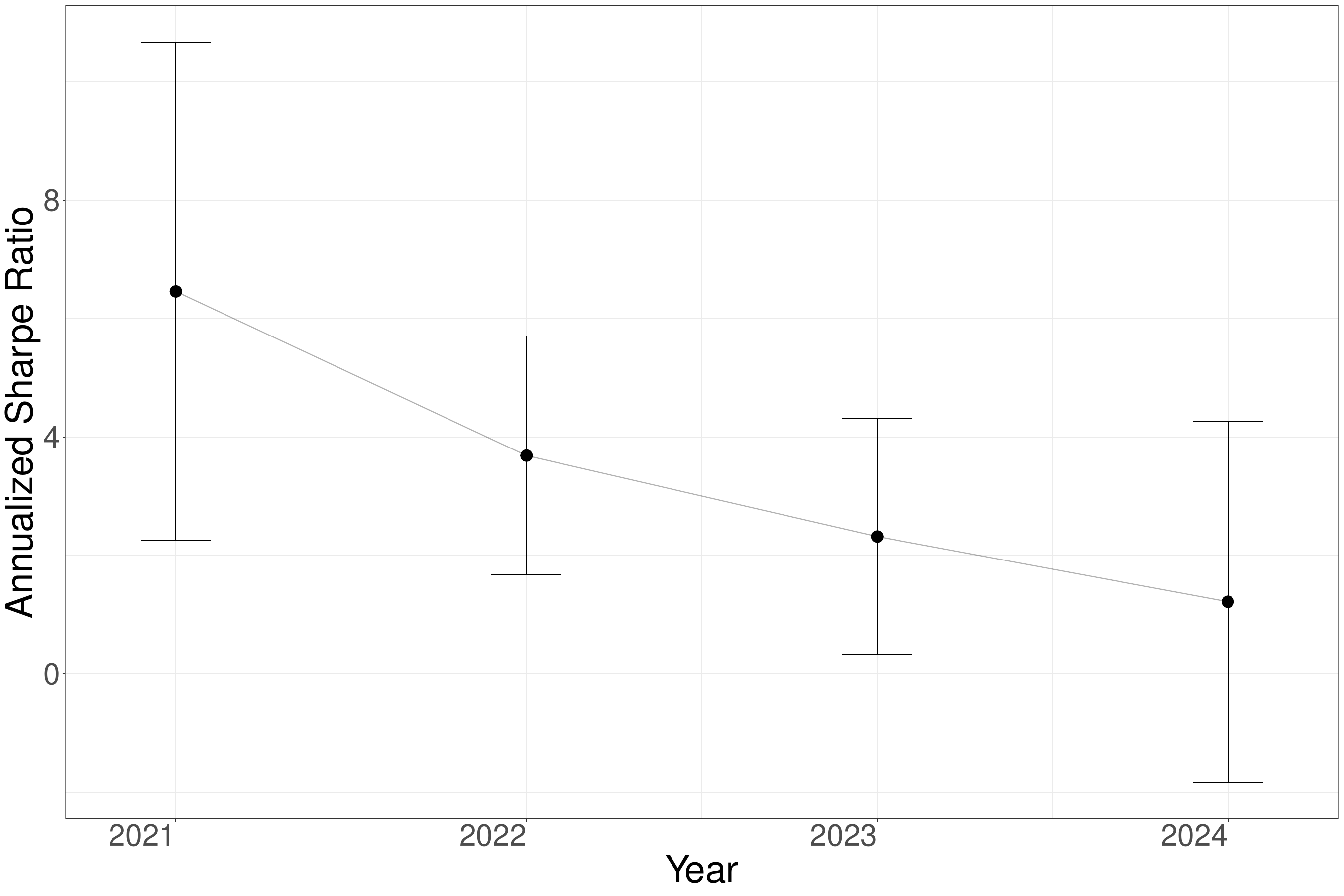}

  \label{fig:yearly_sharpe_ratio}

  \vspace{0.5cm}
\begin{minipage}{15cm}
\small
This figure shows the year-by-year Sharpe ratios of the same long-short strategy based on ChatGPT 4 as in Figure \ref{fig:cumulative_returns}. If a piece of news is released before 9 a.m. on a trading day, we enter the position at the market opening and exit at the close of the same day. If the news is announced after the market closes, we assume we enter the position at the next opening price and exit at the close of the next trading day. All the strategies are rebalanced daily. This figure presents the annualized Sharpe ratios of the strategy for four subperiods of our sample (2021Q4, 2022, 2023, and 2024 Jan.-May) and their 95\% confidence intervals calculated following \citet{lo2002statistics}. 
\end{minipage}
\vspace{0.5cm} 
\end{figure}

As shown in Figure  \ref{fig:yearly_sharpe_ratio}, there is a clear drop in the performance of the ChatGPT-based strategy based on overnight news over our sample period, during which GPT models' capabilities and adoption skyrocketed. Specifically, the annualized Sharpe ratio declines from 6.54 in 2021Q4 to 3.68 in 2022, 2.33 in 2023, and 1.22 over January-May 2024. We view this evidence as consistent with the recent proliferation of LLM technologies being associated with reduced market underreaction and improved market efficiency, as our model would predict if investors are indeed using these technologies.\footnote{For intraday news, drift occurs in two periods as shown in Figure \ref{fig:daily_returns_intraday}: from immediately after news release to market close (same-day) and close-to-close (next-day). The same-day drift strategy, while capturing most of the performance, is impractical to implement because investors cannot accurately anticipate the number of intraday news releases for individual firms, making it challenging to size the portfolio positions. The close-to-close strategy is implementable (as we do in Table \ref{tab:gpt4_portfolio_performance}) but captures only limited drift and is noisier due to the small sample size, resulting in less clear evidence of a decline in the Sharpe ratio. Nevertheless, the close-to-close strategy's Sharpe ratio remains significantly positive in 2022 and 2023, but becomes insignificant in January-May 2024 (see Figure OA2 in the Online Appendix).} While other factors, such as changing market conditions, could also contribute to this decline, the timing coincides remarkably with the proliferation of LLM technologies.

\centerline{\bfseries [Insert Figure \ref{fig:yearly_sharpe_ratio} about here]}

Moreover, we analyze the drift performance by size terciles and find asymmetric drift patterns (see Online Appendix F.6).  Following negative news, large-cap stocks (e.g., the top tercile) exhibit no drift, whereas small-cap stocks continue to drift. Since GPT-4 assessments align with initial reactions across all market sizes, this reflects genuine variation in market efficiency: large-cap markets adjust efficiently, while small-cap markets exhibit underreaction. 

\FloatBarrier

\FloatBarrier
\section{Conclusion}

In this study, we investigate the potential of ChatGPT and other LLMs in assessing the economic implications of news headlines. We document several empirical findings new to the literature. First, LLMs like GPT-4 demonstrate strong capabilities in forecasting stock price movements from news headlines without direct financial training. Using post-knowledge-cutoff headlines, GPT-4 captures initial market reactions with approximately 90\% portfolio hit rates. Second, GPT-4 scores also predict one to two days of return drift, especially for smaller stocks and negative news. Third, our interpretability framework shows that markets react efficiently to some topics (e.g., earnings reports) but underreact to others (e.g., insider transactions). Fourth, LLMs' ability to discern the economic implications of news generally increases with model size. While basic models such as GPT-1, GPT-2, and BERT exhibit limited ability, GPT-4 delivers the highest accuracy in discerning the initial reaction and the highest drift-strategy Sharpe ratio, suggesting that financial reasoning ability is an emerging capacity of complex LLMs. Fifth, we find suggestive evidence that widespread LLM adoption can enhance price efficiency. Notably, LLMs' accessibility distinguishes them from traditional ML methods requiring specialized expertise: their conversational interface substantially lowers barriers to text analysis, enabling previously inattentive investors to better process information and improving price discovery, consistent with the declining predictability we observe over time. Importantly, we develop a theoretical model that incorporates LLM technology, information capacity constraints, underreaction, and limits to arbitrage to explain all of our empirical findings. By demonstrating the value of LLMs in financial economics and uncovering what LLM assessments reveal about how markets process different types of information, our study makes a unique contribution to the literature on the applications of AI and NLP in finance. 

Our research has several implications for future studies. First, it highlights the importance of continued exploration and development of LLMs tailored explicitly for the financial industry. As AI-driven finance evolves, more sophisticated models can be designed to improve the accuracy and efficiency of financial decision-making processes. Second, our findings suggest that future research could focus more on understanding the mechanisms through which LLMs derive their reasoning power in the economics domain. By identifying the factors contributing to the success of models like ChatGPT in understanding the economic content of news and predicting stock returns, researchers can develop more targeted strategies for enhancing these models and maximizing their utility in finance. Third, as LLMs become more prevalent in the financial industry, it is essential to investigate their potential impact on market dynamics, including price formation, information dissemination, and market stability. 
 
Lastly, future studies could explore the integration of LLMs with other machine learning techniques and quantitative models to create hybrid systems that combine the strengths of different approaches. By leveraging the complementary capabilities of various methods, researchers can further enhance the information-processing capacity and predictive power of AI-driven models in financial economics.

%In short, our study demonstrates the value of ChatGPT in predicting stock market returns. It paves the way for future research on the applications and implications of LLMs in the financial industry. As the field of AI-driven finance continues to expand, the insights gleaned from this research can help guide the development of more accurate, efficient, and responsible models that enhance the performance of financial decision-making processes.

\newpage
\FloatBarrier
\printbibliography

\newpage
\FloatBarrier
%\section*{Figures}

%%%%%%%%%%%%%%%%%%%%%%%%%%%%%%%%%%%%

% \begin{figure}[htbp]
%   \centering
%     \caption{Year-by-Year Performance of Overnight News Strategy: Average Daily Returns}
%       \vspace{0.5cm}
%   \includegraphics[width=0.8\textwidth]{Figures/MarketEfficiency/MeanReturn.pdf}

%   \label{fig:yearly_daily_ret}

%   \vspace{0.5cm}
% \begin{minipage}{15cm}
% \small
% This figure shows the year-by-year performance of the same long-short strategy based on ChatGPT 4 as in Figure \ref{fig:cumulative_returns}. If a piece of news is released before 9 a.m. on a trading day, we enter the position at the market opening and exit at the close of the same day. If the news is announced after the market closes, we assume we enter the position at the next opening price and exit at the close of the next trading day. All the strategies are rebalanced daily. This figure presents the average daily returns of the strategy for four subperiods of our sample (2021 Oct.-Dec., 2022, 2023, and 2024 Jan.-May) and their 95\% confidence intervals. 
% \end{minipage}
% \vspace{0.5cm} 
% \end{figure}

\FloatBarrier

\newpage
\FloatBarrier

\newpage
%\section*{Tables}

% Note: Data observation counts are provided in Appendix Table \ref{tab:data_counts}

\begin{table}[htbp]
  \centering
  \caption{Performance of Portfolios based on ChatGPT Assessments}
  \vspace{0.5cm}
  \begin{minipage}{15cm}
  \small
This table reports the performance of daily-rebalanced portfolios based on GPT-4 assessments of news headlines (positive, negative, and neutral sentiment). The analysis aggregates multiple headlines at the firm-day level for portfolio construction. Overnight and intraday news are analyzed separately due to the differences in portfolio formation timing.  %A detailed breakdown of observation counts at each filtering stage is provided in Appendix Table \ref{tab:data_counts}.
The table shows performance measures of Long-Short Portfolio (buying positive predictions and selling negative predictions when both legs have at least two firms; otherwise entering just one leg), Long-Only Portfolio (buying only positive predictions), and Short-Only Portfolio (selling only negative predictions). Note that neutral predictions are included in the total count but do not generate trading positions. Hit Rate shows the percentage of trading days with positive returns. Mean Return shows the average daily return. Sharpe Ratio is annualized and calculated only for drift strategies (initial reactions cannot be traded without advance knowledge of news timing). For overnight news released before 9 a.m. on a trading day or after 4 p.m. of the previous day, the initial reaction is measured as the return from the previous close to the opening of the trading day, and the subsequent drift is measured as the return from the market opening to the close of the same day. For intraday news released between 9:30 a.m. and 4 p.m., the initial reaction is measured as the return from the previous close to 15 minutes after the news release, and the drift is measured as the return from the close of the news day to the close of the next trading day. The bottom two rows show the following summary statistics: Firm-Day Observations (number of firm-day combinations with GPT-4 assessments after aggregating multiple headlines per firm-day, including neutral predictions) and Trading Days (number of days portfolios were evaluated). 

  \end{minipage}
  \vspace{0.5cm}
  \label{tab:gpt4_portfolio_performance}
  \centering
  % Tabular content for Table 1B: GPT-4 Portfolio Trading Performance (All Predictions)
\begin{tabular}{lrrrr}
\toprule
& \multicolumn{2}{c}{Overnight News} & \multicolumn{2}{c}{Intraday News} \\
\cmidrule(lr){2-3} \cmidrule(lr){4-5}
Metric & Initial Reaction & Drift & Initial Reaction & Drift \\
\midrule
\textbf{Long-Short Portfolio} & & & & \\
Hit Rate (\%) &
93.28 & 58.06 & 88.78 & 54.67 \\
Mean Return (\%) &
3.06 & 0.34 & 4.44 & 0.50 \\
Sharpe Ratio &   &
2.97 &   & 2.63 \\
\textbf{Long-Only Portfolio} & & & & \\
Hit Rate (\%) &
83.28 & 50.90 & 76.05 & 46.00 \\
Mean Return (\%) &
1.27 & 0.08 & 1.33 & 0.19 \\
Sharpe Ratio &   &
0.78 &   & 1.37 \\
\textbf{Short-Only Portfolio} & & & & \\
Hit Rate (\%) &
79.40 & 53.58 & 77.57 & 48.67 \\
Mean Return (\%) &
1.79 & 0.26 & 3.11 & 0.31 \\
Sharpe Ratio &   &
2.01 &   & 1.71 \\
\midrule
Firm-Day Observations &
105,742 & 105,742 & 26,109 & 26,109 \\
Trading Days &
670 & 670 & 526 & 526 \\
\bottomrule
\end{tabular}

  \end{table}

\FloatBarrier
  %%%%%%%%%%%%%%%%%%%%%%%%%%%%%%%%%%%%%%%%%%%%%%%%%%%%
  % 2024 Extended Data Regression Tables
  %%%%%%%%%%%%%%%%%%%%%%%%%%%%%%%%%%%%%%%%%%%%%%%%%%%%

  \FloatBarrier  % Force all floats to be placed before landscape
  \newpage
  %\begin{landscape}
  \centering
  \captionof{table}{Regression Analysis: Market Information Processing}
  \label{tab:regression_analysis}
  \vspace{0.5cm}
  \begin{minipage}{15cm}
  \footnotesize
  This table reports regression results comparing market information processing across timing dimensions. Panel A shows results for overnight news, and Panel B shows results for intraday news. Within each panel, the first four columns show initial reactions (\textit{ret\_n}) and the last four columns show price drift (\textit{ret\_o}). Initial reaction and price drift are defined the same way as in Table \ref{tab:gpt4_portfolio_performance}. \emph{Small} is an indicator variable for stocks below the 20th percentile of NYSE size breakpoints the previous day. All regressions follow the form \(r_{i,t} = a_i + b_t + \gamma'x_{i,t} + \varepsilon_{i,t}\), where \(r_{i,t}\) is the return in percentage points, \(a_i\) and \(b_t\) are firm and time fixed effects, and \(x_{i,t}\) contains prediction scores from GPT-4 and RavenPack. The analysis uses the full sample of individual headlines, including all positive, negative, and neutral GPT-4 assessments. Note that the initial reaction columns have fewer observations than the drift columns in Panel B because the initial reaction measure (from previous close to 15 minutes after news release) requires intraday price data from TAQ, different from the drift measure (from close-to-close of the next day). Standard errors are double-clustered by date and firm.
  \end{minipage}
  \vspace{0.5cm}

  % ===== Panel A =====
  \begin{subtable}{\linewidth}
    \centering
    \adjustbox{max width=0.9\linewidth}{%
      % Tabular content for overnight regression table (Panel A)
\resizebox{\textwidth}{!}{%
\begin{tabular}{l cccc cccc}
\hline \hline
& \multicolumn{8}{c}{\textbf{Panel A: Overnight News}} \\
\cmidrule(lr){2-9}
& \multicolumn{4}{c}{Initial Reaction} & \multicolumn{4}{c}{Drift} \\
\cmidrule(lr){2-5} \cmidrule(lr){6-9}
& (1) & (2) & (3) & (4) & (1) & (2) & (3) & (4) \\
\hline
GPT-4 Score &
1.326*** & 1.018*** & 1.083*** & & 0.161*** & 0.158*** & 0.087*** & \\
&
(31.104) & (20.396) & (28.558) & & (6.798) & (5.351) & (4.089) & \\
RavenPack & &
1.141*** & & & & 0.010 & & \\
& &
(10.597) & & & & (0.150) & & \\
Small & & &
0.906*** & & & & 0.816*** & \\
& & &
(4.370) & & & & (4.435) & \\
GPT-4 × Small & & &
1.326*** & & & & 0.404*** & \\
& & &
(9.196) & & & & (4.745) & \\
\hline
Observations & 129,760 & 129,760 & 129,760 & 129,760 & 129,760 & 129,760 & 129,760 & 129,760 \\
Adj. R\textsuperscript{2} & 0.084 & 0.084 & 0.084 & 0.084 & 0.086 & 0.086 & 0.086 & 0.086 \\
\hline \hline
\multicolumn{9}{l}{\footnotesize Standard errors clustered by date and firm. All models include date and firm fixed effects.} \\
\multicolumn{9}{l}{\footnotesize t-statistics in parentheses. *** p$<$0.01, ** p$<$0.05, * p$<$0.1} \\
\end{tabular}%
}
}
  \end{subtable}

  \vspace{0.8cm}

  % ===== Panel B =====
  \begin{subtable}{\linewidth}
    \centering
   \adjustbox{max width=0.9\linewidth}{%
      % Tabular content for intraday regression table (Panel B)
\resizebox{\textwidth}{!}{%
\begin{tabular}{l cccc cccc}
\hline \hline
& \multicolumn{8}{c}{\textbf{Panel B: Intraday News}} \\
\cmidrule(lr){2-9}
& \multicolumn{4}{c}{Initial Reaction} & \multicolumn{4}{c}{Drift} \\
\cmidrule(lr){2-5} \cmidrule(lr){6-9}
& (1) & (2) & (3) & (4) & (1) & (2) & (3) & (4) \\
\hline
GPT-4 Score &
1.748*** & 1.450*** & 1.643*** & & 0.160*** & 0.197*** & 0.049 & \\
&
(13.852) & (8.551) & (18.895) & & (3.544) & (3.054) & (1.343) & \\
RavenPack & &
1.169*** & & & & -0.145 & & \\
& &
(3.535) & & & & (-0.956) & & \\
Small & & &
-1.666** & & & & 0.638 & \\
& & &
(-2.256) & & & & (1.640) & \\
GPT-4 × Small & & &
0.880 & & & & 0.683*** & \\
& & &
(1.165) & & & & (3.293) & \\
\hline
Observations & 24,531 & 24,531 & 24,531 & 24,531 & 29,184 & 29,184 & 29,184 & 29,184 \\
Adj. R\textsuperscript{2} & 0.342 & 0.342 & 0.342 & 0.342 & 0.177 & 0.177 & 0.177 & 0.177 \\
\hline \hline
\multicolumn{9}{l}{\footnotesize Standard errors clustered by date and firm. All models include date and firm fixed effects.} \\
\multicolumn{9}{l}{\footnotesize t-statistics in parentheses. *** p$<$0.01, ** p$<$0.05, * p$<$0.1} \\
\end{tabular}%
}

      }%
  \end{subtable}
  %\end{landscape}

%%%%%%%%%%%%%%%%%%%%%%%%%%
% Intraday - All Stocks
%%%%%%%%%%%%%%%%%%%%%%%%%%
%\begin{landscape}
\newpage    
\begin{table}
\caption{Regression of Intraday News Returns on ChatGPT Scores}
\centering
\vspace{0.5cm}
\begin{minipage}{15cm}
\small
This table reports the results of running regressions of the form \(r_{i,t+1} = a_i + b_t + \gamma'x_{i,t} + \varepsilon_{i,t+1}\), where \(r_{i,t+1}\) is the drift-strategy return based on intraday news, and \(x_{i,t}\) refers to the prediction score from GPT-4. The analysis uses individual intraday headlines (not aggregated at the firm-day level) and requires complete return data across all three time windows. For intraday news released during trading hours, we form three strategies with different entry and exit times, each generating a return over a particular window after the news release. The first strategy ($Return1-15 min$) enters the position one minute after the news announcement and exits 15 minutes after the news; the second one ($Return 15 min - Close$) enters the position 15 minutes after the news and exits at the close of day $t$; the third one ($Return Close-to-close$) enters the position at the close of day $t$ and exits at the next day's close. Standard errors are double clustered by date and firm. All models include firm and time fixed effects. We report t-statistics in parentheses. *** p$<$0.01, ** p$<$0.05, * p$<$0.1.
\end{minipage}
\vspace{0.5cm} 
%\vspace{0.5cm} 
\label{tab:intraday_average_by_score}
\centering

\vspace{0.25cm} 

\begin{tabular}{llll}
\hline
& Return 1-15 min & Return 15 min - Close & Return Close-to-close \\ \hline
GPT-4 Score & 0.027 & 0.189*** & 0.116** \\
& (1.322) & (5.118) & (2.306) \\
Num.Obs. & 24,541 & 24,541 & 24,541 \\
R2 Adj. & 0.205 & 0.249 & 0.177 \\
Std.Errors & by: date \& permno & by: date \& permno & by: date \& permno \\
FE: date & X & X & X \\
FE: permno & X & X & X \\
\hline
\end{tabular}

\end{table}
%\end{landscape}

\newpage
\begin{table}
% \begin{landscape} 
  \centering
  \captionsetup{type=table,skip=10pt} % space below caption
  \caption{Decomposition of Market Information Processing Timeline}
  \label{tab:decomposition_analysis}
  \begin{minipage}{15cm}
  \footnotesize
  This table decomposes market information processing into three components using GPT-4 assessments as a consistent lens. The analysis uses individual overnight news headlines with directional predictions (excluding neutral scores), clustered into topics using BERTopic. The left panel shows raw market returns (Init, Drift, Total) for each news topic. The initial reaction (Init) and the price drift (Drift) are defined the same way as in Table \ref{tab:gpt4_portfolio_performance}, and the total reaction (Total) is the sum of the initial reaction and the subsequent drift. The right panel shows GPT strategy returns. \textit{GPT} shows the average GPT-4 score. \textit{Init×GPT} measures initial market-GPT alignment (GPT score × initial return), capturing immediate sentiment agreement. \textit{Drift×GPT} measures subsequent market adjustment toward or away from GPT-4's view (GPT score × drift return), revealing market learning or correction patterns. \textit{Total×GPT} shows ultimate vindication of GPT-4's assessment over the full cycle. The patterns reveal how different information types are processed by markets in real-time. Standard errors are double-clustered by firm and date. Significance: *** p$<$0.001, ** p$<$0.01, * p$<$0.05, + p$<$0.1.
  \end{minipage}
  \vspace{0.8cm}
  \adjustbox{max width=1.0\textwidth}{
  \begin{tabular}{lcccccccc}
\hline
 & & & \multicolumn{3}{c}{Market Returns (bps)} & \multicolumn{3}{c}{GPT Strategy (bps)} \\ \cline{4-9}
Topic & N & GPT & Init & Drift & Total & Init×GPT & Drift×GPT & Total×GPT \\ \hline
Insider Stock Transactions & 16,790 & 0.39 & 59 & 13 & 72 & 71.4*** & 26.3*** & 97.6*** \\
Earnings and Revenue Announcements & 11,440 & 0.26 & -26 & 4 & -22 & 336.7*** & -4.7 & 332.0*** \\
Executive Stock Disposals &  9,852 & -0.82 & -8 & -2 & -10 & 14.8** & 9.1 & 23.9** \\
Strategic Partnership Announcements &  7,660 & 0.87 & 76 & 0 & 76 & 92.7*** & 4.3 & 97.0*** \\
Dividend Announcements &  6,930 & 0.92 & 38 & 19 & 57 & 45.5*** & 22.3*** & 67.7*** \\
Clinical Trial Announcements &  6,357 & 0.89 & 203 & -16 & 187 & 308.5*** & -11.4 & 297.1*** \\
Stock Offering Announcements &  4,915 & 0.65 & -11 & -24 & -36 & 217.6*** & 13.1 & 230.7*** \\
Executive Leadership Appointments &  4,572 & 0.89 & -5 & -16 & -22 & 29.3** & -4.3 & 25.1* \\
Quarterly Earnings Reports &  4,439 & 0.93 & 80 & -15 & 66 & 125.2*** & -3.2 & 122.1*** \\
Investor Conference Participation &  3,265 & 0.97 & 21 & 13 & 34 & 29.2*** & 11.2 & 40.4** \\
Healthcare Conference Participation &  3,066 & 0.95 & 76 & 21 & 97 & 103.5*** & 34.2* & 137.7*** \\
Stock Rating Adjustments &    889 & 0.31 & 40 & 38 & 78 & 237.6*** & -9.4 & 228.1*** \\
Securities Class Action Lawsuits &    701 & -0.51 & 21 & -3 & 19 & 16.1 & 3.7 & 19.8 \\
Credit Rating Adjustments &    436 & 0.73 & 20 & 8 & 27 & 12.0 & 18.1+ & 30.1* \\
Index Inclusion Announcements &    213 & 0.43 & -28 & 37 & 8 & 59.9* & -34.7 & 25.2 \\
Retail Store Openings &    175 & 1.00 & 16 & 7 & 23 & 15.6* & 7.1 & 22.8+ \\
\hline\hline
\textbf{Overall} & 81,700 & 0.47 & 41 & 2 & 43 & 129.7*** & 9.3*** & 139.0*** \\
\hline
\end{tabular}

  }
  % \end{landscape}
\end{table}
  
%%%%%%%%%%%%%%%%%%%%%%%%%%
% All Stocks
%%%%%%%%%%%%%%%%%%%%%%%%%%

% %%%%%%%%%%%%%%%%%%%%%%%%%%
% % Small Stocks
% %%%%%%%%%%%%%%%%%%%%%%%%%%
% \begin{landscape}
% \begin{table}
% \ContinuedFloat

% \vspace{0.5cm} 
% % \label{tab:average_by_score}
% \centering

% Panel B. Small Stocks 
% \vspace{0.25cm} 

% \input{LatexTables/regression_summary_small}
% \end{table}
% \end{landscape}

% %%%%%%%%%%%%%%%%%%%%%%%%%%
% % Large Stocks
% %%%%%%%%%%%%%%%%%%%%%%%%%%
% \newpage

% \begin{landscape}
% \begin{table}
% \ContinuedFloat
% \vspace{0.5cm} 
% % \label{tab:average_by_score}
% \centering

% Panel C. Non-Small Stocks 
% \vspace{0.25cm} 

% \input{LatexTables/regression_summary_non_small}

% \end{table}
% \end{landscape}

%%%%%%%%%%%%%%%%%%%%%%%%%%%%%%%%%%%%%%%%%%%%%%%%%%%%
% Returns
%%%%%%%%%%%%%%%%%%%%%%%%%%%%%%%%%%%%%%%%%%%%%%%%%%%%
\begin{landscape}
    
\begin{table}[ht]
\caption{Performance Comparison Across LLMs} 
\vspace{0.5cm}
\begin{minipage}{22cm}
\small
This table reports statistics for the portfolios formed based on different LLMs using overnight news headlines. The analysis uses portfolio-level data aggregated at the firm-day level. We provide an overview of the different LLMs in Appendix A of the paper. \emph{HR-I} is the hit rate for initial reaction (the percentage of days with positive returns for initial market reactions to overnight news), and \emph{HR-D} refers to the hit rate for the subsequent drift (the percentage of days with positive returns for post-announcement drift). The column $Sharpe_{LS}$ presents the annualized Sharpe ratio of the long-short portfolio for each model, and the next four columns contain the daily average returns in percentage points for the long-short, long, neutral, and short portfolios, respectively. We also report the average number of stocks in the long ($N_+$) and in the short ($N_-$) legs. Columns $\alpha_{FF5}$, t $\alpha_{FF5}$, and $R^2_{FF5}$ show the daily alpha (in percentage points), t-statistic, and R-squared (in percent) with respect to the 5-factor model of \citet{Fama2015AModel}. The table reports results for all U.S. common stocks with at least one news headline covering the firm. 
\end{minipage}
\vspace{0.5cm} 
\label{tab:average_by_score}
\centering

%Panel A. Full Sample: All Stocks 
\vspace{0.25cm} 

\begin{tabular}{lrrrrrrrrrrrrr}
  \hline
% latex table generated in R 4.5.1 by xtable 1.8-4 package
% Mon Oct 20 13:47:02 2025
Model & HR-I & HR-D & $\text{Sharpe}_{LS}$ &  $\mu_{LS}$ &  $\mu_{+}$ &  $\mu_0$ & $\mu_{-}$ & $N_+$ & $N_-$ & $\alpha_{FF5}$ & t $\alpha_{FF5}$ & $R^2_{FF5}$ (\%) \\ 
  \hline
GPT-4 & 0.93 & 0.58 & 2.97 & 0.34 & 0.08 & -0.20 & -0.26 & 69 & 20 & 0.33 & 4.62 & 0.49 \\ 
  GPT-3.5 & 0.93 & 0.56 & 1.66 & 0.29 & 0.05 & -0.09 & -0.24 & 48 & 6 & 0.28 & 2.65 & 1.05 \\ 
  FinBERT & 0.90 & 0.48 & -0.33 & -0.07 & -0.15 & -0.04 & -0.09 & 21 & 8 & -0.07 & -0.58 & 0.91 \\ 
  DistilBart & 0.87 & 0.56 & 1.26 & 0.14 & -0.03 & 0.04 & -0.17 & 112 & 16 & 0.15 & 2.16 & 1.29 \\ 
  BART-Large & 0.86 & 0.57 & 1.05 & 0.12 & -0.03 & 0.00 & -0.15 & 110 & 19 & 0.14 & 1.97 & 1.62 \\ 
  Llama2-70b & 0.86 & 0.53 & 0.97 & 0.14 & -0.04 & 0.15 & -0.18 & 126 & 7 & 0.15 & 1.58 & 0.76 \\ 
  Llama2-13b & 0.86 & 0.52 & 0.78 & 0.11 & -0.04 & 0.01 & -0.15 & 123 & 8 & 0.11 & 1.27 & 1.07 \\ 
  BERT-Large & 0.70 & 0.50 & 1.05 & 0.16 & -0.06 & -0.05 & -0.22 & 124 & 2 & 0.18 & 2.01 & 5.61 \\ 
  Llama2-7b & 0.66 & 0.48 & -0.43 & -0.05 & -0.05 & 0.03 & -0.01 & 135 & 0 & -0.03 & -0.48 & 35.60 \\ 
  BERT & 0.64 & 0.46 & -0.47 & -0.06 & -0.10 & -0.04 & -0.04 & 32 & 0 & -0.03 & -0.50 & 23.40 \\ 
  GPT-2 & 0.55 & 0.53 & -0.11 & -0.01 & -0.05 & -0.07 & -0.03 & 82 & 18 & -0.01 & -0.20 & 0.48 \\ 
  GPT-1 & 0.53 & 0.48 & -0.64 & -0.07 & -0.05 & -0.11 & 0.02 & 101 & 17 & -0.07 & -1.05 & 0.20 \\ 
   \hline

\end{tabular}

\end{table}
\end{landscape}

\begin{landscape}
\begin{table}[ht]
\centering
\caption{Model Performance by News Complexity and News Type}
\vspace{0.5cm}
\begin{minipage}{22cm}
\small
This table reports portfolio performance metrics for different models using overnight news headlines, decomposed by headline complexity and news types. For each category, we report HR-I (hit rate for initial reaction), HR-D (hit rate for drift), and Sharpe (annualized Sharpe ratio of the long-short portfolio). High Complexity includes headlines above the median Flesch-Kincaid Readability Score, while Low Complexity includes headlines below the median. The median is computed each day for overnight news. News Articles and Press Releases show performance by news source type. Complexity scores and news types are determined at the individual headline level, then portfolios are constructed by aggregating at the firm-day level. Models are sorted by overall HR-I (highest to lowest).
\end{minipage}
\vspace{0.5cm}
\label{tab:sharpe_by_complexity}
\centering
\begin{tabular}{lccccccccccccc}
\hline
 & \multicolumn{3}{c}{High C} & \multicolumn{3}{c}{Low C} & \multicolumn{3}{c}{News Articles} & \multicolumn{3}{c}{Press Releases} \\ \cline{2-4}\cline{5-7}\cline{8-10}\cline{11-13}
Model & HR-I & HR-D & SR & HR-I & HR-D & SR & HR-I & HR-D & SR & HR-I & HR-D & SR \\ \hline
GPT-4 & 0.89 & 0.55 &  1.47 & 0.91 & 0.57 &  2.32 & 0.91 & 0.59 &  2.23 & 0.91 & 0.56 &  1.96 \\
GPT-3.5 & 0.87 & 0.53 &  1.13 & 0.88 & 0.55 &  1.39 & 0.86 & 0.55 &  1.55 & 0.90 & 0.53 &  0.66 \\
FinBERT & 0.84 & 0.51 &  0.29 & 0.81 & 0.46 & -0.50 & 0.74 & 0.49 & -0.13 & 0.88 & 0.48 &  0.20 \\
DistilBart & 0.77 & 0.49 &  0.22 & 0.83 & 0.54 &  1.04 & 0.89 & 0.58 &  1.52 & 0.79 & 0.55 &  0.57 \\
BART-Large & 0.79 & 0.52 &  0.45 & 0.83 & 0.56 &  1.23 & 0.88 & 0.58 &  1.60 & 0.79 & 0.55 &  1.30 \\
Llama2-70b & 0.79 & 0.50 &  1.19 & 0.83 & 0.53 &  0.82 & 0.86 & 0.55 &  1.13 & 0.80 & 0.53 &  1.15 \\
Llama2-13b & 0.76 & 0.51 &  0.14 & 0.82 & 0.52 &  0.86 & 0.84 & 0.55 &  0.89 & 0.79 & 0.52 &  0.64 \\
BERT-Large & 0.66 & 0.51 &  1.26 & 0.67 & 0.49 & -0.36 & 0.71 & 0.52 &  0.41 & 0.66 & 0.48 &  0.72 \\
Llama2-7b & 0.65 & 0.48 &  0.01 & 0.62 & 0.49 & -0.78 & 0.66 & 0.51 &  0.03 & 0.64 & 0.47 & -0.17 \\
BERT & 0.65 & 0.46 & -0.35 & 0.59 & 0.49 & -0.08 & 0.59 & 0.46 &  0.70 & 0.62 & 0.47 & -0.38 \\
GPT-2 & 0.54 & 0.50 & -0.33 & 0.50 & 0.51 & -0.01 & 0.55 & 0.51 &  1.02 & 0.55 & 0.52 & -0.24 \\
GPT-1 & 0.52 & 0.50 & -0.14 & 0.53 & 0.48 & -1.24 & 0.56 & 0.50 & -0.31 & 0.54 & 0.48 & -0.02 \\
\hline
\end{tabular}

\end{table}
\end{landscape}

\newpage

\begin{landscape}
% Analysis: GPT-4 vs GPT-3.5 Decomposition Analysis
\begin{table}[ht]
\centering
\caption{Model Size Comparison Analysis by News Topic}
\centering
\vspace{0.5cm}
\begin{minipage}{22cm}
\small
This table shows the comparative advantage of larger models over smaller ones within the same family across news topics by displaying the difference in their predictive accuracy in basis points. The analysis uses the same filtered dataset as in Table \ref{tab:decomposition_analysis}: individual overnight news headlines with directional predictions only (excluding neutral scores), clustered into topics using BERTopic. For each topic, we show the sample size (N) and the differential performance: positive values indicate the larger model outperforms the smaller one, while negative values indicate the smaller model outperforms the larger one. The left panel shows GPT-4 vs GPT-3.5 differences, while the right panel shows Llama2-70b vs Llama2-7b differences. \textit{Init×(LLM1-LLM2)} measures the difference in initial market reaction prediction between the two LLMs, while \textit{Drift×(LLM1-LLM2)} captures the difference in subsequent price drift prediction. Topic names are generated using GPT-4o for interpretability. Standard errors are double-clustered by firm and date. Significance: *** p$<$0.001, ** p$<$0.01, * p$<$0.05, + p$<$0.1.
\end{minipage}
\vspace{0.5cm}
\label{tab:decomposition_gpt4_vs_gpt3}
\begin{tabular}{lcccccc}
\hline
 & & \multicolumn{2}{c}{GPT-4 vs GPT-3.5 (bp)} & \multicolumn{2}{c}{Llama 70B vs 7B (bp)} \\ \cline{3-4}\cline{5-6}
Topic & N & Init×(G4 - G3) & Drift×(G4 - G3) & Init×(L70 - L7) & Drift×(L70 - L7) \\ \hline
Insider Stock Transactions & 16,790 & 25.2*** & 13.8*** & 7.2** & 4.4* \\
Earnings and Revenue Announcements & 11,440 & 15.4*** & 2.4 & 315.7*** & -6.8 \\
Executive Stock Disposals &  9,852 & 11.3** & 4.8 & 3.2* & 3.0 \\
Strategic Partnership Announcements &  7,660 & 15.4*** & 1.6 & 12.9*** & 3.1+ \\
Dividend Announcements &  6,930 & 0.2 & 0.9 & 6.3** & 1.6 \\
Clinical Trial Announcements &  6,357 & 15.4** & 4.7 & 59.7*** & 0.1 \\
Stock Offering Announcements &  4,915 & -18.5 & 3.8 & 2.6 & 5.2 \\
Executive Leadership Appointments &  4,572 & 9.5+ & -8.7 & 28.3** & 10.5** \\
Quarterly Earnings Reports &  4,439 & 0.7 & -0.4 & 35.4*** & 10.0+ \\
Investor Conference Participation &  3,265 & 11.8* & 4.7 & 5.7* & -1.0 \\
Healthcare Conference Participation &  3,066 & 19.2*** & 20.4+ & 14.9** & 8.6* \\
Stock Rating Adjustments &    889 & 13.9 & -1.9 & 139.6*** & -12.6 \\
Securities Class Action Lawsuits &    701 & -0.5 & -5.2 & -7.8 & 5.1 \\
Credit Rating Adjustments &    436 & 0.3 & -2.1 & -8.1 & 9.7 \\
Index Inclusion Announcements &    213 & 7.7+ & -12.0 & 85.0+ & -76.5 \\
Retail Store Openings &    175 & 1.8*** & 7.4*** & -  &  - \\
\hline\hline
\textbf{Overall} & 81,700 & 12.2*** & 4.9*** & 58.6*** & 2.2*** \\
\hline
\end{tabular}

\end{table}
\end{landscape}

%%%%%%%%%%%%%%%%%%%%%%%%%%%%%%%%%%%%%%%%%%%%%%%%%%%%
% GPT-4 vs Embeddings Focused Analysis (Extended Data)
%%%%%%%%%%%%%%%%%%%%%%%%%%%%%%%%%%%%%%%%%%%%%%%%%%%%
\begin{landscape}
\begin{table}[ht]
  \centering
  \caption{GPT-4 vs Text Embeddings: Long-Short Portfolio Performance}
  \label{tab:gpt4_embeddings_comparison}
  \vspace{0.5cm}
  \begin{minipage}{22cm}
  \small
  This table compares the GPT-4 conversational prompting approach against text embedding models for predicting stock returns following news announcements. The analysis uses portfolio-level data aggregated at the firm-day level for overnight news. We evaluate two embedding approaches: MiniLM and OpenAI embeddings. Both embedding models use multinomial logistic regression with classification to match GPT-4's methodology. Training employs a rolling window with periodic retraining. The analysis focuses on overnight news with separate models trained for initial reaction (\textit{ret\_n}) and post-announcement drift (\textit{ret\_o}). Hit Rate shows the percentage of days with positive portfolio returns. Mean Return is the average daily return. Sharpe Ratio is annualized and shown only for drift strategies, as initial reactions cannot be traded without advance knowledge of news timing. All portfolios are long-short strategies, equally weighted and rebalanced daily.
  \end{minipage}
  \vspace{0.5cm}
  \centering
  % Comprehensive 4-Scenario Table: Overnight vs Intraday, Reaction vs Drift
\begin{tabular}{l|cc|ccc|cc|ccc}
\hline
& \multicolumn{5}{c|}{Overnight News (N=105,742)} & \multicolumn{5}{c}{Intraday News (N=26,109)} \\
& \multicolumn{2}{c|}{Reaction} & \multicolumn{3}{c|}{Drift} & \multicolumn{2}{c|}{Reaction} & \multicolumn{3}{c}{Drift} \\
Model & Hit (\%) & Mean (\%) & Hit (\%) & Mean (\%) & Sharpe & Hit (\%) & Mean (\%) & Hit (\%) & Mean (\%) & Sharpe \\
\hline
GPT-4 & 93.28 & 3.06 & 58.06 & 0.34 & 2.97 & 88.78 & 4.44 & 54.67 & 0.50 & 2.63 \\
MiniLM & 80.15 & 1.12 & 56.57 & 0.11 & 1.42 & 77.76 & 2.58 & 50.17 & 0.13 & 0.79 \\
OpenAI & 84.63 & 1.37 & 59.85 & 0.23 & 2.53 & 83.65 & 3.32 & 52.17 & 0.29 & 1.71 \\
\hline
\end{tabular}

  \end{table}
\end{landscape}

\FloatBarrier

\newpage

\appendix

\newpage
\begin{center}
\section*{Appendices}
\end{center}

\section{Model Summaries}

\label{appendix:summaries}

%\begin{center}
%\section*{Appendix B: Model Summaries} \label{appendix:summaries}  
%\end{center}
\begin{onehalfspacing}
In this section, we present an overview of the 12 different models that we study in this paper.\footnote{We do not have Claude or Gemini in our LLM list because our sample (Oct 2021–May 2024) requires models whose knowledge cutoffs predate the period and can be applied retroactively to all historical headlines; the earliest available Claude 2 (July 2023) and Gemini 1.0 (Dec 2023) versions are no longer accessible via API. While Llama2's knowledge cutoff (September 2022) does not fully predate our sample start, we include it because the Llama2 family (7B, 13B, 70B parameters) provides critical within-family comparisons with identical architecture and training data. Since all three Llama2 variants share the same knowledge cutoff, any potential in-sample bias affects them equally, allowing us to cleanly isolate the effect of model scale on performance.} We order them primarily by their release date.
\hfill \break

\noindent \textbf{Model 1. GPT-1:} Estimated Number of Parameters: 117 million, Release Date: Feb 2018, Website: \url{https://huggingface.co/docs/transformers/model\_doc/openai-gpt}

Generative Pre-trained Transformer 1 (GPT-1) was the first of OpenAI's large language models following Google's invention of the transformer architecture in 2017. It was introduced in February 2018 by OpenAI. GPT-1 had 117 million parameters and significantly improved previous state-of-the-art language models. One of its strengths was its ability to generate fluent and coherent language when given a prompt or context. It was based on the transformer architecture and trained on a large corpus of books.
\hfill \break

\noindent \textbf{Model 2. BERT:} Estimated Number of Parameters: 110 million, Release Date: Nov 2, 2018, Website: \url{https://huggingface.co/bert-base-uncased}

BERT (Bidirectional Encoder Representations from Transformers) is a family of language models introduced in 2018 by researchers at Google. It is based on the transformer architecture and was initially implemented in English at two model sizes: BERT BASE and BERT Large. Both models were pre-trained on the Toronto BookCorpus and English Wikipedia. BERT was pre-trained simultaneously on language modeling and next-sentence prediction. As a result of this training process, BERT learns latent representations of words and sentences in context. It can be fine-tuned with fewer resources on smaller datasets to optimize its performance on specific tasks such as NLP tasks and sequence-to-sequence-based language generation tasks.
\hfill \break

\noindent \textbf{Model 3. BERT-Large:} Estimated Number of Parameters: 336 million, Release Date: Nov 2, 2018, Website: \url{https://huggingface.co/bert-large-uncased}

BERT (Bidirectional Encoder Representations from Transformers) is a family of language models introduced in 2018 by researchers at Google. It is based on the transformer architecture and was initially implemented in English at two model sizes: BERT BASE and BERT Large. Both models were pre-trained on the Toronto BookCorpus and English Wikipedia. BERT was pre-trained simultaneously on language modeling and next-sentence prediction. As a result of this training process, BERT learns latent representations of words and sentences in context. It can be fine-tuned with fewer resources on smaller datasets to optimize its performance on specific tasks such as NLP tasks and sequence-to-sequence-based language generation tasks.

\hfill \break
\noindent \textbf{Model 4. GPT-2:} Estimated Number of Parameters: 124 million, Release Date: Feb 14, 2019, Website: \url{https://huggingface.co/gpt2}

Generative Pre-trained Transformer 2 (GPT-2) is a large language model by OpenAI, the second in their foundational series of GPT models1. It was pre-trained on BookCorpus, a dataset of over 7,000 unpublished fiction books from various genres, and trained on a dataset of 8 million web pages1. GPT-2 was partially released in February 2019. It is a decoder-only transformer model of deep neural networks, which uses attention in place of previous recurrence- and convolution-based architectures. The model demonstrated strong zero-shot and few-shot learning on many tasks. This is the smallest version of GPT-2, with 124M parameters.

\hfill \break
\noindent \textbf{Model 5. BART-Large:} Estimated Number of Parameters: 400 million, Release Date: Oct 29, 2019, Website: \url{https://huggingface.co/facebook/bart-large-mnli}

BART (large-sized model) is a pre-trained model on the English language, introduced in the paper ``BART: Denoising Sequence-to-Sequence Pre-training for Natural Language Generation, Translation, and Comprehension" by Lewis et al. (2019). It uses a standard seq2seq/machine translation architecture with a bidirectional encoder (like BERT) and a left-to-right decoder (similar to GPT). The pre-training task involves randomly shuffling the order of the original sentences and a novel in-filling scheme, where text spans are replaced with a single mask token. BART is particularly effective when fine-tuned for text generation but also works well for comprehension tasks. It matches the performance of RoBERTa with comparable training resources on GLUE and SQuAD. It achieves new state-of-the-art results on a range of abstractive dialogue, question-answering, and summarization tasks, with gains of up to 6 ROUGE. BART (large-sized model) has nearly 400M parameters.
\hfill \break

\noindent \textbf{Model 6. DistilBart-MNLI-12-1:} Estimated Number of Parameters: $<$ 400 million , Release Date: Sep 21, 2020, Website: \url{https://huggingface.co/valhalla/distilbart-mnli-12-1}.

Distilbart-Mnli-12-1 is a distilled version of bart-large-mnli created using the No Teacher Distillation technique proposed for BART summarisation by Huggingface. It was released on September 21, 2020. It copies alternating layers from bart-large-mnli and is fine-tuned more on the same data. The performance drop is minimal compared to the original model.
\hfill \break

\noindent \textbf{Model 7. GPT-3.5:} Estimated Number of Parameters: 175 billion (not officially disclosed; assumed similar to GPT-3), Release Date: Nov 30, 2022, Website:  \url{https://platform.openai.com/docs/models}

GPT-3.5 is a fine-tuned version of the GPT-3 (Generative Pre-Trained Transformer) model. OpenAI has not publicly disclosed the exact parameter count for GPT-3.5, though it is commonly estimated to be similar to GPT-3's 175 billion parameters. The model is trained on a dataset of text and code up to September 2021. GPT-3.5 models can understand and generate natural language or code. For this study, we use the specific model snapshot 
(gpt-3.5-turbo-0301), which has been optimized for chat using the Chat completions API but works well for traditional completions tasks. This version supports a 16,385-token context window and includes improvements in instruction following and reproducible outputs. GPT-3.5 effectively performs various tasks, including text generation, translation, summarization, question answering, code generation, and creative writing.
\hfill \break

\noindent \textbf{Model 8. GPT-4 or ChatGPT 4:} Estimated Number of Parameters: 1.76 trillion (not officially disclosed; based on industry reports), Release Date: Mar 14, 2023, Website: \url{https://platform.openai.com/docs/models}

GPT-4 is a multimodal large language model created by OpenAI and the fourth in its series of GPT foundation models. OpenAI released it on March 14, 2023. As a transformer-based model, GPT-4 uses a paradigm where pre-training using both public data and ``data licensed from third-party providers" is used to predict the next token. After this step, the model was fine-tuned with reinforcement learning feedback from humans and AI for human alignment and policy compliance. OpenAI did not release the technical details of GPT-4; the technical report explicitly refrained from specifying the model size, architecture, or hardware used during either training or inference. The 1.76 trillion parameter estimate is based on leaked industry reports suggesting the model uses a Mixture of Experts architecture with approximately 1.8 trillion parameters across eight sub-models, though this remains unconfirmed by OpenAI.\footnote{See SemiAnalysis (2023), ``GPT-4 Architecture, Infrastructure, Training Dataset, Costs, Vision, MoE,'' available at \url{https://www.semianalysis.com/p/gpt-4-architecture-infrastructure}.} GPT-4 has several capabilities, including generating text that is indistinguishable from human-written text; translating languages with high accuracy; writing different kinds of creative content, such as poems, code, scripts, musical pieces, emails, and letters; and answering questions in an informative way, even if they are open-ended, challenging, or strange. The specific GPT-4 model we use is ``gpt-4-0314," whose pre-training data stops in September 2021.
\hfill \break

\noindent \textbf{Model 9. Llama2 (7B, 13B, or 70B):} Estimated Number of Parameters: 7, 13, or 70 billion, Release Date: July 18, 2023, Website: \url{https://ai.meta.com/llama/}
.

Llama2 (Large Language Model Meta AI) is the second generation of Meta’s open-weight foundation models. Released on July 18, 2023, it includes three pretrained and instruction-tuned variants with 7B, 13B, and 70B parameters, respectively. Compared to Llama1, Llama2 models are trained on a significantly larger dataset with extended context length and improved optimization methods, resulting in stronger performance on a wide range of benchmarks. The instruction-tuned ``chat” models are optimized for dialogue use cases through supervised fine-tuning and reinforcement learning with human feedback. Llama2 was released for both research and commercial use through a partnership with Microsoft and is hosted on platforms such as Hugging Face. The knowledge cutoff date for Llama 2's pre-training data is September 2022.

\hfill \break
\noindent \textbf{Model 10. FinBERT} Estimated Number of Parameters: 110 million, Release Date: 27 Aug 2019, Website: \url{https://huggingface.co/ProsusAI/finbert}

FinBERT, introduced by Dogu Araci, is a pre-trained NLP model fine-tuned for financial sentiment classification. It is based on the BERT-base architecture (with 12 encoder layers, 768 hidden dimensions, and 12 attention heads) and further trained on a large financial corpus, making it effective for sentiment analysis tasks in the financial domain. The model, which relies on Hugging Face's pytorch\_pretrained\_bert library, is available on Hugging Face's model hub and their GitHub repository.

 \hfill \break

%\noindent \textbf{Model 11. RavenPack} Estimated Number of Parameters: NA, Release Date: NA, Website: https://www.ravenpack.com/ 

%RavenPack is a leading provider of news analytics data. For each news item in their database, RavenPack Analytics generates an event sentiment score using their proprietary algorithms. 

 \end{onehalfspacing}

%%%%%%%%%%%%
%Theoretical Model
%%%%%%%%%%%%%

\newpage
\section{Theoretical Model}  \label{appendix:theory}
To provide a conceptual framework for understanding how Large Language Models (LLMs) transform information processing in financial markets, we develop an economic model. Our approach builds on established noisy rational expectations frameworks \citep{grundy1989trade, brown1989technical, gromb2010limits} but distinctly introduces LLMs as a new class of information processor with varying capabilities (e.g., model size). This allows us to move beyond simply documenting return predictability and instead explore the underlying economic mechanisms: How does the interaction between LLM information processing, bounded human rationality (represented by attentive and inattentive agents), and limits to arbitrage shape market outcomes? What conditions determine the efficacy of LLMs, and what are the broader consequences for market efficiency as these technologies proliferate?

Our model yields several key insights into these questions. First, it formally establishes the concept of a critical threshold in LLM sophistication (parameterized by model size relative to news complexity) required to generate genuinely informative signals, thus providing a theoretical basis for the observed heterogeneity in performance across different LLMs. Second, it demonstrates how return predictability arises from the temporary divergence between LLM-based assessments and the slower-reacting beliefs of human agents, with the magnitude of this predictability moderated by market structure (e.g., the proportion of attentive agents, arbitrage costs). Crucially, the model also elucidates the dynamic impact of LLM adoption on market efficiency: as more agents utilize increasingly capable LLMs, information is incorporated into prices more rapidly, potentially reducing certain forms of mispricing, even if some predictability persists due to fundamental frictions. These theoretical results offer a structured way to interpret our empirical findings and consider the evolving landscape of AI in finance. Detailed derivations and proofs for all propositions are presented in Section D of the Online Appendix.
\subsection{Agents}
We have two types of agents, attentive and inattentive, indexed by A and I, respectively, an asset with uncertain dividends and news revealed at the beginning of each period. Both agents have CARA utility with the same risk aversion ($\alpha$) and are price-takers.  The asset is in zero net supply so that we can focus on expectations rather than risk premiums. Incorporating two types of agents in the model allows us to examine how the adoption of LLMs by different market participants influences return predictability and overall market efficiency.

The total measure of agents paying attention to the asset is denoted by $V > 0$. $V$ closely resembles the potential volume of traders and should be lower in smaller stocks and, by definition, in illiquid stocks. Attentive agents are a fraction $\pi_A \in (0,1)$ of this population and have a total measure of $\pi_A V$. Inattentive agents a fraction of $\pi_I = 1 - \pi_A$ and a total population of $\pi_I V$. We model $V$ as fixed for a given market so that $\pi_A$ can be determined in equilibrium based on the costs of participating as an attentive investor and the attractiveness of speculation.

\subsection{Time Dynamics}
The model features three periods that model short trading horizons.  In period one, unexpected news about the company is released. Attentive agents incorporate this information in their forecasts better than inattentive investors. We assume inattentive agents only update their forecasts partially (because of information capacity constraints) and do not consider the price when updating since they have just enough information capacity to understand a fraction of the news implications.  In period two, both agents update their expectations entirely. Finally, in period three, the asset pays off a dividend, $\Tilde{d}$, and the economy ends. There is no intertemporal discount for notational ease since we are modeling very short horizons.

Attentive agents know that inattentive agents do not update entirely in the first period but will do so in the second. Hence, they can trade in the first period to take advantage of the price discrepancy. Without an additional source of uncertainty, this would be a riskless trade, which is unrealistic. We assume that when trading with a short horizon, nonfundamental trades may move the price further away from its fundamental value (e.g., \citet{de1990noise}). That is, there is nonfundamental risk stemming from noise traders, normally distributed with an expected value of zero and variance of $\sigma_u^2$.\footnote{Note that we only add noise traders in the last period for tractability.}

\vspace{1cm}
\begin{tikzpicture}[>=Stealth, node distance=3cm, font=\tiny]
    \node (period0) {Period 0};
    % \node (start) {}  ;
    % \node[right=of start] (period0) {Period 0};
    \node[right=of period0] (period1) {Period 1};
    \node[right=of period1] (period2) {Period 2};
    \node[right=of period2] (period3) {Period 3};
    % \node[right=of period3] (end) {};
    
    % \draw[->] (start) -- (period0);
    \draw[->] (period0) -- (period1) node[midway,above] {News};
    \draw[->] (period1) -- (period2) node[midway,above] { };
    \draw[->] (period2) -- (period3) node[midway,above] {};
    % \draw[->] (period3) -- (end);
    
    \node[above=0.3cm of period0] (assets) {};
    \node[below=0.3cm of period0] (agents) {};
    \node[above=0.3cm of period1] (info) {Information revealed};
    \node[below=0.3cm of period1] (type1) {Inattentive partially updates};
    \node[below=0.3cm of type1] (typeA) {Attentive fully updates};
    \node[below=0.3cm of period2] (type2) {Inattentive fully update};
    \node[above=0.3cm of period3] (info) {Asset Pays};
    % \node[below=0.3cm of period3] (both) {Both agents fully update};
\end{tikzpicture}

\subsection{Equilibrium Before News}

The asset payoff is random and is given by
\begin{equation}
    \tilde{d} = \mu_d + \sigma_{\xi} \tilde{\xi},\ \tilde{\xi} \sim N(0,1). 
\end{equation}

We refer to the expected value of the asset payoff, $\mu_d$, as the asset fundamentals. There are two sources of uncertainty. First, agents do not perfectly know the fundamentals, $\mu_d$. Second, there is real variation in the asset's payoff given by $\sigma_{\xi} \tilde{\xi}$. We assume independence between the uncertainty about the fundamentals and the real variation. While agents do not know the fundamentals, they have a correct prior. The prior is correct in the sense that it is equal to the population distribution:
\begin{equation}
    \mu_d \sim N(\bar{d}, \sigma_d^2).
\end{equation}

We let $x_{j, t+1}$ denote the gross demand of the asset for each type of investor in period $t$.  Without unexpected news, the demand of both agents is given by: 
\begin{align}
x_{1, j} = \frac{\bar{d} -  p_0 }{\alpha  (\sigma_d^2+ \sigma_{\xi}^2)}.
\end{align}

The asset is in zero net supply and we obtain the equilibrium price by the market clearing condition:
\begin{align}
V\frac{\bar{d} -  p_0 }{\alpha  (\sigma_d^2+ \sigma_{\xi}^2)}= 0
\end{align}

Hence, the price at period zero is given by its unconditional expected value:
\begin{align}
p_0 = \bar{d}.
\end{align}

\subsection{Information Structure}

Consider a scenario where news about a company gets released unexpectedly. Agents interpret this news as a signal with information about the fundamentals. The amount of information they use is related to their information processing capacity and the news complexity.

We assume the news' maximum information content about the fundamental contains a total precision of $\tau_S$.
\begin{equation}
s = \mu_d+ \varepsilon, \quad \varepsilon \sim N(0, \sigma_S^2 = \frac{1}{\tau_S})
\end{equation}

All agents, including attentive, inattentive, and LLMs, observe the same signal realization $s$, representing the company's news.\footnote{Hence, our work differs from the literature studying how private information gets incorporated into prices (e.g., \citet{kyle1985continuous,shleifer1997limits}).} However, they process this signal differently due to their varying information processing capacities. We implicitly assume agents are not biased when processing the information by assuming everyone observes the same signal. We could incorporate bias at the cost of additional notation.\footnote{For example,  different signal components for the signal of inattentive agents at the first and second period and the attentive agents' signal ($s_{I,1}$, $s_{I,2}$, and $s_A$) would allow for reversal return dynamics at the cost of a more complex model structure.}

The precisions for each type of agent for the new fundamental are given by:
\begin{align}
\tau_A = \gamma_A \tau_S , \
\tau_I =  \omega \tau_A, \
\tau_{L} = \lambda(c, k) \tau_S,
\end{align}

\noindent where $\gamma_A \in (0,1)$ is attentive agents' information capacity, and $\omega \in (0,1)$, describes how good inattentive agents' capacity is relative to attentive agents. A central innovation in our framework is the modeling of LLM information capacity, $\lambda(c, k):\mathbb{R}^+_2 \to (0,1]$. This function captures the idea that an LLM's ability to extract information from news is not fixed but depends critically on both its inherent sophistication (model size $k$) and the nature of the information itself (news complexity $c$). This formulation allows us to analyze how the interplay between LLM quality and task difficulty affects its economic usefulness, a crucial aspect for understanding the real-world impact of AI technologies.

We assume $\lambda(c,k)$ is twice differentiable with $\frac{\partial \lambda}{\partial c} \leq 0$ (more complex news is harder to process), $\frac{\partial \lambda}{\partial k} > 0$ (larger models are better), $\frac{\partial^2 \lambda}{\partial c^2} < 0$ (increasing difficulty with complexity), $\frac{\partial^2 \lambda}{\partial k^2} < 0$ (diminishing returns to size), and $\frac{\partial^2 \lambda}{\partial c \partial k} > 0$ (larger models are better at handling more complex news). Furthermore, $\lambda(c, 0) = 0 \ \forall \ c \geq 0$ (a model with no size has no capacity), and $\lim_{k\to\infty} \lambda(c, k) = 1 \  \forall \ c \geq 0$ (a sufficiently large model can, in principle, extract all information). These properties reflect stylized facts about current LLM technologies and provide a microfoundation for heterogeneity in LLM performance. When there is no confusion given the context, we use $\lambda$ to denote $\lambda(c,k)$.

Attentive agents process the signal with precision $\tau_A$ in the first period. In contrast, inattentive agents process the signal with precision $\omega \tau_A$ in the first period and the remaining precision $(1 - \omega) \tau_A$ in the second period. That is, inattentive agents only process a fraction of the information of attentive agents given by $\omega$ in the first period and the remaining fraction in the second. Intuitively, inattentive investors see some news, have an initial reaction, and trade on it. After some time, say a few hours later, these inattentive investors will better understand the news' full implications and can trade again. The signal inattentive investors processed is entirely contained in the signal the attentive investors processed. 

Attentive agents only process a fraction $\gamma_A$ of the total information. When $\gamma_A < 1$, even attentive agents cannot process the complete information, resulting in return predictability from the perspective of an agent who can process the entire signal. We analyze this case later by incorporating a hypothetical LLM with a large enough model size, such that $\lambda(c, k) > \gamma_A$.

LLMs process the signal with precision $\tau_{L}$ in the first period. We assume that if $\tau_{L} < \tau_A$, then the rest of the precision $\tau_A - \tau_L$ will be processed in the second period (intuitively, one could think that a human would revise the output); otherwise, there is just one update for LLMs in the first period. 

We assume that inattentive investors do not condition their forecasts on prices. Attentive investors have the complete information set available in the economy, so they do not need to learn from the price unless there are LLMs with information capacity greater than theirs and attentive agents are not using LLMs.

\subsection{Information Update}

Upon observing the signal $s$, each agent type $j \in \{A,I, LLM\}$ updates their beliefs about the asset's fundamental value $\mu_d$ using standard Bayesian rules. This results in a posterior mean expectation $\mu_{j|s} = E[\tilde{d}|s_j]$ and a posterior variance of the fundamental $Var[\mu_d|s_j] = \sigma_{\mu, j|s}^2$. The total conditional variance of the asset payoff for agent $j$ is then $\sigma_{d, j|s}^2 = \sigma_{\mu, j|s}^2 + \sigma_\xi^2$. The explicit formulas for these updated moments are provided in Online Appendix D.

\subsection{LLMs for Forecasting}

We first focus on the case where agents are not using LLMs to trade and are only available to the econometrician. Intuitively, we are modeling the case of how LLMs react (out-of-sample) to the information, closely aligning with the post-September 2021 period, the knowledge cutoff date for ChatGPT-4, but before March 2023, when ChatGPT-4 was released. We first characterize the price and return dynamics and then explore how LLMs forecast. In a later subsection, we make LLMs available to agents and analyze the resulting dynamics.

Attentive agents understand there is an initial underreaction. Hence, in the first period, they will trade to take advantage of the difference between the future and current prices. In the second period, they will trade considering the dividend. Because of the dynamic nature of the problem, it is easier to solve the model starting from the second period. 

\subsubsection{Second Period}

In the second period, there are no further signals for the attentive agent, and the inattentive agent fully updates. Consequently, both agent types share the same fundamental expectation, $\mu_{A|s} \equiv E[\tilde{d}|s_A]  =  \frac{\bar{d}  \tau_d + s \tau_A }{\tau_d + \tau_A}$, and conditional variance $\sigma_{\mu, A|s}^2 = (\tau_d + \tau_A)^{-1}$. Given their CARA utility, agents' optimal demands lead to an equilibrium price:
\begin{equation}
    p_2 = \mu_{A|s}  + u_2\frac{\alpha \sigma_{d, A|s}^2}{V}.
\end{equation}
This price consists of the consensus fundamental expectation plus a shock from noise trading $u_2$, scaled by risk aversion $\alpha$, total asset variance $\sigma_{d, A|s}^2 = \sigma_{\mu, A|s}^2 + \sigma_\xi^2$, and inversely by the measure of traders $V$. 

For attentive agents in period 1, the expected price in period 2 is $E_{1,A}[p_2] = \mu_{A|s}$, as noise trader demand $u_2$ is zero on average. The variance of this future price, $Var_{1,A}[p_2]$, arises solely from the noise trader risk and is given by:
\begin{align}
   \sigma_p^2 \equiv Var_{1,A}[p_2] = \alpha^2 \frac{\sigma_u^2}{V^2} (\sigma_{d, A|s}^2)^2.
\end{align}
This variance is amplified by fundamental asset risk and risk aversion, and dampened by the total volume of traders.

\subsubsection{Period One}

In period 1, attentive agents understand the news and trade to profit from the differences in future and current prices. In contrast, inattentive agents will trade to get the final period payoff, not realizing they do not fully understand the impact of news until the second period.

\paragraph{Attentive agents}

Let $\tilde{w_3} = w_2 + x (\tilde{d} - p_2)$ and $\tilde{w_2} = w_1 + x_2 (\tilde{p_2} - p_1)$. Given the demand in period 2 and the equilibrium prices, the value function of period 2 for attentive agents is:
\begin{align}
    V2(w_2) & = \max_{x}  E[U(\tilde{w_3})] =  \max_{x_3}  -exp\{ -\alpha E[\tilde{w_3}] + \frac{\alpha^2}{2} Var(\tilde{w_3})\}
\end{align}

In period 1, attentive agents maximize the expected period-2 value function:
\begin{align}
    V1(w_1) &=  \max_{x_2} E_1[V2(\tilde{w_2})]
\end{align}

In Section D of the Online Appendix, we show that we can write the optimal demand function as  
\begin{align}
    x_{2,A} &= \frac{(\mu_{A|s} - p_1)}{\alpha (\sigma_{d, A|s}^2)} + \frac{(\mu_p - p_1)}{\alpha\sigma_p^2}.
\end{align}

The first part corresponds to the traditional demand from CARA investors for an asset that pays a dividend, and the second part is due to the opportunity to take advantage of temporary mispricing. The first part arises because, given the second period's desired demand, there is an opportunity to obtain a potentially lower price and hold the asset until the dividend is realized.

Another interpretation is viewing agents as more confident regarding their signals. Let $\tau_p = \frac{1}{\sigma_p^2}$ and $\tau_{d, A|s} = \frac{1}{\sigma_{d, A|s}^2}$. We can rewrite the demand as
\begin{align}
    x_{2,A} &= (\tau_{d, A|s} + \tau_p)\frac{\mu_{A|s} - p_1}{\alpha } 
    \equiv \tau_{p,d} \frac{\mu_{A|s} - p_1}{\alpha },
\end{align}

\noindent where $\tau_{p,d} = \tau_{d, A|s} + \tau_p$ is the total precision.

\paragraph{Inattentive agents}
Inattentive agents' demand is given by
\begin{equation}
    x_{2,I} = \frac{(\mu_{I|s} - p_1)}{\alpha (\sigma_{I|s}^2+ \sigma_{\xi}^2)} = \tau_{d,I|s} \frac{(\mu_{I|s} - p_1)}{\alpha}.
\end{equation}

\paragraph{Equilibrium Price}

If we define weighted precisions as $\tau_{\pi_A} = \pi_A \tau_{p,d}$, $\tau_{\pi_I} = \pi_I \tau_{d,I|s}$,  then we can express the the first-period equilibrium price as 
\begin{align}
p_1 &=\frac{ \mu_{A|s} \pi_A \tau_{p,d} + \mu_{I|s} \pi_I \tau_{d,I|S}}{\pi_A  \tau_{p,d} + \pi_I  \tau_{d,I|S} } 
= \frac{\mu_{A|s} \tau_{\pi_A} +  \mu_{I|s} \tau_{\pi_I}}{\tau_{\pi_A} + \tau_{\pi_I}} \\
    &\equiv \mu_E,
\end{align}

\noindent where $\mu_E =\frac{\mu_{A|s} \tau_{\pi_A} +  \mu_{I|s} \tau_{\pi_I}}{\tau_{\pi_A} + \tau_{\pi_I}} $ is the economy-wide expectation. The price is the dividend expectation weighted by each type of agent's (weighted) precisions.

To determine the equilibrium proportion of informed agents, we can assume, as in \citet{grossman1980impossibility}, that there is a fixed cost of participating in this market as an attentive investor of $c_A \geq 0$, and this cost is higher when the investor wants to short. $\pi_A$ adjusts until attentive agents are indifferent between being attentive and paying $c_A$ or not participating in the market. We model $c_A$ as an exogenous constant, which is higher when participating as an attentive agent that can short stocks. These costs can stem from trading fees and investment restrictions, as only certain participants can short stocks.\footnote{The equilibrium exists because attentive agents' expected utility (and profits) are decreasing in the proportion of attentive agents. If $c_A$ is too high, there may not be any attentive investors. There is no paradox in this case since we assume news is produced exogenously within the short horizon.}

The higher participation cost for shorting captures well-documented real-world frictions. Institutional constraints substantially limit who can short: many mutual funds and retail investors face regulatory restrictions or outright bans on short-selling, making it more costly or even impossible for them to participate as attentive agents when negative news arrives. These asymmetric participation barriers have important implications for information incorporation: when negative news arrives, fewer attentive agents find it profitable to pay the higher $c_A$ to participate, resulting in a lower equilibrium proportion of attentive agents ($\pi_A$). With fewer sophisticated traders correcting the initial underreaction, prices adjust more slowly toward fundamentals, generating stronger predictability for negative news compared to positive news where long positions face lower barriers to entry. This mechanism generates Proposition \ref{prop:cond_pred}'s prediction of asymmetric predictability across news sentiment.

\paragraph{Period 1 Unconditional Returns}The immediate (unconditional) expected jump from the non-news price to the news price is 
\begin{align}
    E[r_1] &= E[p_1] - E[p_0] = E[\mu_E] - \bar{d} \\ 
    &= \bar{d} - \bar{d} = 0,
\end{align}

\noindent which corresponds to the returns we would observe in high-frequency data. These returns are unconditionally 0 as the agents are not biased regarding the signal. For this simple economy, it means that the econometrician needs to condition a signal to study the properties of period 1 returns.\footnote{For simplicity, we assume there are no nonfundamental traders in the first period, which implies there is always momentum due to the delay in attention within the model, and this assumption can be relaxed at the cost of more algebra.}

\paragraph{Period 1 to Period 2 Conditional Returns}

The expected price of period 2 is simply $\mu_{A|s}$. Hence, the expected (dollar) returns, conditional on period 1's information set (not necessarily observable to the econometrician), between period 2 and period 1 are given by: 
\begin{align}
    E_1[r_2] = E_1[p_2 - p_1] &=\mu_{A|s}  - \mu_E.
\end{align}

The unconditional period 1 to period 2 return is zero on average since the signals are unbiased. That is, without taking a stance on whether the news is good or bad, it is not possible to take a profitable strategy.

\paragraph{Market Efficiency}

We define pricing errors, $\alpha_M$, as the difference between the expected value of the dividend conditional on the signal, and the price at period 1.

\begin{equation}
    \alpha_M \equiv E[\tilde{d}|s] - E[p_1|s].
\end{equation}

And we define mispricing as the expected square value of the pricing errors, where the expectation is taken with respect to the signal s:

\begin{align}
   E[\alpha_M^2] & \equiv E[(E[\tilde{d}|s] - E[p_1|s])^2].
\end{align}

\begin{proposition}
\label{prop:mispricing}
    Mispricing is 
    \begin{enumerate}
        \item Decreasing in the proportion of attentive agents, the information capacity of either type of agent, and the total volume of traders.
        \item Increasing in the risk aversion and the noise trader risk. 
    \end{enumerate}
\end{proposition}

Mispricing is lower when there are more attentive trades, either because of more attentive agents or lower risks to trade, such as when the volume of traders is higher, causing the impact of noise traders to be smaller. Moreover, we define markets as more efficient if there is less mispricing. However, we cannot measure mispricing directly because we do not have access to the subjective information set. Fortunately, LLMs provide an imperfect proxy for this information set. 

\subsubsection{Predictability with LLMs}

An econometrician (or an agent) using an LLM forms an expectation $\mu_{L|s}$ about the dividend based on signal $s$ and the LLM's specific information processing capacity $\lambda(c,k)$ (as per the formulas in Online Appendix D). The optimal CARA demand for an agent acting on this LLM-based expectation would be $x_{2,L} = \tau_{d, L|s} \frac{(\mu_{L|s} - p_1)}{\alpha}$, where $\tau_{d,L|s}$ is the precision of the LLM's forecast.

The crucial measure for predictability is the actual profit this strategy yields, which depends on the true expectation of attentive agents, $\mu_{A|s}$, that ultimately influences future price resolution. The actual expected profit (unconditional on the signal $s$, without price impact) of an LLM-based strategy is:
\begin{equation} \label{eq:profits_llm}
    E[Profits_{LLM}] = E\left[x_{2,L} (\mu_{A|s} -  p_1)\right] = E\left[\tau_{d, L|s} \frac{(\mu_{L|s} -  p_1)(\mu_{A|s} -  p_1)}{\alpha}\right].
\end{equation}
This expression captures that profits are realized if the LLM's assessment $(\mu_{L|s} - p_1)$ correctly signs the subsequent price movement towards the attentive agents' more informed valuation $(\mu_{A|s} - p_1)$. This formulation directly maps to our empirical strategy of trading on LLM signals. Theorem \ref{theorem:llm_profitability} characterizes the conditions under which these profits are positive.

\begin{theorem}
\label{theorem:llm_profitability}
In the case when only the econometrician has access to LLMs, the profitability of LLM-based strategies is increasing in the LLM's model size ($k$). This result highlights that higher quality LLMs are more likely to generate valuable signals. More significantly, for a fixed set of market parameters and a given level of news complexity ($c$), there exists a unique threshold of LLM model size, $k^*$, such that only LLMs with $k > k^*$ can predict returns profitably on average. This threshold $k^*$ formalizes the crucial economic insight that a minimum level of AI sophistication is necessary for an LLM to overcome inherent market noise and information processing frictions to provide economically valuable signals. An LLM that falls below this quality threshold, despite processing information, will not yield profitable strategies. The threshold $k^*$ is:

\begin{enumerate}
    \item Increasing in inattentive agents information capacity $\omega$, attentive agents information capacity $\gamma_A$, the proportion of attentive agents $\pi_A$, and the total volume of agents V.
    \item Decreasing in agents' risk aversion, $\alpha$, and the noise trader variance $\sigma_u^2$.
\end{enumerate}

\end{theorem}

Conditional on the strategy being profitable, the profitability is decreasing in inattentive agents' information capacity $\omega$, the proportion of attentive agents $\pi_A$, and increasing in the noise trader variance $\sigma_u^2$ and agents' risk aversion. Hence, we have the following propositions.

\begin{proposition}\label{prop:cond_pred}
    If LLM-return predictability is profitable, LLM-return predictability is larger in markets with a lower proportion of attentive market participants, such as smaller or illiquid stocks, and with negative news.
\end{proposition}

\begin{proposition}
    \label{prop:llm_mispricing}
    Holding model size and news complexity constant, and if the LLM-based strategy is profitable, a decline in its returns due to better information processing of inattentive agents, higher volume of traders, a higher proportion of attentive market participants, or lower noise traders' risk corresponds to a reduction in mispricing.
\end{proposition}

These propositions help us know when using LLM return predictability should be lower and when we can use this decline as a proxy for decreased mispricing.

%To get the intuition of some of the results, Figure \ref{fig:LLM_figures_main} presents how the expected profitability of LLM-based strategies differ when changing one of the following parameters: the information capacity of the LLM technology ($\lambda$) or the information capacity of the inattentive agents ($\omega$).\footnote{We change directly the parameter $\lambda$ instead of $k$ or $c$ for illustration purposes.} The plots show that LLM-return predictability is monotonically increasing in LLM's information capacity and decreasing in inattentive agents' information capacity if the predictability is positive.
%The values used in the figure are the proportion of attentive agents ($\pi_A = 0.5$), the proportion of inattentive agents ($\pi_I = 0.5$), the precision of the dividend ($\tau_D = 1$), the precision of the signal ($\tau_S = 1$), the signal realization ($s = 2$), the mean dividend ($\bar{d} = 1$), the attentive agents' information capacity ($\gamma_A = 0.75$), the inattentive agents' relative information capacity ($\omega = 0.5$), the precision of nonfundamental demand ($\tau_U = 1$), the risk aversion ($\alpha = 1$), and the LLM information capacity ($\lambda = 0.7$).

%\centerline{\bfseries [Insert Figure \ref{fig:LLM_figures_main} about here]}

\subsection{LLMs Available for Trading}

The previous section assumes only the econometrician has access to LLMs and hence abstracts from the price impact associated with trading based on LLMs' signals.  Now we model the case where LLMs are better than inattentive investors but not better than attentive investors, $\gamma_I < \lambda(c,k) \leq \gamma_A$. Implicitly, we assume the technology is widely available, as is the case for the current best LLMs.

\subsubsection{LLMs Better Than Inattentive Agents}

Let's assume a fraction of inattentive agents, $\theta$, start processing the information with LLMs, and the remaining information will be processed in the second period. Then the new price in the first period, $p_{1|L,I}$ is given by

\begin{align}
p_{1|L, I} &=\frac{ \mu_{A|s} \pi_A \tau_{p,d} + (1 - \theta)\mu_{I|s} \pi_I \tau_{d,I|S} + \theta \pi_I \mu_{L|s}  \tau_{d, L|s}}{\pi_A  \tau_{p,d} + (1 - \theta)\pi_I  \tau_{d,I|S}  + \theta \pi_I  \tau_{d, L|s} }.
\end{align}

In this case, mispricing will decline as markets become more efficient. Moreover, all things equal, mispricing will decline more with more advanced LLMs.

\begin{theorem}
\label{theorem:market_efficiency}
If LLMs are more capable than inattentive traders (i.e., $\lambda(c,k) > \omega \gamma_A$), then broader adoption of these LLMs by inattentive agents (increasing $\theta$) or improvements in LLM quality (increasing $k$) leads to a reduction in market mispricing. This highlights a key channel through which AI can enhance market efficiency: by augmenting the information processing capacity of less informed market participants, thereby accelerating the incorporation of news into prices.
\end{theorem}

Intuitively, agents using LLMs that are better than them at correctly inferring the news' impact will have better expectations. The more agents use these advanced LLMs, the more the price will reflect the fundamental information.

\begin{proposition}\label{prop:inattentive_llm}
    If LLMs have a sufficiently large model size (approaching attentive agent capabilities, i.e., $\lambda(c,k) \approx \gamma_A$) and all inattentive agents use them ($\theta=1$), then return predictability based on the initial LLM signal diminishes significantly, potentially vanishing if LLM capabilities match those of attentive agents. This suggests a potential trajectory where widespread adoption of high-quality LLMs by the broader market erodes the very predictability that such LLMs initially uncover.
\end{proposition}

When all inattentive agents use LLMs, and LLMs have the same capacity as the attentive agents, return predictability completely disappears due to the price impact.

Similar to the case where LLMs are only available to the econometrician, for a sufficient size of LLMs, there is positive return predictability. In contrast to before, however, this return predictability is not typically monotonic in the size of the LLM. It is also not monotonic in the proportion of inattentive agents using LLMs.

\subsubsection{LLMs Better than Attentive Agents}

We now focus on the case where LLMs become better than attentive agents at processing information (i.e., $\lambda(c,k) > \gamma_A$), and all attentive agents switch to this superior LLM technology. This scenario explores how markets change when the most sophisticated human agents are further empowered by AI.

\begin{proposition}
\label{prop:attentive_efficiency}
    If attentive agents adopt LLMs that surpass their own (unaided) information processing capabilities, overall market efficiency improves as LLM model size $k$ increases. This illustrates a distinct channel for AI-driven efficiency gains: enhancing the abilities of already sophisticated traders, leading to even faster and more accurate price discovery.
\end{proposition}

More complex scenarios, such as partial LLM adoption by attentive agents where non-adopters may learn from prices, can also be considered. While introducing richer dynamics, these cases generally reinforce the notion that LLM adoption tends to enhance market efficiency (see Section H of Online Appendix D for the formal proof).

\newpage

\section{Prompts for Other LLMs}

\label{appendix:GPT1_GPT2}
%\begin{center}
%\section*{Appendix C: Prompts for Other LLMs} \label{appendix:GPT1_GPT2}
%\end{center}

\onehalfspacing
\indent This appendix provides details on the prompts for other LLMs. While the central focus of our paper is on ChatGPT, we compare its results with those of more basic models, such as BERT, GPT-1, and GPT-2. We employ a different strategy because those models cannot follow instructions or answer specific questions. 

GPT-1 and GPT-2 are autocomplete models. Hence, we use the following sentence that the models complete:

\begin{quote}
    News: + headline + f"Will this increase or decrease the stock price of {firm}? This will make {firm}'s stock price go "
\end{quote}

The usual response is ``up,'' ``down,'' followed by a brief sentence fragment. The answers are usually not fully legible but include positive and negative words. We count the positive words against the negative words and assign a $+1$ for every positive and a $-1$ for every negative. We then consider the sentiment positive if the sum is positive and vice versa. The positive words are `up,' `high,' `sky,' `top,' `increase,' `stratosphere,' `boom,' `roof,' `skyrocket,' `soar,' `surge,' `climb,' `rise,' `rising,' `expand,' `flourish.' The negative words are `down,' `low,' `bottom,' `decrease,' `back,' `under,' `plummet,' `drop,' `decline,' `tumble,' `fall,' `contract,' `struggle.'

BERT is only able to complete one word out of a sentence. Hence, we ask it to complete the following sentence:

\begin{quote}
    Headline: {headline} This is [MASK] news for {firm}'s stock price in the short-term
\end{quote}

Where [MASK] is the corresponding word that BERT will input. The answers set consists of `good,' `the,' `big,' and `bad.' We classify `good' as +1, 'bad' as -1, and the others as zero.

The BART model is capable of zero-shot classification. This means it can classify text according to predefined categories without seeing examples of what corresponds to a good category. We provide each headline and then classify it into one of the following categories:

\begin{enumerate}
    \item good news for the stock price of {firm} in the short term
    \item bad news for the stock price of {firm} in the short term
    \item not news for the stock price of {firm} in the short term
\end{enumerate}

We then assign a numerical score of +1 for good, -1 for bad, and 0 for not.

\newpage

%\section{Appendix Tables}

%\subfile{AppendixTables}

%\section{Appendix Figures}

%\subfile{AppendixFigures}

\newpage

\section*{\centering ONLINE APPENDIX}
\vspace{0.5cm}
\subsection*{\centering Can ChatGPT Forecast Stock Price Movements? \\ Return Predictability and Large Language Models}
\vspace{0.5cm}
\subsubsection*{\centering This Version: \today}
\setcounter{table}{0}
\renewcommand{\thetable}{OA\arabic{table}}
\newpage

\appendix

\section*{\centering Online Appendix A: Sample Characteristics}
\label{online_appendix:sample}
%\subsection{Sample Characteristics}

This appendix compares the characteristics of firms in our news sample to the broader CRSP universe to assess the representativeness of the firms in our sample. Understanding these differences is important for interpreting our results, as news coverage is not random and may be concentrated among certain types of firms. In general, firms with news are larger and more liquid, although their returns during news periods are more volatile.

\subsubsection*{1. News Sources}

Our sample draws from comprehensive coverage of major news agencies and newswire services. Table \ref{tab:news_sources} presents the distribution of news stories by source. The sample includes 153,892 unique news stories from 17 distinct sources over the period October 2021 to May 2024.\footnote{The number of unique news stories (153,892) is lower than the total number of headline observations in our analysis sample (159,137) because some news stories are relevant to multiple firms on the same date. For example, industry-wide news or market updates may appear for several companies simultaneously. The analysis uses all headline-firm-date observations, while this table characterizes the underlying news coverage.}

The four largest sources account for 94.2\% of news coverage: Dow Jones Newswires (33.8\%), Business Wire (23.1\%), GlobeNewswire (21.6\%), and PR Newswire (15.6\%). Additional sources include MarketWatch (2.7\%), Wall Street Journal (0.2\%), and various international newswire services. The top ten sources collectively represent 99.9\% of news stories in our sample.

This distribution reflects the institutional structure of corporate news dissemination, where companies primarily use major newswire services (Business Wire, GlobeNewswire, PR Newswire) to distribute press releases, while financial news outlets (Dow Jones Newswires, MarketWatch, Wall Street Journal) provide additional coverage and analysis. The broad coverage across multiple major sources ensures our sample captures the primary channels through which market-moving information reaches investors.

% latex table generated in R 4.5.1 by xtable 1.8-4 package
% Tue Oct 21 22:05:36 2025
\begin{table}[ht]
\centering
\caption{Distribution of News Stories by Source} 
\label{tab:news_sources}
\begin{tabular}{lrrr}
  \toprule
Source & N & Percent (\%) & Cumulative (\%) \\ 
  \midrule
Dow Jones Newswires & 52,028 & 33.8 & 33.8 \\ 
  Business Wire & 35,613 & 23.1 & 56.9 \\ 
  GlobeNewswire & 33,289 & 21.6 & 78.6 \\ 
  PR Newswire & 24,007 & 15.6 & 94.2 \\ 
  MarketWatch &  4,231 & 2.7 & 96.9 \\ 
  DJ Global Press Release Wire &  3,245 & 2.1 & 99.0 \\ 
  Canadian News Wire &    714 & 0.5 & 99.5 \\ 
  Wall Street Journal &    313 & 0.2 & 99.7 \\ 
  LSE Regulatory News Service (RNS) &    191 & 0.1 & 99.8 \\ 
  DGAP News &     89 & 0.1 & 99.9 \\ 
   \bottomrule
\end{tabular}
\end{table}

\subsubsection*{2. Methodology}

We compare firm-day observations from two samples over the period October 2021 to May 2024:

\begin{itemize}
    \item \textbf{CRSP Universe:} All U.S. common stocks (share codes 10 or 11) trading on NYSE, NASDAQ, and AMEX exchanges (exchange codes 1-3) with available return data. This represents the full population of eligible U.S. equity securities during our sample period, applying the same filters used throughout our analysis.
    \item \textbf{News Sample:} The subset of CRSP firms with at least one news headline covered by major news media or newswires, matched with RavenPack and used in our analysis. This sample inherits the same filters as the CRSP universe (exchange codes 1-3, share codes 10 or 11, non-missing returns). After news matching, our final sample contains only share code 11 companies, as share code 10 firms are rare in recent years and did not have news coverage during our sample period.
\end{itemize}

For each sample, we calculate summary statistics at the firm-day level for three key characteristics: daily returns (in percentage points), market capitalization (in millions of USD), and trading volume (in thousands of shares). Market capitalization is calculated as price times shares outstanding divided by 1,000 to convert to millions. Statistics include the mean, standard deviation, 25th percentile, median, and 75th percentile.

\subsubsection*{3. Results}

Table \ref{tab:sample_vs_crsp_desc} presents the comparison of firm characteristics between the CRSP universe and our news sample. The results reveal differences consistent with news coverage being concentrated among larger, more actively traded stocks:

\textbf{Firm Size:} The news sample has substantially larger firms than the CRSP universe. The median market capitalization is \$1.99 billion for the news sample compared to \$0.56 billion for the CRSP universe—approximately 3.6 times larger. This difference is even more pronounced in the mean (\$36.91B vs \$10.24B) and 75th percentile (\$10.33B vs \$3.24B), indicating that news coverage disproportionately focuses on large-cap stocks.

\textbf{Trading Activity:} Firms in the news sample exhibit significantly higher trading volume. The median volume is 827.47 thousand shares for the news sample versus 265.32 thousand for the CRSP universe—approximately 3.1 times higher. This pattern persists across all quantiles, suggesting that news media preferentially cover more liquid stocks with greater investor interest.

\textbf{Return Characteristics:} The news sample shows higher return volatility than the CRSP universe. The standard deviation of daily returns is 14.12\% for the news sample compared to 5.70\% for the CRSP universe—approximately 2.5 times higher. Mean returns are slightly higher for the news sample (0.21\% vs -0.03\%), though both are close to zero as expected for daily returns.

\subsubsection*{4. Implications}

These differences have several implications for interpreting our results:

First, our sample has larger, more liquid stocks relative to the typical CRSP firm. This selection is economically sensible: news media naturally focus on companies of greater interest to investors and the broader public. 

Second, the higher volatility in the news sample is partly mechanical; firms experience news events precisely when unusual developments warrant media coverage, and such events naturally coincide with larger price movements. 

Finally, we note that while the news sample is not representative of the average CRSP stock, it represents a large fraction of overall market capitalization and trading activity given the concentration of both among larger firms. Our results therefore apply to a significant portion of the equity market.

\begin{landscape}
    
\begin{table}[!htbp] \centering
  \caption{Comparison of Sample Firm Characteristics to CRSP Universe}
  \label{tab:sample_vs_crsp_desc}
  \begin{tabular}{@{\extracolsep{5pt}}lccccc}
  \\[-1.8ex]\hline
  \hline \\[-1.8ex]
   & Mean & SD & P25 & Median & P75 \\
  \hline \\[-1.8ex]
  \multicolumn{6}{l}{\textit{Panel A: CRSP Universe}} \\
  \hline \\[-1.8ex]
  Market Cap (\$M) & 10239.39 & 71759.13 & 112.78 & 559.78 & 3238.66 \\
  Volume ('000s) & 1534.16 & 8449.77 & 52.81 & 265.32 & 944.87 \\
  Daily Return (\%) & -0.03 & 5.70 & -1.65 & 0.00 & 1.41 \\
  \hline \\[-1.8ex]
  \multicolumn{6}{l}{\textit{Panel B: News Sample}} \\
  \hline \\[-1.8ex]
  Market Cap (\$M) & 36905.60 & 182467.80 & 332.65 & 1993.62 & 10328.16 \\
  Volume ('000s) & 4943.53 & 18253.04 & 220.23 & 827.47 & 3033.95 \\
  Daily Return (\%) & 0.21 & 14.12 & -2.12 & 0.00 & 2.15 \\
  \hline \\[-1.8ex]
  \end{tabular}
  \\[1ex]
  \parbox{\textwidth}{\footnotesize \textit{Note:} Summary statistics are calculated at the firm-day level. Market Cap is in millions of USD. Volume is in thousands of shares. Daily Return is in percentages.}
\end{table}

\end{landscape}
\FloatBarrier

\newpage

\section*{\centering Online Appendix B: Additional Figures and Tables}

% Descriptive Statistics of News Data
\begin{table}[htbp]
\centering
\caption{Descriptive Statistics of News Data}
\vspace{0.5cm}
\begin{minipage}{15cm}
\small
This table presents descriptive statistics of the news dataset used in our analysis. Panel A shows the distribution of news headlines by year, including the number of unique firms covered, trading days, and average coverage intensity. Panel B shows the distribution by timing, distinguishing between overnight news (released between market close and next open) and intraday news (released during trading hours). Panel C shows the distribution by news type (press releases vs. full articles). Panel D shows the distribution by event group, displaying the top 10 event categories in our sample, with earnings-related news (27.6\%) and insider trading disclosures (19.3\%) as the two most prevalent categories.
\end{minipage}
\vspace{0.5cm}
\label{tab:news_descriptive_stats}
\centering
% Appendix Table: Descriptive Statistics of News Data
\begin{tabular}{lrrrr@{\hskip 0.3in}r}
\toprule
\multicolumn{6}{l}{\textbf{Panel A: News Distribution by Year}} \\
\midrule
Year & Headlines & Firms & Days & Per Day & Per Firm \\
\midrule
2021 & 14,374 & 3,482 &  64 & 224.6 & 4.1 \\
2022 & 57,886 & 3,824 & 251 & 230.6 & 15.1 \\
2023 & 60,119 & 3,771 & 250 & 240.5 & 15.9 \\
2024 & 26,897 & 3,398 & 105 & 256.2 & 7.9 \\
\midrule
\multicolumn{6}{l}{\textbf{Panel B: News Distribution by Timing}} \\
\midrule
Timing & Headlines & \% Total & Firms & Per Day & \\
\midrule
Overnight & 129,953 & 81.6 & 4,072 & 194.0 & \\
Intraday &  29,184 & 18.3 & 3,651 & 48.6 & \\
\midrule
\multicolumn{6}{l}{\textbf{Panel C: News Distribution by Type}} \\
\midrule
Type & Headlines & \% Total & Firms & & \\
\midrule
PRESS-RELEASE & 107,381 & 67.5 & 4,065 & & \\
FULL-ARTICLE &  51,756 & 32.5 & 3,838 & & \\
\midrule
\multicolumn{6}{l}{\textbf{Panel D: News Distribution by Event Group (Top 10)}} \\
\midrule
Event Group & Headlines & \% Total & & & \\
\midrule
earnings & 44,033 & 27.6 & & & \\
insider-trading & 30,686 & 19.3 & & & \\
marketing & 12,121 & 7.6 & & & \\
revenues & 11,759 & 7.4 & & & \\
products-services &  9,936 & 6.2 & & & \\
labor-issues &  9,351 & 5.9 & & & \\
dividends &  8,950 & 5.6 & & & \\
investor-relations &  7,640 & 4.8 & & & \\
equity-actions &  5,579 & 3.5 & & & \\
partnerships &  4,922 & 3.1 & & & \\
\bottomrule
\end{tabular}

\end{table}

\newpage

\begin{table}[ht]
\centering
\caption{ChatGPT Predictions: Answering First or Reasoning First} 
\label{table:reasoning_first}
\vspace{0.5cm}
\begin{minipage}{15cm}
\small
This table tabulates the distribution (\%) of predictive recommendations by different ways of prompting ChatGPT 4 for 1,000 news headlines randomly selected from our sample. In the rows, we use the standard prompt mentioned in Section \ref{sec:prompting} of the paper and ask the model, ChatGPT 4, to provide the answer first and then provide the reasoning for the recommendation. In the columns, we ask the model to reason first and then provide a recommendation. 
\end{minipage}

\vspace{0.5cm} 
\adjustbox{width=0.45\textwidth}{
\begin{tabular}{r|rrr}
  \hline 
& \multicolumn{3}{c}{\text{Reason First}} \\
Answer First  & NO & UNKNOWN & YES \\ 
  \hline
 NO & 16.80 & 0.20 & 0.00 \\ 
    UNKNOWN & 1.00 & 34.70 & 0.90 \\ 
    YES & 0.60 & 3.80 & 42.00 \\ 
   \hline
\end{tabular}
}
\end{table}
%
%%%%%%%%%%%%%%%%%%%%%%%%%%%%%%%%%
\begin{landscape}
\begin{table}[ht]
\caption{Average Next Day's Return by Prediction Score: Small vs NonSmall Stocks} 
\vspace{0.1cm}
\begin{minipage}{22cm}
\small
This table repeats the analysis of different LLMs in Table \ref{tab:average_by_score} for small and non-small stocks separately. Panel A analyzes the sample of small stocks (below the 20th percentile NYSE market capitalization), and Panel B analyzes the remaining non-small stocks. All the statistics are the same as in Table 5. We provide an overview of the different LLMs in Section A of the Appendix.  
\end{minipage}
\vspace{0.1cm} 
\label{tab:average_by_score_small}
\centering

\vspace{0.1cm} 
\centering

Panel A. Small Stocks 
\vspace{0.1cm} 

\begin{tabular}{lrrrrrrrrrrrrr}
  \hline
% latex table generated in R 4.5.1 by xtable 1.8-4 package
% Mon Oct 20 13:47:18 2025
Model & HR-I & HR-D & $\text{Sharpe}_{LS}$ &  $\mu_{LS}$ &  $\mu_{+}$ &  $\mu_0$ & $\mu_{-}$ & $N_+$ & $N_-$ & $\alpha_{FF5}$ & t $\alpha_{FF5}$ & $R^2_{FF5}$ (\%) \\ 
  \hline
GPT-4 & 0.84 & 0.53 & 2.38 & 0.63 & -0.04 & -0.48 & -0.67 & 18 & 3 & 0.64 & 3.68 & 0.86 \\ 
  GPT-3.5 & 0.83 & 0.48 & 0.26 & 0.09 & -0.16 & -0.32 & -0.24 & 13 & 1 & 0.11 & 0.58 & 2.08 \\ 
  FinBERT & 0.76 & 0.40 & -0.46 & -0.17 & -0.51 & -0.19 & -0.34 & 5 & 1 & -0.17 & -0.73 & 1.80 \\ 
  Llama2-70b & 0.73 & 0.47 & 0.38 & 0.09 & -0.25 & -0.09 & -0.34 & 31 & 1 & 0.12 & 0.77 & 2.36 \\ 
  Llama2-13b & 0.72 & 0.47 & 0.20 & 0.05 & -0.27 & -0.10 & -0.32 & 30 & 1 & 0.08 & 0.49 & 2.01 \\ 
  DistilBart & 0.72 & 0.47 & 0.32 & 0.07 & -0.26 & -0.11 & -0.34 & 29 & 3 & 0.11 & 0.76 & 1.45 \\ 
  BART-Large & 0.71 & 0.47 & 0.22 & 0.05 & -0.26 & -0.21 & -0.31 & 28 & 3 & 0.08 & 0.55 & 1.11 \\ 
  Llama2-7b & 0.68 & 0.40 & -1.85 & -0.28 & -0.28 & -0.00 & 0.00 & 32 & 0 & -0.24 & -2.94 & 27.17 \\ 
  BERT-Large & 0.67 & 0.41 & -1.38 & -0.23 & -0.29 & -0.18 & -0.06 & 30 & 0 & -0.20 & -2.09 & 18.59 \\ 
  BERT & 0.65 & 0.41 & -1.18 & -0.34 & -0.37 & -0.21 & -0.03 & 7 & 0 & -0.29 & -1.68 & 8.82 \\ 
  GPT-1 & 0.58 & 0.45 & -1.12 & -0.33 & -0.27 & -0.38 & 0.06 & 24 & 4 & -0.32 & -1.77 & 0.39 \\ 
  GPT-2 & 0.55 & 0.49 & -0.52 & -0.23 & -0.31 & -0.17 & -0.09 & 20 & 4 & -0.23 & -0.90 & 0.49 \\ 
   \hline

\end{tabular}

\vspace{0.1cm} 

Panel B. Non-Small Stocks 

\vspace{0.10cm} 

\begin{tabular}{lrrrrrrrrrrrrr}
  \hline
% latex table generated in R 4.5.1 by xtable 1.8-4 package
% Mon Oct 20 13:47:14 2025
Model & HR-I & HR-D & $\text{Sharpe}_{LS}$ &  $\mu_{LS}$ &  $\mu_{+}$ &  $\mu_0$ & $\mu_{-}$ & $N_+$ & $N_-$ & $\alpha_{FF5}$ & t $\alpha_{FF5}$ & $R^2_{FF5}$ (\%) \\ 
  \hline
GPT-4 & 0.92 & 0.58 & 2.77 & 0.25 & 0.12 & 0.07 & -0.13 & 50 & 16 & 0.25 & 4.35 & 0.53 \\ 
  GPT-3.5 & 0.89 & 0.56 & 1.54 & 0.23 & 0.12 & 0.04 & -0.10 & 34 & 5 & 0.22 & 2.46 & 1.59 \\ 
  BART-Large & 0.87 & 0.61 & 3.32 & 0.26 & 0.11 & 0.10 & -0.15 & 78 & 15 & 0.27 & 5.31 & 1.11 \\ 
  DistilBart & 0.87 & 0.60 & 2.95 & 0.24 & 0.10 & 0.16 & -0.14 & 80 & 13 & 0.24 & 4.72 & 1.27 \\ 
  FinBERT & 0.86 & 0.49 & -0.13 & -0.03 & 0.04 & 0.06 & 0.06 & 15 & 6 & -0.03 & -0.20 & 0.68 \\ 
  Llama2-70b & 0.85 & 0.53 & 1.45 & 0.17 & 0.07 & 0.21 & -0.09 & 94 & 6 & 0.17 & 2.33 & 1.56 \\ 
  Llama2-13b & 0.85 & 0.53 & 1.42 & 0.15 & 0.07 & 0.09 & -0.08 & 92 & 6 & 0.15 & 2.28 & 1.01 \\ 
  BERT-Large & 0.64 & 0.53 & 1.66 & 0.22 & 0.06 & 0.07 & -0.16 & 92 & 1 & 0.24 & 3.22 & 9.84 \\ 
  Llama2-7b & 0.57 & 0.52 & 0.45 & 0.04 & 0.07 & 0.02 & 0.02 & 102 & 0 & 0.06 & 1.32 & 45.67 \\ 
  BERT & 0.57 & 0.51 & 0.77 & 0.08 & 0.05 & 0.07 & -0.04 & 25 & 0 & 0.10 & 1.78 & 29.57 \\ 
  GPT-2 & 0.50 & 0.52 & 0.20 & 0.02 & 0.07 & 0.06 & 0.05 & 61 & 13 & 0.02 & 0.47 & 0.63 \\ 
  GPT-1 & 0.50 & 0.50 & 0.07 & 0.01 & 0.06 & 0.08 & 0.06 & 75 & 13 & 0.00 & 0.01 & 1.05 \\ 
   \hline

\end{tabular}

\end{table}
\end{landscape}

\FloatBarrier

\begin{landscape}
\begin{table}[ht]
\centering
\caption{News Topic Categories}
\label{table:news_topics}
\vspace{0.15cm}
\begin{minipage}{22cm}
\small
This table presents the 15 news topic categories identified through BERTopic clustering of overnight headlines. Topics were generated using K-Means clustering (K=50) followed by hierarchical merging to 15 final categories. The ``Label'' column shows GPT-4o-generated topic names. ``Top Keywords'' displays the five most representative terms from each topic's c-TF-IDF representation. ``Representative Headline'' shows an example headline near each topic's centroid. The sample includes 81,700 overnight news headlines with directional GPT-4 predictions from 2021-2024. See Section E of the Online Appendix  %~\ref{appendix:topic}
for detailed methodology.
\end{minipage}

\vspace{0.15cm}

%\scriptsize{
%\input{LatexTables/TopicsHeadline_Updated}
%}
\scalebox{0.731}{
  \begin{tabular}{clrrp{10cm}}
\toprule
Topic & Label & Count & Top Keywords & Representative Headline \\
\midrule
0 & Insider Stock Transactions & 16,790 & holdings corp, corp dir, financial corp, sells 3000, sells 10000 & 'Dir Trainor Sells 14700 Of Paymentus Holdings Inc \\
1 & Earnings and Revenue Announcements & 11,440 & 1q profit, 1q revenue, 2q profit, earnings beat, 3q profit & 'Burlington Stores Q3 adj. EPS 98 cents vs. 43 cents a year ago, tops the FactSet consensus 97 cents... \\
2 & Executive Stock Disposals & 9,852 & corp cfo, surrenders, cfo, financial corp, holding corp & 'CFO Palatnik Surrenders 2494 Of Coherent Inc \\
3 & Strategic Partnership Announcements & 7,660 & present goldman sachs, present goldman, goldman sachs, financial services conference, goldman & 'KKR to Present at the Goldman Sachs US Financial Services Conference 2021 \\
4 & Dividend Announcements & 6,930 & quarterly cash dividend, declares cash dividend, declares quarterly dividend, announces quarterly dividend, cash dividend & 'Waterstone Financial Declares Regular Quarterly Cash Dividend \\
5 & Clinical Trial Announcements & 6,357 & phase clinical trial, clinical trial, therapeutics announces, therapeutics present, phase clinical & '4D Molecular Therapeutics Announces First Patient Dosed in Phase 1/2 Clinical Trial of 4D-150, a Du... \\
6 & Stock Offering Announcements & 4,915 & offering common stock, public offering common, announces pricing public, pricing public offering, common stock & 'ATEC Announces Proposed Public Offering of Common Stock \\
7 & Executive Leadership Appointments & 4,572 & chief financial officer, chief financial, chief executive officer, corporation appoints, financial officer & 'Kirby Corporation Names Raj Kumar Executive Vice President and Chief Financial Officer', ""Ocwen Fi... \\
8 & Quarterly Earnings Reports & 4,439 & 2023 financial results, reports quarter 2023, quarter 2023 results, quarter 2023 financial, 2022 financial results & 'Hims \& Hers Health, Inc. Reports Second Quarter 2023 Financial Results and Raises Full Year 2023 Ou... \\
9 & Investor Conference Participation & 3,265 & upcoming investor conferences, investor conferences, upcoming investor conference, investors conference, investor conference & 'Appian to Participate in Upcoming Investor Conferences \\
10 & Healthcare Conference Participation & 3,066 & morgan healthcare conference, annual morgan healthcare, annual healthcare conference, morgan healthcare, healthcare conference & 'Generation Bio to Present at the 41st Annual J.P. Morgan Healthcare Conference \\
11 & Stock Rating Adjustments & 889 & stock price target, price target, target price, therapeutics stock, stock & 'NVR started at buy with \$7000 stock price target at Seaport Research \\
12 & Securities Class Action Lawsuits & 701 & lawsuit, assertio holdings, litigation, law firm, shareholder & 'PEGASYSTEMS ALERT: Bragar Eagel \& Squire, P.C. Announces that a Class Action Lawsuit Has Been Filed... \\
13 & Credit Rating Adjustments & 436 & fitch, affirms, announces quarterly, quarterly, declares quarterly & ""Fitch Affirms Athene's Ratings; Outlook Stable"", ""Fitch Affirms Tempur Sealy's IDR at 'BB+'; Out... \\
14 & Index Inclusion Announcements & 213 & russell, index, joins, join, 3000 & 'ReposiTrak Set to Join Russell 3000(R) Index \\
15 & Retail Store Openings & 175 & ross stores, ross, new store, stores, store & 'Ross Dress for Less to Open a New Store in Troy, Alabama \\
\bottomrule
\end{tabular}

}
\end{table}
\end{landscape}

\FloatBarrier
\newpage

%%%%%%%%%%%%%%%%%%%%%%%%%%%%%%%%%%%%%%%%%%%%%%%%%%%%
% Transaction Cost Analysis Tables
%%%%%%%%%%%%%%%%%%%%%%%%%%%%%%%%%%%%%%%%%%%%%%%%%%%%

\begin{table}[htbp]
  \centering
  \caption{Transaction Cost Analysis: Overnight News Drift Strategies}
  \label{tab:tcosts_overnight}
  \vspace{0.5cm}
  \begin{minipage}{15cm}
  \small
  This table presents transaction cost analysis for GPT-4 based drift strategies using overnight news. The strategy trades on ret\_o (open-to-close returns). Panel A shows Sharpe ratios, Panel B shows mean daily returns (in percent), Panel C shows average daily one-way turnover and average gross exposure (turnover / gross exposure), and Panel D shows daily return volatility (in percent) for different rebalancing intensities. Turnover is measured as sum of absolute weight changes divided by 2, representing the fraction of portfolio value traded. Gross exposure is the sum of absolute long and short positions. Lower rebalancing percentages reduce both turnover and exposure (e.g., 50\% rebalancing achieves $\sim$100\% gross exposure vs $\sim$200\% for 100\% rebalancing). Transaction costs range from 0 to 20 basis points. Portfolio formation requires more than 1 stock in each leg (long/short). Note that volatility remains nearly constant across transaction cost levels as costs only affect mean returns.
  \end{minipage}
  \vspace{0.5cm}
  \begin{tabular}{llcccc}
\toprule
 & & \multicolumn{4}{c}{Transaction Costs (basis points)} \\
\cmidrule(lr){3-6}
Weighting & Rebalance & 0 bp & 5 bp & 10 bp & 20 bp \\
\midrule
\multicolumn{6}{l}{\textit{Panel A: Sharpe Ratios}} \\
Equal-weighted & 100\% & 2.971 & 2.130 & 1.289 & -0.394 \\
 & 50\% & 2.946 & 2.129 & 1.312 & -0.322 \\
 & 25\% & 2.885 & 2.110 & 1.336 & -0.213 \\
\addlinespace
Value-weighted & 100\% & 0.559 & -0.398 & -1.357 & -3.274 \\
 & 50\% & 0.408 & -0.484 & -1.376 & -3.162 \\
 & 25\% & 0.287 & -0.489 & -1.266 & -2.820 \\
\midrule
\multicolumn{6}{l}{\textit{Panel B: Mean Daily Returns (\%)}} \\
Equal-weighted & 100\% & 0.336 & 0.241 & 0.146 & -0.045 \\
 & 50\% & 0.169 & 0.122 & 0.075 & -0.018 \\
 & 25\% & 0.085 & 0.063 & 0.040 & -0.006 \\
\addlinespace
Value-weighted & 100\% & 0.054 & -0.039 & -0.131 & -0.317 \\
 & 50\% & 0.021 & -0.024 & -0.069 & -0.159 \\
 & 25\% & 0.008 & -0.014 & -0.035 & -0.079 \\
\midrule
\multicolumn{6}{l}{\textit{Panel C: Turnover (\%) and Gross Exposure (\%)}} \\
Equal-weighted & 100\% & 190.3 / 196 & \multicolumn{3}{c}{---} \\
 & 50\% & 93.6 / 101 & \multicolumn{3}{c}{---} \\
 & 25\% & 45.9 / 54 & \multicolumn{3}{c}{---} \\
\addlinespace
Value-weighted & 100\% & 185.4 / 196 & \multicolumn{3}{c}{---} \\
 & 50\% & 89.7 / 105 & \multicolumn{3}{c}{---} \\
 & 25\% & 43.3 / 59 & \multicolumn{3}{c}{---} \\
\midrule
\multicolumn{6}{l}{\textit{Panel D: Daily Return Volatility (\%)}} \\
Equal-weighted & 100\% & 1.796 & 1.796 & 1.796 & 1.796 \\
 & 50\% & 0.910 & 0.909 & 0.909 & 0.909 \\
 & 25\% & 0.470 & 0.470 & 0.470 & 0.470 \\
\addlinespace
Value-weighted & 100\% & 1.536 & 1.536 & 1.536 & 1.535 \\
 & 50\% & 0.799 & 0.798 & 0.798 & 0.798 \\
 & 25\% & 0.443 & 0.443 & 0.443 & 0.443 \\
\bottomrule
\end{tabular}

\end{table}

\newpage

\begin{table}[htbp]
  \centering
  \caption{Transaction Cost Analysis: Intraday News Drift Strategies}
  \label{tab:tcosts_intraday}
  \vspace{0.5cm}
  \begin{minipage}{15cm}
  \small
  This table presents transaction cost analysis for GPT-4 based drift strategies using intraday news. The strategy trades on ret\_o (close-to-close returns). Panel A shows Sharpe ratios, Panel B shows mean daily returns (in percent), Panel C shows average daily one-way turnover and average gross exposure (turnover / gross exposure), and Panel D shows daily return volatility (in percent) for different rebalancing intensities. Turnover is measured as sum of absolute weight changes divided by 2, representing the fraction of portfolio value traded. Gross exposure is the sum of absolute long and short positions. Lower rebalancing percentages reduce both turnover and exposure (e.g., 50\% rebalancing achieves $\sim$100\% gross exposure vs $\sim$200\% for 100\% rebalancing). Transaction costs range from 0 to 20 basis points. Portfolio formation requires more than 1 stock in each leg (long/short). Note that volatility remains nearly constant across transaction cost levels as costs only affect mean returns.
  \end{minipage}
  \vspace{0.5cm}
  \begin{tabular}{llcccc}
\toprule
 & & \multicolumn{4}{c}{Transaction Costs (basis points)} \\
\cmidrule(lr){3-6}
Weighting & Rebalance & 0 bp & 5 bp & 10 bp & 20 bp \\
\midrule
\multicolumn{6}{l}{\textit{Panel A: Sharpe Ratios}} \\
Equal-weighted & 100\% & 2.631 & 2.172 & 1.712 & 0.792 \\
 & 50\% & 2.584 & 2.138 & 1.692 & 0.799 \\
 & 25\% & 2.556 & 2.123 & 1.690 & 0.823 \\
\addlinespace
Value-weighted & 100\% & 1.283 & 0.685 & 0.087 & -1.110 \\
 & 50\% & 1.294 & 0.720 & 0.145 & -1.003 \\
 & 25\% & 1.329 & 0.801 & 0.273 & -0.783 \\
\midrule
\multicolumn{6}{l}{\textit{Panel B: Mean Daily Returns (\%)}} \\
Equal-weighted & 100\% & 0.499 & 0.412 & 0.325 & 0.150 \\
 & 50\% & 0.249 & 0.206 & 0.163 & 0.077 \\
 & 25\% & 0.126 & 0.104 & 0.083 & 0.040 \\
\addlinespace
Value-weighted & 100\% & 0.184 & 0.098 & 0.012 & -0.159 \\
 & 50\% & 0.095 & 0.053 & 0.011 & -0.074 \\
 & 25\% & 0.052 & 0.031 & 0.011 & -0.031 \\
\midrule
\multicolumn{6}{l}{\textit{Panel C: Turnover (\%) and Gross Exposure (\%)}} \\
Equal-weighted & 100\% & 174.4 / 178 & \multicolumn{3}{c}{---} \\
 & 50\% & 86.2 / 91 & \multicolumn{3}{c}{---} \\
 & 25\% & 42.7 / 48 & \multicolumn{3}{c}{---} \\
\addlinespace
Value-weighted & 100\% & 171.2 / 178 & \multicolumn{3}{c}{---} \\
 & 50\% & 84.2 / 94 & \multicolumn{3}{c}{---} \\
 & 25\% & 41.5 / 51 & \multicolumn{3}{c}{---} \\
\midrule
\multicolumn{6}{l}{\textit{Panel D: Daily Return Volatility (\%)}} \\
Equal-weighted & 100\% & 3.010 & 3.009 & 3.009 & 3.009 \\
 & 50\% & 1.533 & 1.532 & 1.532 & 1.531 \\
 & 25\% & 0.781 & 0.781 & 0.781 & 0.781 \\
\addlinespace
Value-weighted & 100\% & 2.271 & 2.271 & 2.271 & 2.271 \\
 & 50\% & 1.164 & 1.164 & 1.164 & 1.164 \\
 & 25\% & 0.623 & 0.623 & 0.623 & 0.623 \\
\bottomrule
\end{tabular}

\end{table}

\FloatBarrier
\newpage

%%%%%%%%%%%%%%%%%%%%%%%%%%%%%%%%%%%%%%%%%%%%%%%%%%%%
% Size Breakdown Analysis Tables
%%%%%%%%%%%%%%%%%%%%%%%%%%%%%%%%%%%%%%%%%%%%%%%%%%%%

\begin{table}[htbp]
  \centering
  \caption{Stock Returns by Market Capitalization: Initial Reaction and Drift}
  \label{tab:size_reaction_drift}
  \vspace{0.5cm}
  \begin{minipage}{15cm}
  \small
  This table decomposes stock-level returns into initial reaction (ret\_n) and subsequent drift (ret\_o) for overnight news, stratified by market capitalization terciles. Panel A shows stocks with positive GPT-4 signals (long positions), and Panel B shows stocks with negative signals (short positions). Size groups are defined as Small (bottom tercile, approximately bottom 33\%), Medium (middle tercile, 33\%-67\%), and Large (top tercile, top 33\%), based on lagged market capitalization (me\_lag). Tercile breakpoints are recalculated every 6 months using an expanding window approach that uses only data from the beginning of the sample through each rebalancing date, avoiding look-ahead bias while adapting to changes in market cap distribution over time (October 2021 - May 2024). The critical finding is asymmetric drift patterns by size: for stocks with negative news, large-cap stocks show mean reversion (drift of +0.06\%), while small-cap stocks show continued decline (drift of -0.70\%). This size-dependent drift asymmetry explains why equal-weighted portfolios dramatically outperform value-weighted portfolios---the short side of the strategy is only profitable for smaller stocks.
  \end{minipage}
  \vspace{0.5cm}
  \begin{tabular}{lccccc}
\toprule
Size Group & N & Initial Reaction (\%) & Drift (\%) & Total Return (\%) & SD Drift (\%) \\
\midrule
\multicolumn{6}{l}{\textbf{Panel A: Positive Signals (Long Positions)}} \\
Small & 17,179 & 1.99 & 0.06 & 2.05 & 8.64 \\
Medium & 14,679 & 0.88 & 0.11 & 0.99 & 4.34 \\
Large & 17,953 & 0.46 & 0.05 & 0.51 & 2.81 \\
\midrule
\multicolumn{6}{l}{\textbf{Panel B: Negative Signals (Short Positions)}} \\
Small & 3,819 & -2.33 & -0.70 & -3.03 & 7.10 \\
Medium & 5,080 & -1.34 & -0.18 & -1.52 & 4.30 \\
Large & 5,452 & -1.12 & 0.06 & -1.06 & 2.92 \\
\bottomrule
\end{tabular}

\end{table}

\newpage

\begin{table}[htbp]
  \centering
  \caption{Portfolio Performance by Market Capitalization}
  \label{tab:size_portfolio}
  \vspace{0.5cm}
  \begin{minipage}{15cm}
  \small
  This table shows equal-weighted portfolio performance stratified by market capitalization terciles for overnight news drift strategies. Size groups are defined as Small (bottom tercile, approximately bottom 33\%), Medium (middle tercile, 33\%-67\%), and Large (top tercile, top 33\%), based on lagged market capitalization with breakpoints recalculated every 6 months using an expanding window. Each size group forms separate daily portfolios based on GPT-4 signals, requiring at least 2 stocks in each leg (long/short). Returns are daily percentages. The striking pattern is that short-only portfolios are highly profitable for small caps (Sharpe ratio 3.17) but unprofitable for large caps (Sharpe ratio -0.61), while long-only portfolios perform reasonably well across all sizes. This asymmetry directly explains the dramatic difference between equal-weighted (Sharpe 2.97) and value-weighted (Sharpe 0.56) portfolio performance documented in the transaction cost analysis: value-weighting overweights large-cap stocks where the short strategy fails due to mean reversion rather than drift continuation.
  \end{minipage}
  \vspace{0.5cm}
  \begin{tabular}{lcccccc}
\toprule
& \multicolumn{3}{c}{Initial Reaction} & \multicolumn{3}{c}{Drift} \\
\cmidrule(lr){2-4} \cmidrule(lr){5-7}
Size Group & Hit Rate (\%) & Sharpe & Return (\%) & Hit Rate (\%) & Sharpe & Return (\%) \\
\midrule
Small & 86.85 & 11.40 & 4.53 & 57.85 & 3.09 & 0.73 \\
Medium & 84.33 & 10.10 & 2.23 & 57.91 & 2.09 & 0.28 \\
Large & 84.48 & 10.02 & 1.48 & 48.96 & 0.06 & 0.00 \\
\bottomrule
\end{tabular}

\end{table}

\FloatBarrier
\newpage

\begin{figure}[htbp]
  \centering
    \caption*{Figure OA1: Cumulative Returns of Investing \$1 in the Long-Short Strategy for Intraday News}
          \vspace{0.5cm}
  \includegraphics[width=0.8\textwidth]{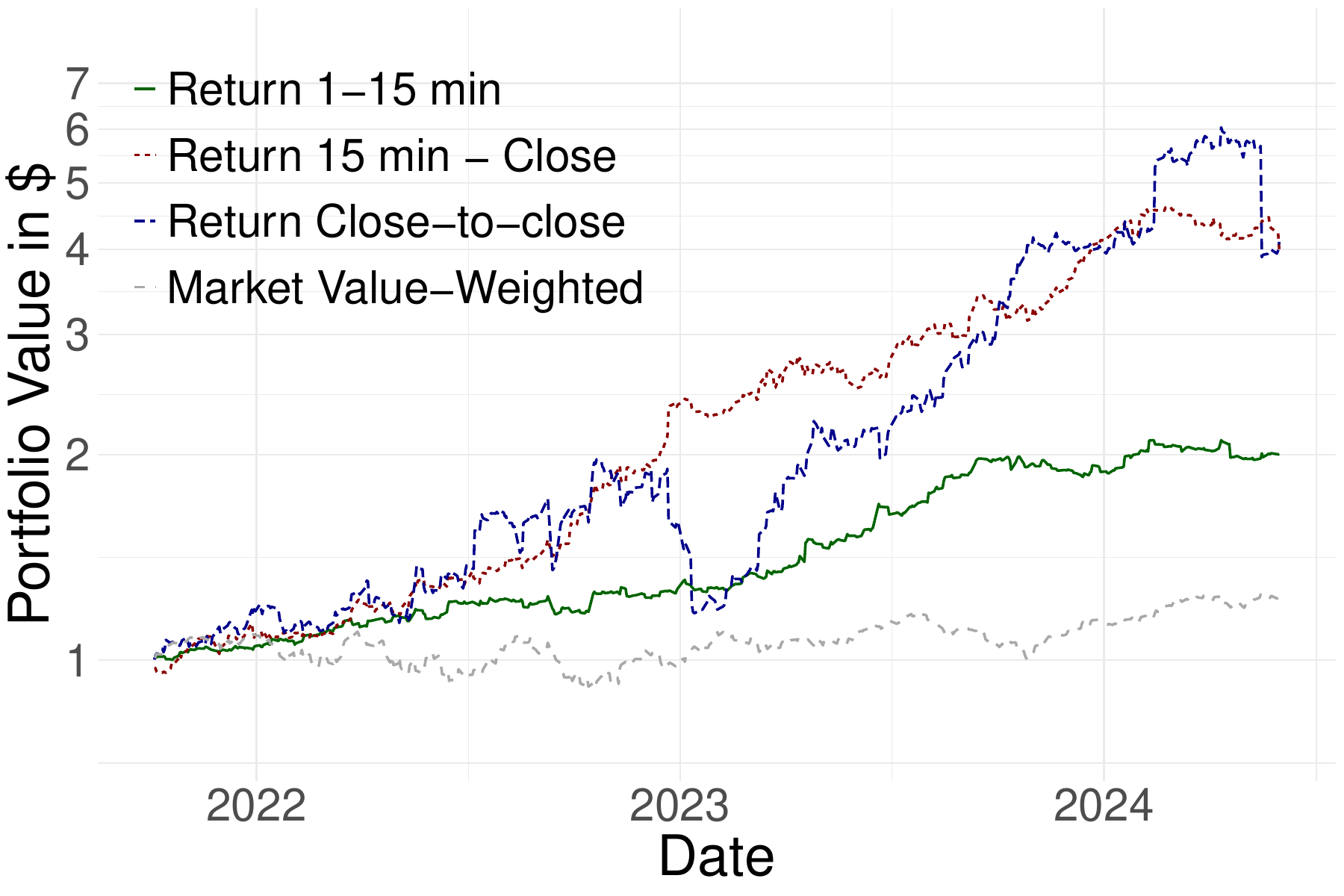}

  \label{fig:cumulative_returns_entry}

  \vspace{0.5cm}

\begin{minipage}{15cm}
\small
This figure shows the performance of trading strategies based on ChatGPT 4's prediction scores of intraday news, without considering transaction costs. For intraday news released between 9:30 a.m. and 4 p.m. of trading day $t$, we form three strategies with different entry times, each creating an equal-weighted long-short portfolio that buys companies with good news and short-sells companies with bad news according to ChatGPT 4. The first strategy (the green line) enters the position one minute after the news announcement and exits 15 minutes after the news announcement on day $t$; the second one (the red line) enters the position 15 minutes after the news announcement and exits at the market close of day $t$; the third one (the blue line) enters the position at the market close of day $t$ and exits at the market close of day $t$+1. All the strategies are rebalanced daily. The grey line corresponds to a value-weighted market portfolio without transaction costs.
\end{minipage}
\vspace{0.5cm} 
\end{figure}

\begin{figure}[htbp]
  \centering
    \caption*{Figure OA2: Year-by-Year Performance of Intraday News Strategy: Sharpe Ratios}
      \vspace{0.5cm}
  \includegraphics[width=0.8\textwidth]{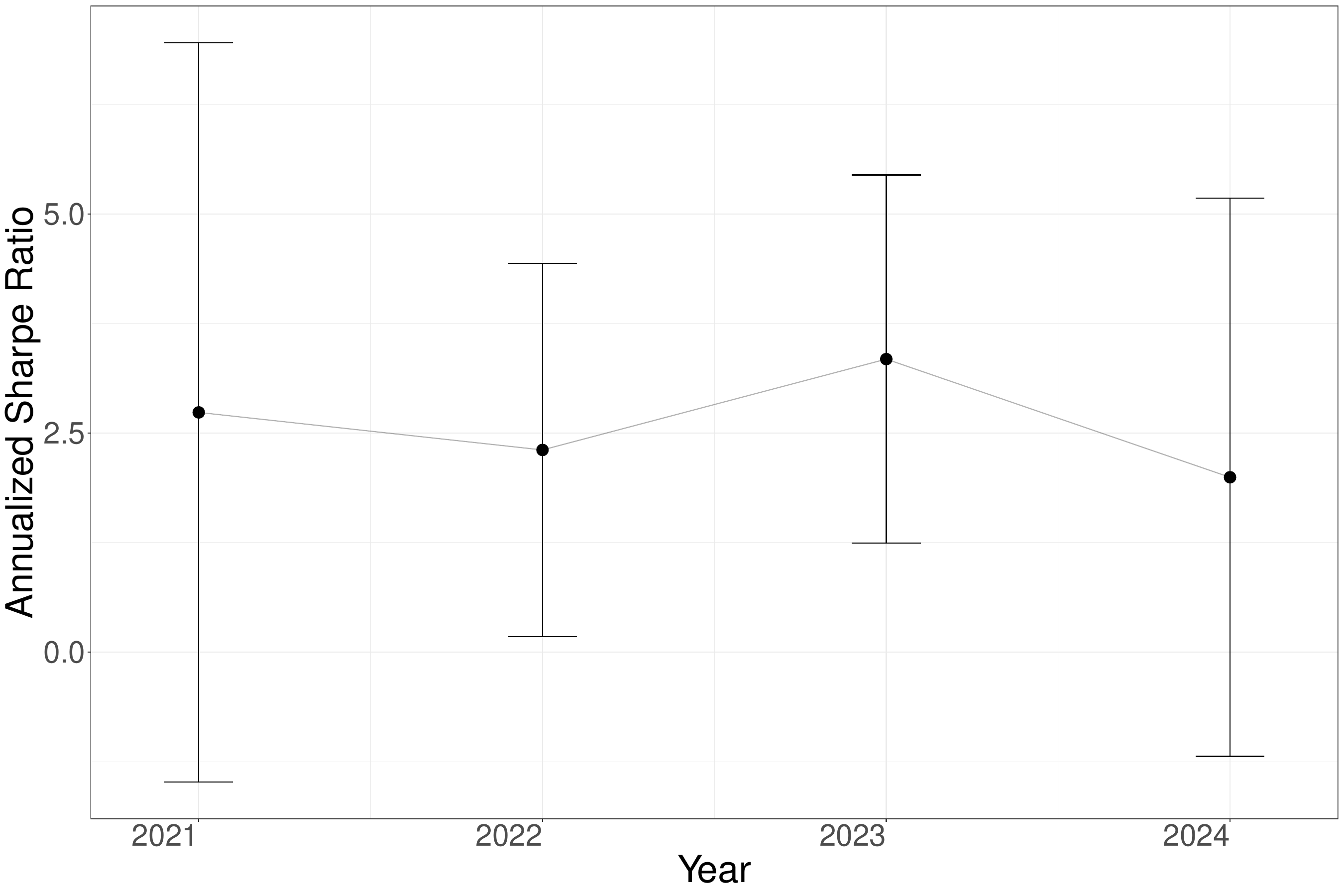}

  \label{fig:yearly_sharpe_ratio_intraday}

  \vspace{0.5cm}
\begin{minipage}{15cm}
\small
This figure shows the year-by-year Sharpe ratios of the long-short strategy based on ChatGPT 4 for intraday news. For intraday news released between 9:30 a.m. and 4 p.m. on trading day $t$, we enter the position at the close of day $t$ and exit at the close of day $t$+1. All the strategies are rebalanced daily. This figure presents the annualized Sharpe ratios of the strategy for four subperiods of our sample (2021 Oct.-Dec., 2022, 2023, and 2024 Jan.-May) and their 95\% confidence intervals calculated following \citet{lo2002statistics}.
\end{minipage}
\vspace{0.5cm}
\end{figure}

\begin{figure}[htbp]
  \centering
    \caption*{Figure OA3: Time Variation in Portfolio Composition}
      \vspace{0.5cm}
  \includegraphics[width=0.8\textwidth]{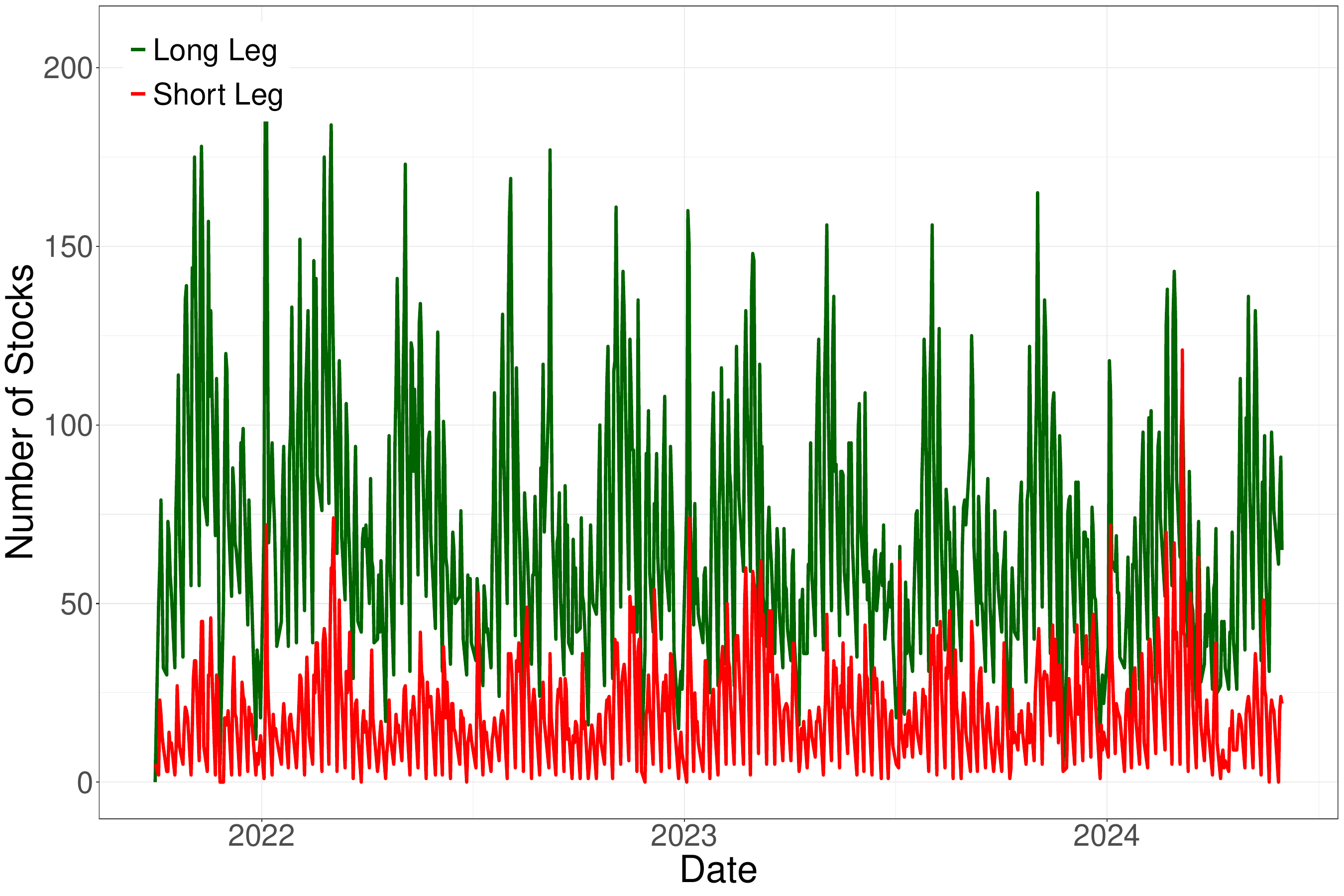}

  \label{fig:portfolio_composition}

  \vspace{0.5cm}
\begin{minipage}{15cm}
\small
This figure shows the daily number of stocks held in the long and short legs of the overnight news strategy over the sample period (October 2021 - May 2024). The long leg (green) consists of stocks with positive GPT-4 scores, while the short leg (red) consists of stocks with negative GPT-4 scores. On average, the strategy holds 74 stocks in the long leg and 21 stocks in the short leg per day. The variation in portfolio composition reflects fluctuations in the daily availability of news and the proportion of positive vs negative news events.
\end{minipage}
\vspace{0.5cm}
\end{figure}

\begin{figure}[htbp]
  \centering
    \caption*{Figure OA4: Daily Portfolio Turnover}
      \vspace{0.5cm}
  \includegraphics[width=0.8\textwidth]{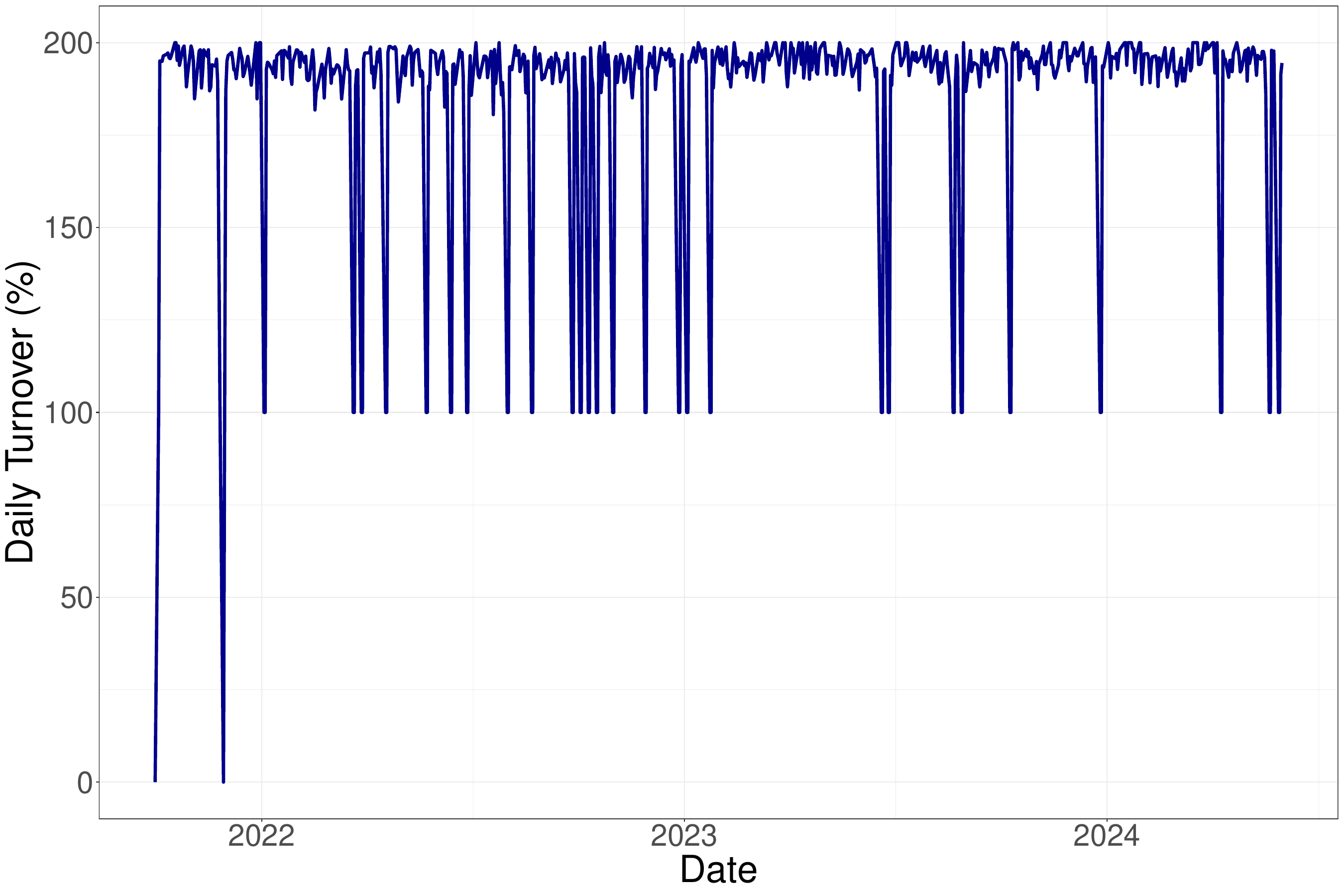}

  \label{fig:daily_turnover}

  \vspace{0.5cm}
\begin{minipage}{15cm}
\small
This figure shows the daily turnover of the overnight news strategy over the sample period (October 2021 - May 2024). Turnover is measured as one-way turnover: the sum of absolute weight changes across all stocks divided by 2, representing the fraction of portfolio value traded. The strategy exhibits very high turnover, averaging 186\% daily, which reflects the fact that most news events are single-day occurrences and stocks enter and exit the portfolio rapidly. This high turnover makes the strategy highly sensitive to transaction costs, as documented in the main text.
\end{minipage}
\vspace{0.5cm}
\end{figure}

\newpage
\begin{center}
\section*{Online Appendix C: GPT-4 vs. GPT-3.5 Performance Over Time}
\end{center}

\label{appendix:gpt4vsgpt3}

\begin{onehalfspacing}

%\subsection*{GPT-4 vs. GPT-3.5 Performance Over Time}

One potential concern with our GPT-4 results is lookahead bias. While OpenAI's documentation states that GPT-4's training data cutoff is September 2021, the model may have been exposed to additional data during fine-tuning or through other means. If GPT-4 had somehow ``seen" news from 2022-2024 during its training process, this could artificially inflate its performance on our sample period. 

To further address this concern, we compare GPT-4's performance to GPT-3.5 (which has a training cutoff of June 2021) across our sample period. If GPT-4 suffers from lookahead bias, we would expect its advantage over GPT-3.5 to \textit{decline} over time, as any memorized information becomes increasingly stale. Instead, we find the opposite pattern.\footnote{Relatedly, recent work by \citet{lopez2025memorization} provides a comprehensive set of evidence that ChatGPT can accurately recall economic and financial information, including macroeconomic variables, stock returns, news headlines, and earnings call content, before the model's knowledge cutoff, and finds no evidence of lookahead bias or memorization after the cutoff.}

\hfill \break

\noindent \textbf{Methodology:} We calculate the difference in performance between GPT-4 and GPT-3.5 at the headline level. For each news headline, we compute:
\begin{equation*}
\text{Performance}_i = \text{Score}_i \times \text{Return}_i \times 10{,}000
\end{equation*}
where $\text{Score}_i \in \{-1, 0, 1\}$ is the model's sentiment classification and $\text{Return}_i$ is the stock's return following the news. We decompose returns into two components: (1) the initial market reaction (measured from news release to market close) and (2) the overnight drift (measured from close to next day's open). We then calculate the mean difference (GPT-4 minus GPT-3.5) in basis points for each year.

\hfill \break

\noindent \textbf{Results:} Figure OA5 %\ref{fig:gpt4_vs_gpt3_init} 
shows GPT-4's advantage in predicting the initial market reaction. The advantage remains relatively stable across years, ranging from 5.1 to 10.2 basis points. Figure OA6 %\ref{fig:gpt4_vs_gpt3_drift} 
shows GPT-4's advantage in predicting overnight drift. Here we observe a striking pattern: GPT-4's advantage \textit{increases} from essentially zero in 2021 (-0.4 bp) to 8.9 basis points in 2023, before moderating to 3.7 basis points in 2024.

\hfill \break

\noindent \textbf{Interpretation:} The increasing advantage in drift prediction does not seem consistent with lookahead bias driving the performance. If GPT-4 had memorized information about 2022-2024 news during training, we would expect its advantage to \textit{decline} as this knowledge becomes outdated. The observed pattern---particularly the increase from -0.4 basis points in 2021 to 8.9 basis points in 2023 (t-statistic = 3.35)---suggests that GPT-4's superior performance may stem from genuine advances in language understanding. The pattern is particularly pronounced for overnight drift, which requires understanding subtle implications of news that may not be immediately reflected in prices.

\begin{figure}[H]
    \centering
    \includegraphics[width=0.85\textwidth]{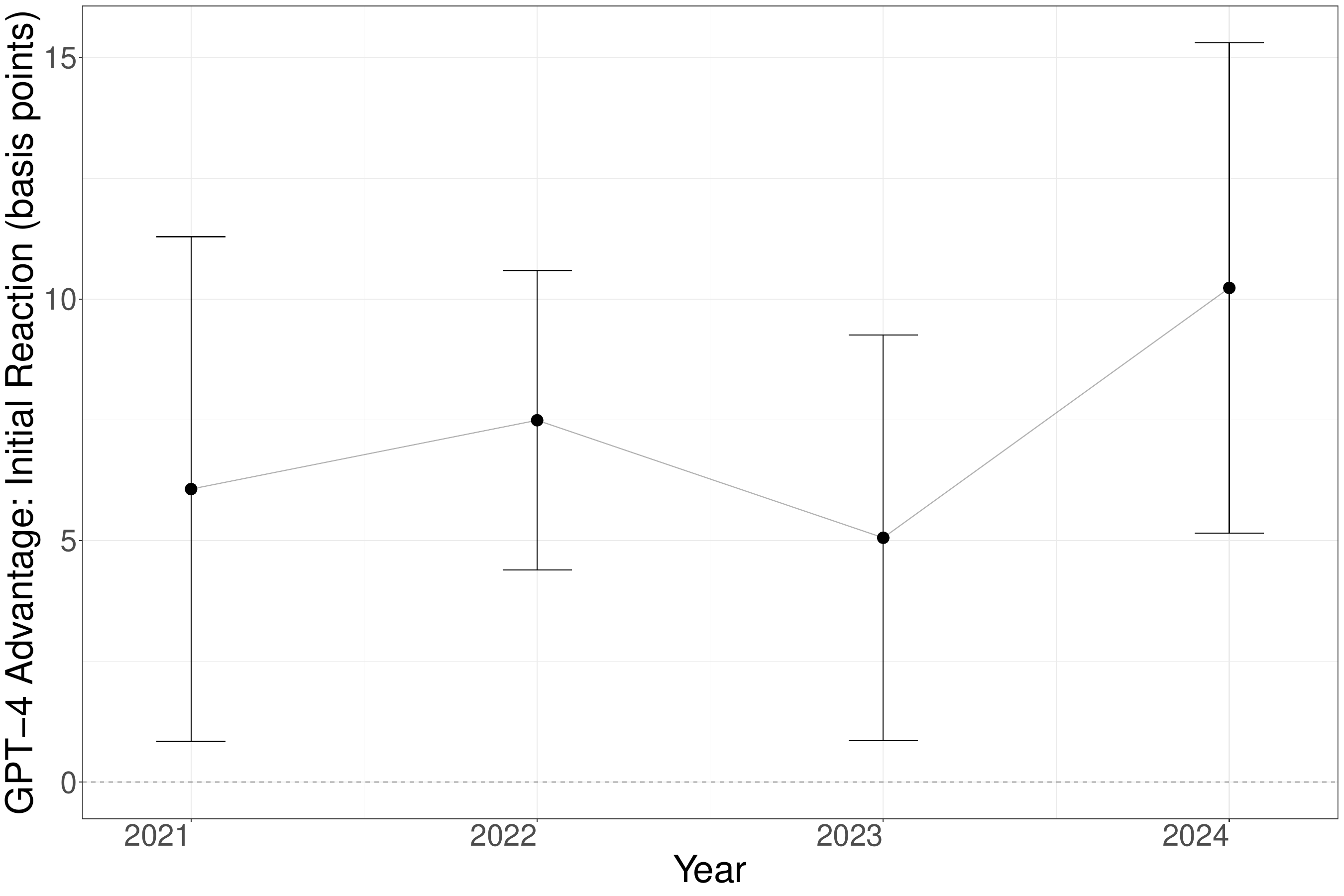}
    \caption*{Figure OA5: GPT-4 Advantage Over GPT-3.5: Initial Reaction}
    \label{fig:gpt4_vs_gpt3_init}
    \begin{minipage}{0.95\textwidth}
        \footnotesize
        \textit{Notes:} This figure shows the annual mean difference in performance (GPT-4 minus GPT-3.5) for predicting initial market reactions to news, measured in basis points. Each point represents the average across all headlines in that year, with error bars showing 95\% confidence intervals. The initial reaction is measured from news release to market close. Sample includes all news headlines from October 2021 to May 2024 with both GPT-4 and GPT-3.5 sentiment scores.
    \end{minipage}
\end{figure}

\begin{figure}[H]
    \centering
    \includegraphics[width=0.85\textwidth]{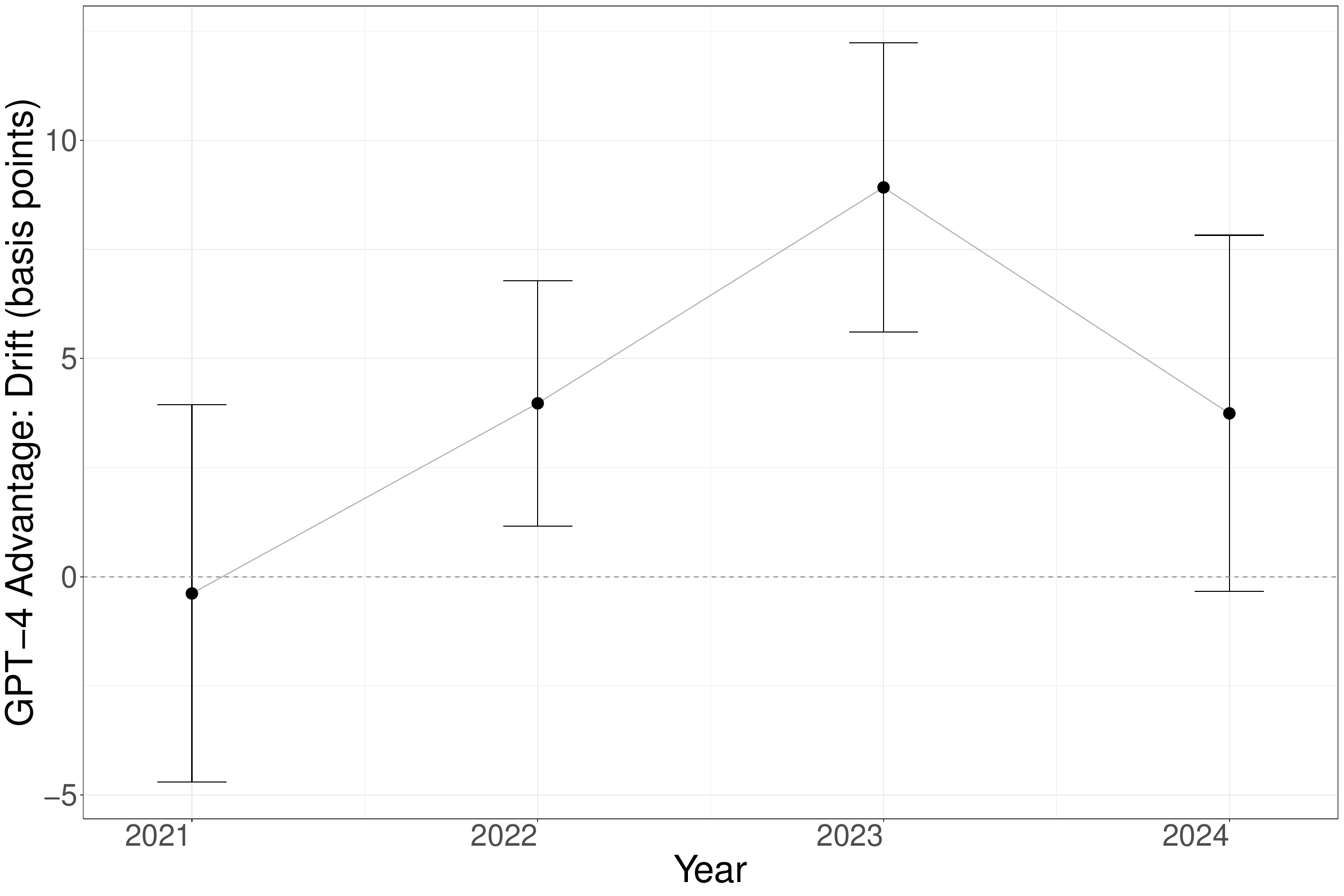}
    \caption*{Figure OA6: GPT-4 Advantage Over GPT-3.5: Overnight Drift}
    \label{fig:gpt4_vs_gpt3_drift}
    \begin{minipage}{0.95\textwidth}
        \footnotesize
        \textit{Notes:} This figure shows the annual mean difference in performance (GPT-4 minus GPT-3.5) for predicting overnight drift following news, measured in basis points. Each point represents the average across all headlines in that year, with error bars showing 95\% confidence intervals. Overnight drift is measured from market close on the news day to the next day's open. The increasing advantage over time provides evidence against lookahead bias, as memorized information would become less valuable in later years. Sample includes all news headlines from October 2021 to May 2024 with both GPT-4 and GPT-3.5 sentiment scores.
    \end{minipage}
\end{figure}

\end{onehalfspacing}

\newpage
\begin{landscape}

\begin{center}
\section*{Online Appendix D: Model Proofs} \label{appendix:proofs}
\end{center}

In this section, we provide the detailed proofs for the theoretical results in Section B of the Appendix in the paper. We use Mathematica 14.0 to verify all the sign inequalities, derivatives, and expectations. 
\vspace{0.8cm}

\section{Equilibrium Price in Period 2 ($p_2$)}

In period 2, both attentive and inattentive agents have the same conditional expectation $\mu_{A|s} = E[\tilde{d}|s_A]$ and conditional variance of the asset payoff $\sigma_{d,A|s}^2 = Var[\tilde{d}|s_A]$. 
The demand for the risky asset by any such agent $j$ is given by the standard CARA demand:
\begin{equation}
    x_{3,j} = \frac{E_j[\tilde{d}] - p_2}{\alpha Var_j[\tilde{d}]} = \frac{\mu_{A|s} - p_2}{\alpha \sigma_{d, A|s}^2}.
\end{equation}
The total measure of these agents is $V$. The model incorporates a noise trading shock $u_2 \sim N(0, \sigma_u^2)$ which affects the price. This shock can be interpreted as the net demand from noise traders that needs to be absorbed by other market participants, or directly as a shock to the price.
If the aggregate demand from the informed (attentive/inattentive) agents, $V x_{3,j}$, must offset the scaled impact of noise traders, leading to the equilibrium price $p_2$ as given in the main text:
\begin{equation}
    p_2 = \mu_{A|s}  + u_2\frac{\alpha \sigma_{d, A|s}^2}{V}.
\end{equation}
To verify consistency, substituting this $p_2$ back into the aggregate demand $D_3 = V x_{3,j}$:
\begin{equation}
    D_3 = V \frac{\mu_{A|s} - (\mu_{A|s}  + u_2\frac{\alpha \sigma_{d, A|s}^2}{V})}{\alpha \sigma_{d, A|s}^2} = V \frac{- u_2\frac{\alpha \sigma_{d, A|s}^2}{V}}{\alpha \sigma_{d, A|s}^2} = -u_2.
\end{equation}
This confirms that the informed agents collectively absorb the noise trading shock $u_2$ (with a negative sign indicating they trade against it if $u_2$ is defined as pushing the price in its direction).

\section{Bayesian Information Update Formulas}

As described in the main text, upon observing the signal $s$, agents update their beliefs. The prior belief about the fundamental $\mu_d$ is $\mu_d \sim N(\bar{d}, \sigma_d^2)$, so the prior precision is $\tau_d = 1/\sigma_d^2$. The signal is $s = \mu_d + \varepsilon$, where $\varepsilon \sim N(0, \sigma_S^2)$, and $\tau_S = 1/\sigma_S^2$ is the total precision of the news.

For an agent type $j \in \{A,I, LLM\}$ processing the signal $s$ with effective precision $\tau_j$ (where $\tau_A = \gamma_A \tau_S$, $\tau_I = \omega \tau_A$, and $\tau_L = \lambda(c,k) \tau_S$):

The updated posterior expectation of the fundamental is given by the standard Bayesian formula:
\begin{equation}
     \mu_{j|s} \equiv  E[\mu_d|s_j]   =  \frac{\bar{d}  \tau_d + s \tau_{j} }{\tau_d + \tau_{j}}.
\end{equation}

The updated posterior variance of the fundamental is:
\begin{equation}
   Var[\mu_d|s_j] = \sigma_{\mu, j|s}^2 \equiv \frac{1}{\tau_{\mu,j|s}} \equiv \frac{1}{\tau_d + \tau_{j}}.
\end{equation}

Given that the asset payoff is $\tilde{d} = \mu_d + \sigma_{\xi} \tilde{\xi}$ with $\tilde{\xi} \sim N(0,1)$ and $\sigma_\xi^2$ being the variance of the real shock, the total conditional variance of the asset payoff for agent $j$ is:
\begin{equation}
    \sigma_{d, j|s}^2 \equiv Var[\tilde{d}|s_j] = E[Var[\tilde{d}|\mu_d, s_j]] + Var[E[\tilde{d}|\mu_d, s_j]|s_j] = \sigma_\xi^2 + \sigma_{\mu, j|s}^2.
\end{equation}

\section{Attentive Agent's Demand}

Substituting the attentive agent's optimal demand, the value function can be written as:

\begin{align}
    V_2(w_2) & = \max_{x_3}  -exp\{ -\alpha E[\tilde{w_3}] + \frac{\alpha^2}{2} Var(\tilde{w_3})\}\\
    &= \max_{x_3} -exp\{ -\alpha(w_2 + x_3 (\mu_{A|s} - p_2)) + \frac{\alpha^2}{2} x^2  \sigma_{d, A|s}^2 \}\\
 &= -exp\{ -\alpha(w_2 + \frac{\mu_{A|s} - p_2}{\alpha \sigma_{d, A|s}^2} (\mu_{A|s} - p_2)) + \frac{\alpha^2}{2} (\frac{\mu_{A|s} - p_2}{\alpha \sigma_{d, A|s}^2})^2  \sigma_{d, A|s}^2    \}\\
 & = -exp\{ -\alpha(w_2) - \frac{(\mu_{A|s} - p_2)^2}{ \sigma_{d, A|s}^2} + \frac{1}{2} \frac{(\mu_{A|s} - p_2)^2}{ \sigma_{d, A|s}^2}     \}\\
 & = -exp\{ -\alpha(w_2) - \frac{1}{2} \frac{(\mu_{A|s} - p_2)^2}{ \sigma_{d, A|s}^2} \}.
\end{align}

In period 1, attentive agents maximize the expected period 2 value function:

\begin{align}
    V1(w_1) &=  \max_{x_2} E_1[V2(\tilde{w_2})] \\
    & \tilde{w_2} = w_1 + x_2 (\tilde{p_2} - p_1),
\end{align}

which we can rewrite as

\begin{align}
    V1(w_1) &=  \max_{x} E_1[V_2( \tilde{w_2})] \\
    &= \max_{x} -E_1[exp\{ -\alpha(w_2) - \frac{1}{2} \frac{(\mu_{A|s} - \tilde{p_2})^2}{ \sigma_{d, A|s}^2} \}]\\
    &= \max_{x} -E_1[exp\{ -\alpha (w_1 + x (\tilde{p_2} - p_1)) - \frac{1}{2} \frac{(\mu_{A|s} - \tilde{p_2})^2}{ \sigma_{d, A|s}^2} \}] \\
     &= \max_x  -E_1[exp\{ a(x) + b(x) \tilde{p_2} + c \tilde{p_2}^2 \}]
\end{align}

with $a = a(x)= \alpha p_1 x-\frac{\mu_{A|s}^2}{2 \sigma_{d, A|s}^2}-\alpha w_1$, $b = b(x) = \frac{\mu_{A|s}}{\sigma_{d, A|s}^2}-\alpha x$, and $c = -\frac{1}{2 \sigma_{d, A|s}^2}$.
\vspace{0.5cm}

To get a closed-form solution, we use the following Lemma:

\begin{lemma}\label{lem:mgf_quadratic}
    Let $z \sim N(\mu, \sigma^2)$. Then, if  $1-2c\sigma^2 >  0$,
    \begin{align}
        E[\exp(a + bz + cz^2)] = \frac{\exp\left(-\frac{\sigma^2\left(b^2-4ac\right)+2(a+\mu(b+c\mu))}{4c\sigma^2-2}\right)}{\sqrt{1-2c\sigma^2}}.
    \end{align}
\begin{proof}
    The omitted proof follows from completing the square in the probability density function.
\end{proof}
\end{lemma}

\begin{align}
    G(w, x) = \frac{\exp\left(\frac{-4ac \sigma_p ^2+2a+b^2 \sigma_p ^2+2b \mu_p +2c \mu_p ^2}{2-4c \sigma_p ^2}\right)}{\sqrt{1-2c \sigma_p ^2}}.
\end{align}

Substituting the values of a, b, and c, we obtain:

\begin{align}
    G(w, x) \equiv -\frac{\exp\left(-\frac{2\left(\alpha p_1 x-\frac{ \mu_{A|s} ^2}{2 \sigma_{d, A|s}^2}+ \mu_{A|s} \left(\frac{ \mu_{A|s} }{2 \sigma_{d, A|s}^2}-\alpha x\right)-\alpha w \right)+\frac{\alpha^2 \sigma_{d, A|s} ^4 \sigma_u ^2}{V^2}\left(\frac{2\left(\alpha p_1 x-\frac{ \mu_{A|s} ^2}{2 \sigma_{d, A|s}^2}-\alpha w \right)}{ \sigma_{d, A|s}^2}+\left(\frac{ \mu_{A|s} }{ \sigma_{d, A|s}^2}-\alpha x\right)^2\right)}{-\frac{2\alpha^2 \sigma_{d, A|s}^2 \sigma_u ^2}{V^2}-2}\right)}{\sqrt{\frac{\alpha^2 \sigma_{d, A|s}^2 \sigma_u ^2}{V^2}+1}}.
\end{align}

We then use the first-order condition:

\begin{align}
    \frac{\partial G}{\partial x}|_{x=x_{2,A}}
 =& -\frac{\left(2(\alpha p_1-\alpha \mu_{A|s})+\frac{\alpha^2 \sigma_{d, A|s}^4 \sigma_u^2}{V^2}\left(\frac{2\alpha p_1}{ \sigma_{d, A|s}^2}-2\alpha\left(\frac{ \mu_{A|s}}{ \sigma_{d, A|s}^2}-\alpha x_{2,A}\right)\right)\right)G(w_1, x_{2,A})}{\left(-\frac{2\alpha^2 \sigma_{d, A|s}^2 \sigma_u^2}{V^2}-2\right)\sqrt{\frac{\alpha^2 \sigma_{d, A|s}^2 \sigma_u^2}{V^2}+1}} &= 0\\
    \Leftrightarrow& \\
    & 2(\alpha p_1-\alpha \mu_{A|s})+\frac{\alpha^2 \sigma_{d, A|s}^4 \sigma_u^2}{V^2}\left(\frac{2\alpha p_1}{ \sigma_{d, A|s}^2}-2\alpha\left(\frac{ \mu_{A|s}}{ \sigma_{d, A|s}^2}-\alpha x_{2,A}\right)\right) &= 0\\
    \Rightarrow& \\
   x_{2,A} &=  \frac{(\mu_{A|s} - p_1)\left(\frac{\alpha^2 \sigma_{d, A|s}^2 \sigma_u^2}{V^2}+1\right)}{\frac{\alpha^3 \sigma_{d, A|s}^4 \sigma_u^2}{V^2}}
\end{align}

Notice that we can write the demand function as  
\begin{align}
    x_{2,A} &= \frac{(\mu_{A|s} - p_1)\left(\frac{\alpha^2 \sigma_{d, A|s}^2 \sigma_u^2}{V^2}+1\right)}{\frac{\alpha^3 \sigma_{d, A|s}^4 \sigma_u^2}{V^2}} \\ 
    &= \frac{(\mu_{A|s} - p_1)\left(\frac{\alpha^2 \sigma_{d, A|s}^2 \sigma_u^2}{V^2}\right)}{\frac{\alpha^3 \sigma_{d, A|s}^4 \sigma_u^2}{V^2}} + \frac{(\mu_{A|s} - p_1)}{\frac{\alpha^3 \sigma_{d, A|s}^4 \sigma_u^2}{V^2}}\\
    & = \frac{(\mu_{A|s} - p_1)}{\alpha \sigma_{d, A|s}^2} + \frac{(\mu_p - p_1)}{\alpha\sigma_p^2}.
\end{align}

\section{Proposition \ref{prop:mispricing} - Mispricing}

Mispricing is defined as:

\begin{align}
   E[\alpha_M^2] & \equiv E[(E[\tilde{d}|s] - E[p_1|s])^2].
\end{align}

Evaluating the expectation, we get:

\begin{tiny}\begin{align}
  E[\alpha_M^2] =&\\
  &\frac{\tau _D^2 \tau _S \left(\alpha ^2 \sigma _u^2 \left(\sigma _{\xi }^2 \left((\tau_{\mu,A|s})\right)+1\right)
   \left(-\pi _A \gamma _A-\left(\left((-\pi_I)\right) \tau _D \sigma _{\xi }^2\right)+\left((-\pi_I)\right) \gamma _A
   \left(\omega  \tau _D \sigma _{\xi }^2-\sigma _{\xi }^2 \tau _S+\omega \right)+\left((-\pi_I)\right) \omega  \gamma _A^2
   \sigma _{\xi }^2 \tau _S+1\right)-\pi _A V^2 \left(\gamma _A-1\right) \left(\tau_A+\tau_D\right)\right){}^2}{\left(\tau _D+\tau _S\right){}^2 \left(\alpha ^2 \sigma _u^2 \left(\sigma _{\xi }^2 \left(\gamma _A
   \tau _S+\tau _D\right)+1\right) \left(\left((-\pi_I)\right) \tau _D^2 \sigma _{\xi }^2+\tau _D \left(\left((-\pi_I)\right)
   (\omega +1) \gamma _A \sigma _{\xi }^2 \tau _S-1\right)+\tau_A \left(\left((-\pi_I)\right) \omega  \left(\gamma _A
   \sigma _{\xi }^2 \tau _S+1\right)-\pi _A\right)\right)-\pi _A V^2 \left((\tau_{\mu,A|s})\right){}^2\right){}^2}
\end{align}\end{tiny}

The derivative w.r.t. the quantity of attentive agents, $\pi_A$, is:

\begin{tiny}\begin{align}
   \frac{\partial E[\alpha_M^2]}{\partial \pi_A}&=\\
   &-(2 \alpha ^2 (\omega -1) \gamma _A \tau _D^2 \tau _S \sigma _u^2 \times\\
   & \left(\sigma _{\xi }^2 \left(\tau_A+\tau
   _D\right)+1\right){}^2 \times\\
   & \left(\alpha ^2 \sigma _u^2 \left(\sigma _{\xi }^2 \left((\tau_{\mu,A|s})\right)+1\right)+V^2
   \left((\tau_{\mu,A|s})\right)\right)\times \\
   &\left(\alpha ^2 \sigma _u^2 \left(\sigma _{\xi }^2 \left(\tau_A+\tau
   _D\right)+1\right) \left(-\pi _A \gamma _A-\left(\left((-\pi_I)\right) \tau _D \sigma _{\xi }^2\right)+\left((-\pi_I)\right)
   \gamma _A \left(\omega  \tau _D \sigma _{\xi }^2-\sigma _{\xi }^2 \tau _S+\omega \right)+\left((-\pi_I)\right) \omega  \gamma
   _A^2 \sigma _{\xi }^2 \tau _S+1\right)-\pi _A V^2 \left(\gamma _A-1\right) \left(\tau_A+\tau
   _D\right)\right))/\\
   & (\left(\tau _D+\tau _S\right) \left(\alpha ^2 \sigma _u^2 \left(\sigma _{\xi }^2 \left(\gamma _A \tau
   _S+\tau _D\right)+1\right) \left(\left((-\pi_I)\right) \tau _D^2 \sigma _{\xi }^2+\tau _D \left(\left((-\pi_I)\right) (\omega
   +1) \gamma _A \sigma _{\xi }^2 \tau _S-1\right)+\tau_A \left(\left((-\pi_I)\right) \omega  \left(\gamma _A \sigma
   _{\xi }^2 \tau _S+1\right)-\pi _A\right)\right)-\pi _A V^2 \left((\tau_{\mu,A|s})\right){}^2\right){}^3).
\end{align}\end{tiny}

This derivative is negative since every term in the numerator is negative, and the denominator is positive.

The derivative w.r.t. the information capacity of attentive agents, $\gamma_A$, is:

\begin{tiny}\begin{align}
   \frac{\partial E[\alpha_M^2]}{\partial \gamma_A}&=\\
   &(2 \tau _D^2 \tau _S \left(\alpha ^2 \sigma _u^2 \left(\sigma _{\xi }^2 \left((\tau_{\mu,A|s})\right)+1\right)
   \left(-\pi _A \gamma _A-\left(\left((-\pi_I)\right) \tau _D \sigma _{\xi }^2\right)+\left((-\pi_I)\right) \gamma _A
   \left(\omega  \tau _D \sigma _{\xi }^2-\sigma _{\xi }^2 \tau _S+\omega \right)+\left((-\pi_I)\right) \omega  \gamma _A^2
   \sigma _{\xi }^2 \tau _S+1\right)-\pi _A V^2 \left(\gamma _A-1\right) \left((\tau_{\mu,A|s})\right)\right) \times \\
   & (\alpha ^4 \sigma _u^4 \left(\sigma _{\xi }^2 \left((\tau_{\mu,A|s})\right)+1\right){}^2 \left(\left(\pi
   _A-1\right) \left(\omega  \left(\left((-\pi_I)\right) \sigma _{\xi }^4 \left((\tau_{\mu,A|s})\right){}^2+\sigma
   _{\xi }^2 \left(\left(\pi _A-2\right) \tau _D-2 \tau_A\right)-1\right)-\pi _A \tau _D \sigma _{\xi }^2\right)+\pi
   _A\right) -\\
   & \alpha ^2 \pi _A V^2 \sigma _u^2 \left(\sigma _{\xi }^2 \left((\tau_{\mu,A|s})\right)+1\right)
   \left(\left((-\pi_I)\right) (\omega +1) \tau _D^2 \sigma _{\xi }^2+\tau _D \left(\left((-\pi_I)\right) (3 \omega +1) \gamma
   _A \sigma _{\xi }^2 \tau _S+\pi _A \omega -\pi _A-\omega -1\right)+2 \tau_A \left(\left((-\pi_I)\right) \omega 
   \gamma _A \sigma _{\xi }^2 \tau _S-1\right)\right)+\pi _A^2 V^4 \left((\tau_{\mu,A|s})\right){}^2))/\\
   & (\left(\tau
   _D+\tau _S\right) \left(\alpha ^2 \sigma _u^2 \left(\sigma _{\xi }^2 \left((\tau_{\mu,A|s})\right)+1\right)
   \left(\left((-\pi_I)\right) \tau _D^2 \sigma _{\xi }^2+\tau _D \left(\left((-\pi_I)\right) (\omega +1) \gamma _A \sigma _{\xi
   }^2 \tau _S-1\right)+\tau_A \left(\left((-\pi_I)\right) \omega  \left(\gamma _A \sigma _{\xi }^2 \tau
   _S+1\right)-\pi _A\right)\right)-\pi _A V^2 \left((\tau_{\mu,A|s})\right){}^2\right){}^3).
\end{align}\end{tiny}

This derivative is negative since every term in the numerator is negative, and the denominator is positive.

The derivative w.r.t. the information capacity of inattentive agents, $\omega$, is:

\begin{tiny}\begin{align}
   \frac{\partial E[\alpha_M^2]}{\partial \omega}&=\\
   &(2 \alpha ^2 \left((-\pi_I)\right) \gamma _A \tau _D^2 \tau _S \sigma _u^2 \left(\sigma _{\xi }^2 \left(\gamma _A \tau
   _S+\tau _D\right)+1\right){}^2 \times \\
   & \left(\alpha ^2 \sigma _u^2 \left(\sigma _{\xi }^2 \left(\tau_A+\tau
   _D\right)+1\right) \left(\left((-\pi_I)\right) \sigma _{\xi }^2 \left((\tau_{\mu,A|s})\right)-1\right)-\pi _A V^2
   \left((\tau_{\mu,A|s})\right)\right) \times \\
   & \left(\alpha ^2 \sigma _u^2 \left(\sigma _{\xi }^2 \left(\tau_A+\tau
   _D\right)+1\right) \left(-\pi _A \gamma _A-\left(\left((-\pi_I)\right) \tau _D \sigma _{\xi }^2\right)+\left((-\pi_I)\right)
   \gamma _A \left(\omega  \tau _D \sigma _{\xi }^2-\sigma _{\xi }^2 \tau _S+\omega \right)+\left((-\pi_I)\right) \omega  \gamma
   _A^2 \sigma _{\xi }^2 \tau _S+1\right)-\pi _A V^2 \left(\gamma _A-1\right) \left(\tau_A+\tau
   _D\right)\right))/\\
   &(\left(\tau _D+\tau _S\right) \left(\alpha ^2 \sigma _u^2 \left(\sigma _{\xi }^2 \left(\gamma _A \tau
   _S+\tau _D\right)+1\right) \left(\left((-\pi_I)\right) \tau _D^2 \sigma _{\xi }^2+\tau _D \left(\left((-\pi_I)\right) (\omega
   +1) \gamma _A \sigma _{\xi }^2 \tau _S-1\right)+\tau_A \left(\left((-\pi_I)\right) \omega  \left(\gamma _A \sigma
   _{\xi }^2 \tau _S+1\right)-\pi _A\right)\right)-\pi _A V^2 \left((\tau_{\mu,A|s})\right){}^2\right){}^3).
\end{align}\end{tiny}

This derivative is negative since the last term in the numerator is negative, while all other terms are positive.

The derivative w.r.t. the total volume of traders, $V$, is also negative and given by:

\begin{tiny}\begin{align}
   \frac{\partial E[\alpha_M^2]}{\partial V}=&\\
   &(4 \alpha ^2 \left((-\pi_I)\right) \pi _A V (\omega -1) \gamma _A \tau _D^2 \tau _S \sigma _u^2 \left(\tau_A+\tau
   _D\right) \times \\
   &\left(\sigma _{\xi }^2 \left((\tau_{\mu,A|s})\right)+1\right){}^2 \left(\alpha ^2 \sigma _u^2 \left(\sigma
   _{\xi }^2 \left((\tau_{\mu,A|s})\right)+1\right) \left(-\pi _A \gamma _A-\left(\left((-\pi_I)\right) \tau _D \sigma _{\xi
   }^2\right)+\left((-\pi_I)\right) \gamma _A \left(\omega  \tau _D \sigma _{\xi }^2-\sigma _{\xi }^2 \tau _S+\omega
   \right)+\left((-\pi_I)\right) \omega  \gamma _A^2 \sigma _{\xi }^2 \tau _S+1\right)-\pi _A V^2 \left(\gamma _A-1\right)
   \left((\tau_{\mu,A|s})\right)\right))/\\
   &(\left(\tau _D+\tau _S\right) \left(\alpha ^2 \sigma _u^2 \left(\sigma _{\xi }^2
   \left((\tau_{\mu,A|s})\right)+1\right) \left(\left((-\pi_I)\right) \tau _D^2 \sigma _{\xi }^2+\tau _D \left(\left(\pi
   _A-1\right) (\omega +1) \gamma _A \sigma _{\xi }^2 \tau _S-1\right)+\tau_A \left(\left((-\pi_I)\right) \omega 
   \left(\gamma _A \sigma _{\xi }^2 \tau _S+1\right)-\pi _A\right)\right)-\pi _A V^2 \left(\tau_A+\tau
   _D\right){}^2\right){}^3.
\end{align}\end{tiny}

The derivative w.r.t. the risk aversion, $\alpha$, is:

\begin{tiny}\begin{align}
    \frac{\partial E[\alpha_M^2]}{\partial \alpha}=&\\
   &-(4 \alpha  \left((-\pi_I)\right) \pi _A V^2 (\omega -1) \gamma _A \tau _D^2 \tau _S \sigma _u^2 \left(\tau_A+\tau
   _D\right) \times \\
   & \left(\sigma _{\xi }^2 \left((\tau_{\mu,A|s})\right)+1\right){}^2 \left(\alpha ^2 \sigma _u^2 \left(\sigma
   _{\xi }^2 \left((\tau_{\mu,A|s})\right)+1\right) \left(-\pi _A \gamma _A-\left(\left((-\pi_I)\right) \tau _D \sigma _{\xi
   }^2\right)+\left((-\pi_I)\right) \gamma _A \left(\omega  \tau _D \sigma _{\xi }^2-\sigma _{\xi }^2 \tau _S+\omega
   \right)+\left((-\pi_I)\right) \omega  \gamma _A^2 \sigma _{\xi }^2 \tau _S+1\right)-\pi _A V^2 \left(\gamma _A-1\right)
   \left((\tau_{\mu,A|s})\right)\right))/\\
   & (\left(\tau _D+\tau _S\right) \left(\alpha ^2 \sigma _u^2 \left(\sigma _{\xi }^2
   \left((\tau_{\mu,A|s})\right)+1\right) \left(\left((-\pi_I)\right) \tau _D^2 \sigma _{\xi }^2+\tau _D \left(\left(\pi
   _A-1\right) (\omega +1) \gamma _A \sigma _{\xi }^2 \tau _S-1\right)+\tau_A \left(\left((-\pi_I)\right) \omega 
   \left(\gamma _A \sigma _{\xi }^2 \tau _S+1\right)-\pi _A\right)\right)-\pi _A V^2 \left(\tau_A+\tau
   _D\right){}^2\right){}^3).
\end{align}\end{tiny}

This derivative is positive since every term in the numerator is negative, and the denominator is positive.

The derivative w.r.t. the noise trader volatility, $\sigma_u$, is:

\begin{tiny}\begin{align}
   \frac{\partial E[\alpha_M^2]}{\partial \sigma_u}&=\\
   & -(4 \alpha ^2 \left((-\pi_I)\right) \pi _A V^2 (\omega -1) \gamma _A \tau _D^2 \tau _S \sigma _u \left(\tau_A+\tau
   _D\right) \times \\
   & \left(\sigma _{\xi }^2 \left((\tau_{\mu,A|s})\right)+1\right){}^2 \left(\alpha ^2 \sigma _u^2 \left(\sigma
   _{\xi }^2 \left((\tau_{\mu,A|s})\right)+1\right) \left(-\pi _A \gamma _A-\left(\left((-\pi_I)\right) \tau _D \sigma _{\xi
   }^2\right)+\left((-\pi_I)\right) \gamma _A \left(\omega  \tau _D \sigma _{\xi }^2-\sigma _{\xi }^2 \tau _S+\omega
   \right)+\left((-\pi_I)\right) \omega  \gamma _A^2 \sigma _{\xi }^2 \tau _S+1\right)-\pi _A V^2 \left(\gamma _A-1\right)
   \left((\tau_{\mu,A|s})\right)\right))/\\
   &(\left(\tau _D+\tau _S\right) \left(\alpha ^2 \sigma _u^2 \left(\sigma _{\xi }^2
   \left((\tau_{\mu,A|s})\right)+1\right) \left(\left((-\pi_I)\right) \tau _D^2 \sigma _{\xi }^2+\tau _D \left(\left(\pi
   _A-1\right) (\omega +1) \gamma _A \sigma _{\xi }^2 \tau _S-1\right)+\tau_A \left(\left((-\pi_I)\right) \omega 
   \left(\gamma _A \sigma _{\xi }^2 \tau _S+1\right)-\pi _A\right)\right)-\pi _A V^2 \left(\tau_A+\tau
   _D\right){}^2\right){}^3).
\end{align}\end{tiny}

This derivative is positive.

\section{Theorem \ref{theorem:llm_profitability}}

The unconditional expectation of the profits using the LLM's signal is given by 
\begin{tiny}\begin{align}
    \text{Profits}_{LLM} &=  E[x_{2,L} (\mu_{A|s} -  p_1)] = \\
    &(\alpha  \left((-\pi_I)\right) (\omega -1) \gamma _A \tau _D^2 \tau _S \sigma _u^2 \left(\sigma _{\xi }^2 \left(\gamma _A \tau
   _S+\tau _D\right)+1\right){}^2 \times \\
   & (\lambda (c,k) \left(\pi _A V^2 \left((\tau_{\mu,A|s})\right)-\alpha ^2 \sigma _u^2
   \left(\sigma _{\xi }^2 \left((\tau_{\mu,A|s})\right)+1\right) \left(\left((-\pi_I)\right) \sigma _{\xi }^2 \left(\gamma
   _A \tau _S+\tau _D\right)-1\right)\right)+\\
   & \gamma _A \left(\alpha ^2 \sigma _u^2 \left(\sigma _{\xi }^2 \left(\gamma _A \tau
   _S+\tau _D\right)+1\right) \left(\left((-\pi_I)\right) \omega  \left(\sigma _{\xi }^2 \left(\tau_A+\tau
   _D\right)+1\right)-\pi _A\right)-\pi _A V^2 \left((\tau_{\mu,A|s})\right)\right)))/\\
   &(\left(\tau_A+\tau
   _D\right) \left(\alpha ^2 \sigma _u^2 \left(\sigma _{\xi }^2 \left((\tau_{\mu,A|s})\right)+1\right) \left(\left(\pi
   _A-1\right) \tau _D^2 \sigma _{\xi }^2+\tau _D \left(\left((-\pi_I)\right) (\omega +1) \gamma _A \sigma _{\xi }^2 \tau
   _S-1\right)+\tau_A \left(\left((-\pi_I)\right) \omega  \left(\gamma _A \sigma _{\xi }^2 \tau _S+1\right)-\pi
   _A\right)\right)-\pi _A V^2 \left((\tau_{\mu,A|s})\right){}^2\right){}^2).
\end{align}\end{tiny}

To see that the profits increase with LLM model size, the derivative of the profits with respect to the model size is given by 

\begin{tiny}\begin{align}
\frac{\partial\text{Profits}_{LLM}}{\partial k} = &\\
    &\frac{\alpha  \left((-\pi_I)\right) (\omega -1) \gamma _A \tau _D^2 \tau _S \sigma _u^2 \frac{\partial \lambda}{\partial k} \left(\sigma _{\xi }^2
   \left((\tau_{\mu,A|s})\right)+1\right){}^2 \left(\pi _A V^2 \left((\tau_{\mu,A|s})\right)-\alpha ^2 \sigma _u^2
   \left(\sigma _{\xi }^2 \left((\tau_{\mu,A|s})\right)+1\right) \left(\left((-\pi_I)\right) \sigma _{\xi }^2 \left(\gamma
   _A \tau _S+\tau _D\right)-1\right)\right)}{\left((\tau_{\mu,A|s})\right) \left(\alpha ^2 \sigma _u^2 \left(\sigma _{\xi
   }^2 \left((\tau_{\mu,A|s})\right)+1\right) \left(\left((-\pi_I)\right) \tau _D^2 \sigma _{\xi }^2+\tau _D \left(\left(\pi
   _A-1\right) (\omega +1) \gamma _A \sigma _{\xi }^2 \tau _S-1\right)+\tau_A \left(\left((-\pi_I)\right) \omega 
   \left(\gamma _A \sigma _{\xi }^2 \tau _S+1\right)-\pi _A\right)\right)-\pi _A V^2 \left(\tau_A+\tau
   _D\right){}^2\right){}^2},
\end{align}\end{tiny}

\noindent where both the numerator and the denominator are positive.

Moreover, the profits are positive if and only if

\begin{tiny}\begin{align}
    \lambda(c,k) &> \text{Threshold}_\lambda \\
    & \equiv \frac{\gamma _A \left(\alpha ^2 \sigma _u^2 \left(\sigma _{\xi }^2 \left((\tau_{\mu,A|s})\right)+1\right) \left(\left(\pi
   _A-1\right) \omega  \left(\sigma _{\xi }^2 \left((\tau_{\mu,A|s})\right)+1\right)-\pi _A\right)-\pi _A V^2 \left(\gamma
   _A \tau _S+\tau _D\right)\right)}{\alpha ^2 \sigma _u^2 \left(\sigma _{\xi }^2 \left((\tau_{\mu,A|s})\right)+1\right)
   \left(\left((-\pi_I)\right) \sigma _{\xi }^2 \left((\tau_{\mu,A|s})\right)-1\right)-\pi _A V^2 \left(\gamma _A \tau
   _S+\tau _D\right)}.
\end{align}\end{tiny}

For a fixed news complexity level $c$, let $\lambda_c(k) = \lambda(c, k)$. $\lambda_c(k)$ is a strictly increasing function of model size $k$ and hence it is invertible. Thus, if we define

\begin{tiny}\begin{align}
    k^* \equiv \lambda^{-1}_c (\text{Threshold}_\lambda),
\end{align}\end{tiny}

\noindent the profits of the LLM strategy are positive if and only if $k > k^*$. Since the inverse function is strictly increasing, it suffices to focus on the properties of $\text{Threshold}_{\lambda}$ to characterize its behavior.

\subsection{Threshold Properties}

For inattentive agents' information capacity $\omega$, the derivative of the threshold is positive and given by

\begin{tiny}\begin{align}
    \frac{\partial\text{Threshold}_{\lambda}}{\partial \omega}  &= \\
    & \frac{\alpha ^2 \left((-\pi_I)\right) \gamma _A \sigma _u^2 \left(\sigma _{\xi }^2 \left(\tau_A+\tau
   _D\right)+1\right){}^2}{\alpha ^2 \sigma _u^2 \left(\sigma _{\xi }^2 \left((\tau_{\mu,A|s})\right)+1\right)
   \left(\left((-\pi_I)\right) \sigma _{\xi }^2 \left((\tau_{\mu,A|s})\right)-1\right)-\pi _A V^2 \left(\gamma _A \tau
   _S+\tau _D\right)}.
\end{align}\end{tiny}

For attentive agents' information capacity $\gamma_A$, the derivative of the threshold is also positive and given by

\begin{tiny}\begin{align}
    \frac{\partial\text{Threshold}_{\lambda}}{\partial \gamma_A}  &= \\
    & (\alpha ^4 \sigma _u^4 \left(\sigma _{\xi }^2 \left((\tau_{\mu,A|s})\right)+1\right){}^2 \times \\
    & \left(\left((-\pi_I)\right)
   \left(\omega  \left(\left((-\pi_I)\right) \sigma _{\xi }^4 \left((\tau_{\mu,A|s})\right){}^2+\sigma _{\xi }^2
   \left(\left(\pi _A-2\right) \tau _D-2 \tau_A\right)-1\right)-\pi _A \tau _D \sigma _{\xi }^2\right)+\pi
   _A\right) -\\
   & \alpha ^2 \pi _A V^2 \sigma _u^2 \left(\sigma _{\xi }^2 \left((\tau_{\mu,A|s})\right)+1\right) \times \\
   & \left(\left(\pi
   _A-1\right) (\omega +1) \tau _D^2 \sigma _{\xi }^2+\tau _D \left(\left((-\pi_I)\right) (3 \omega +1) \gamma _A \sigma _{\xi }^2
   \tau _S+\pi _A \omega -\pi _A-\omega -1\right)+2 \tau_A \left(\left((-\pi_I)\right) \omega  \gamma _A \sigma _{\xi }^2
   \tau _S-1\right)\right)+\pi _A^2 V^4 \left((\tau_{\mu,A|s})\right){}^2)/\\
   & (\left(\alpha ^2 \sigma _u^2 \left(\sigma _{\xi
   }^2 \left((\tau_{\mu,A|s})\right)+1\right) \left(\left((-\pi_I)\right) \sigma _{\xi }^2 \left(\tau_A+\tau
   _D\right)-1\right)-\pi _A V^2 \left((\tau_{\mu,A|s})\right)\right){}^2).
\end{align}\end{tiny}

Moreover, for the proportion of attentive agents, $\pi_A$, the derivative of the threshold is also positive and given by

\begin{tiny}\begin{align}
\frac{\partial\text{Threshold}_{\lambda}}{\partial \pi_A}  &= \\
    & -\frac{\alpha ^2 (\omega -1) \gamma _A \sigma _u^2 \left(\sigma _{\xi }^2 \left((\tau_{\mu,A|s})\right)+1\right){}^2
   \left(\alpha ^2 \sigma _u^2 \left(\sigma _{\xi }^2 \left((\tau_{\mu,A|s})\right)+1\right)+V^2 \left(\gamma _A \tau
   _S+\tau _D\right)\right)}{\left(\alpha ^2 \sigma _u^2 \left(\sigma _{\xi }^2 \left((\tau_{\mu,A|s})\right)+1\right)
   \left(\left((-\pi_I)\right) \sigma _{\xi }^2 \left((\tau_{\mu,A|s})\right)-1\right)-\pi _A V^2 \left(\gamma _A \tau
   _S+\tau _D\right)\right){}^2}.
\end{align}\end{tiny}

A similar positive result holds with respect to the total volume of traders V:

\begin{tiny}\begin{align}
\frac{\partial \text{Threshold}_{\lambda}}{\partial V}  &= \\
   & \frac{2 \alpha ^2 \left((-\pi_I)\right) \pi _A V (\omega -1) \gamma _A \sigma _u^2 \left((\tau_{\mu,A|s})\right)
   \left(\sigma _{\xi }^2 \left((\tau_{\mu,A|s})\right)+1\right){}^2}{\left(\alpha ^2 \sigma _u^2 \left(\sigma _{\xi }^2
   \left((\tau_{\mu,A|s})\right)+1\right) \left(\left((-\pi_I)\right) \sigma _{\xi }^2 \left(\tau_A+\tau
   _D\right)-1\right)-\pi _A V^2 \left((\tau_{\mu,A|s})\right)\right){}^2}.
\end{align}\end{tiny}

On the negative side, for risk aversion and noise trader standard deviation, the derivatives are negative and given by

\begin{tiny}\begin{align}
\frac{\partial \text{Threshold}_{\lambda}}{\partial \alpha}  &= \\
   & -\frac{2 \alpha  \left((-\pi_I)\right) \pi _A V^2 (\omega -1) \gamma _A \left((\tau_{\mu,A|s})\right) \left(\sigma _{\xi }^2
   \sigma _u \left((\tau_{\mu,A|s})\right)+\sigma _u\right){}^2}{\left(\alpha ^2 \sigma _u^2 \left(\sigma _{\xi }^2
   \left((\tau_{\mu,A|s})\right)+1\right) \left(\left((-\pi_I)\right) \sigma _{\xi }^2 \left(\tau_A+\tau
   _D\right)-1\right)-\pi _A V^2 \left((\tau_{\mu,A|s})\right)\right){}^2}.
\end{align}\end{tiny}

\begin{tiny}\begin{align}
\frac{\partial \text{Threshold}_{\lambda}}{\partial \sigma_u}  &= \\
    &-\frac{2 \left((-\pi_I)\right) \pi _A V^2 (\omega -1) \gamma _A \sigma _u \left((\tau_{\mu,A|s})\right) \left(\alpha +\alpha
    \sigma _{\xi }^2 \left((\tau_{\mu,A|s})\right)\right){}^2}{\left(\alpha ^2 \sigma _u^2 \left(\sigma _{\xi }^2
   \left((\tau_{\mu,A|s})\right)+1\right) \left(\left((-\pi_I)\right) \sigma _{\xi }^2 \left(\tau_A+\tau
   _D\right)-1\right)-\pi _A V^2 \left((\tau_{\mu,A|s})\right)\right){}^2}.
\end{align}\end{tiny}.

\subsection{Proposition \ref{prop:cond_pred}}

If we restrict to the space of positive profitability with $k > k^*$, the predictability is lower in the presence of more attentive agents, as shown by its derivative.

\begin{tiny}\begin{align}
k > k^*&  \Rightarrow \frac{\partial\text{Profits}_{LLM}}{\partial \pi_A}  = \\
    &(\alpha  (\omega -1) \gamma _A \tau _D^2 \tau _S \sigma _u^2 \left(\sigma _{\xi }^2 \left(\tau_A+\tau
   _D\right)+1\right){}^2 \times \\
   & \left(\alpha ^2 \sigma _u^2 \left(\sigma _{\xi }^2 \left((\tau_{\mu,A|s})\right)+1\right)+V^2
   \left((\tau_{\mu,A|s})\right)\right) \times \\
   & (\lambda (c,k) \left(\alpha ^2 \sigma _u^2 \left(\sigma _{\xi }^2 \left(\gamma
   _A \tau _S+\tau _D\right)+1\right) \left(\left((-\pi_I)\right) \tau _D^2 \sigma _{\xi }^2+\tau _D \left(-\left(\left(\pi
   _A-1\right) (\omega -3) \gamma _A \sigma _{\xi }^2 \tau _S\right)-1\right)+\tau_A \left(-\left(\left((-\pi_I)\right)
   (\omega -2) \gamma _A \sigma _{\xi }^2 \tau _S\right)-\pi _A \omega +\pi _A+\omega -2\right)\right)-\pi _A V^2 \left(\gamma _A
   \tau _S+\tau _D\right){}^2\right) +\\
   & \gamma _A \left(\pi _A V^2 \left((\tau_{\mu,A|s})\right){}^2-\alpha ^2 \sigma _u^2
   \left(\sigma _{\xi }^2 \left((\tau_{\mu,A|s})\right)+1\right) \left(\left((-\pi_I)\right) (2 \omega -1) \tau _D^2 \sigma
   _{\xi }^2+\tau _D \left(\left((-\pi_I)\right) (3 \omega -1) \gamma _A \sigma _{\xi }^2 \tau _S+2 \pi _A \omega -2 \pi _A-2 \omega
   +1\right)+\tau_A \left(\left((-\pi_I)\right) \omega  \left(\gamma _A \sigma _{\xi }^2 \tau _S+1\right)-\pi
   _A\right)\right)\right)))/\\
   & (\left((\tau_{\mu,A|s})\right) \left(\alpha ^2 \sigma _u^2 \left(\sigma _{\xi }^2
   \left((\tau_{\mu,A|s})\right)+1\right) \left(\left((-\pi_I)\right) \tau _D^2 \sigma _{\xi }^2+\tau _D \left(\left(\pi
   _A-1\right) (\omega +1) \gamma _A \sigma _{\xi }^2 \tau _S-1\right)+\tau_A \left(\left((-\pi_I)\right) \omega 
   \left(\gamma _A \sigma _{\xi }^2 \tau _S+1\right)-\pi _A\right)\right)-\pi _A V^2 \left(\tau_A+\tau
   _D\right){}^2\right){}^3) < 0.
\end{align}\end{tiny}

Conversely, any market where it is more costly to be an attentive agent will have a lower proportion of attentive agents.

\subsection{Proposition \ref{prop:llm_mispricing}}

We have already shown above that an increase in the proportion of attentive agents, $\pi_A$, implies a decline in LLM profitability (if the profitability is positive) and in mispricing. In addition, if $k > k^*$, then higher volume, better information processing of inattentive agents, or lower noise risk reduces return predictability and mispricing, as shown by the following derivatives:

\begin{tiny}\begin{align}
k > k^*&  \Rightarrow \frac{\partial \text{Threshold}_{\lambda}}{\partial V}  = \\
    &(2 \alpha  \left((-\pi_I)\right) \pi _A V (\omega -1) \gamma _A \tau _D^2 \tau _S \sigma _u^2 \left(\sigma _{\xi }^2
   \left((\tau_{\mu,A|s})\right)+1\right){}^2 \times \\
   & (\lambda (c,k) \left(\pi _A V^2 \left(\tau_A+\tau
   _D\right){}^2-\alpha ^2 \sigma _u^2 \left(\sigma _{\xi }^2 \left((\tau_{\mu,A|s})\right)+1\right) \left(\left(\pi
   _A-1\right) \tau _D^2 \sigma _{\xi }^2+\tau _D \left(-\left(\left((-\pi_I)\right) (\omega -3) \gamma _A \sigma _{\xi }^2 \tau
   _S\right)-1\right)+\tau_A \left(-\left(\left((-\pi_I)\right) (\omega -2) \gamma _A \sigma _{\xi }^2 \tau _S\right)-\pi
   _A \omega +\pi _A+\omega -2\right)\right)\right) +\\
   & \gamma _A \left(\alpha ^2 \sigma _u^2 \left(\sigma _{\xi }^2 \left(\gamma _A \tau
   _S+\tau _D\right)+1\right) \left(\left((-\pi_I)\right) (2 \omega -1) \tau _D^2 \sigma _{\xi }^2+\tau _D \left(\left(\pi
   _A-1\right) (3 \omega -1) \gamma _A \sigma _{\xi }^2 \tau _S+2 \pi _A \omega -2 \pi _A-2 \omega +1\right)+\tau_A
   \left(\left((-\pi_I)\right) \omega  \left(\gamma _A \sigma _{\xi }^2 \tau _S+1\right)-\pi _A\right)\right)-\pi _A V^2 \left(\gamma
   _A \tau _S+\tau _D\right){}^2\right)))/\\
   &(\left(\alpha ^2 \sigma _u^2 \left(\sigma _{\xi }^2 \left(\tau_A+\tau
   _D\right)+1\right) \left(\left((-\pi_I)\right) \tau _D^2 \sigma _{\xi }^2+\tau _D \left(\left((-\pi_I)\right) (\omega +1) \gamma
   _A \sigma _{\xi }^2 \tau _S-1\right)+\tau_A \left(\left((-\pi_I)\right) \omega  \left(\gamma _A \sigma _{\xi }^2 \tau
   _S+1\right)-\pi _A\right)\right)-\pi _A V^2 \left((\tau_{\mu,A|s})\right){}^2\right){}^3)\\
   < 0.
\end{align}\end{tiny}

\begin{tiny}\begin{align}
k > k^*&  \Rightarrow \frac{\partial \text{Threshold}_{\lambda}}{\partial \sigma_u}  = \\
    &(2 \alpha  \left((-\pi_I)\right) \pi _A V^2 (\omega -1) \tau_A \sigma _u \left(\tau _D \sigma _{\xi }^2 \left(\gamma
   _A \tau _S+\tau _D\right)+\tau _D\right){}^2 \times \\
   & (\lambda (c,k) (\alpha ^2 \sigma _u^2 \left(\sigma _{\xi }^2 \left(\gamma
   _A \tau _S+\tau _D\right)+1\right) \times \\
   & \left(\left((-\pi_I)\right) \tau _D^2 \sigma _{\xi }^2+\tau _D \left(-\left(\left(\pi
   _A-1\right) (\omega -3) \gamma _A \sigma _{\xi }^2 \tau _S\right)-1\right)+\tau_A \left(-\left(\left((-\pi_I)\right)
   (\omega -2) \gamma _A \sigma _{\xi }^2 \tau _S\right)-\pi _A \omega +\pi _A+\omega -2\right)\right)-\\
   & \pi _A V^2 \left(\gamma _A
   \tau _S+\tau _D\right){}^2)+\gamma _A \times \\
   & \left(\pi _A V^2 \left((\tau_{\mu,A|s})\right){}^2-\alpha ^2 \sigma _u^2
   \left(\sigma _{\xi }^2 \left((\tau_{\mu,A|s})\right)+1\right) \left(\left((-\pi_I)\right) (2 \omega -1) \tau _D^2 \sigma
   _{\xi }^2+\tau _D \left(\left((-\pi_I)\right) (3 \omega -1) \gamma _A \sigma _{\xi }^2 \tau _S+2 \pi _A \omega -2 \pi _A-2 \omega
   +1\right)+\tau_A \left(\left((-\pi_I)\right) \omega  \left(\gamma _A \sigma _{\xi }^2 \tau _S+1\right)-\pi
   _A\right)\right)\right)))/\\
   &(\left(\alpha ^2 \sigma _u^2 \left(\sigma _{\xi }^2 \left((\tau_{\mu,A|s})\right)+1\right)
   \left(\left((-\pi_I)\right) \tau _D^2 \sigma _{\xi }^2+\tau _D \left(\left((-\pi_I)\right) (\omega +1) \gamma _A \sigma _{\xi }^2
   \tau _S-1\right)+\tau_A \left(\left((-\pi_I)\right) \omega  \left(\gamma _A \sigma _{\xi }^2 \tau _S+1\right)-\pi
   _A\right)\right)-\pi _A V^2 \left((\tau_{\mu,A|s})\right){}^2\right){}^3)\\
   > 0.
\end{align}\end{tiny}

\begin{tiny}\begin{align}
k > k^*&  \Rightarrow \frac{\partial \text{Threshold}_{\lambda}}{\partial \omega}  = \\
    &(\alpha  \left((-\pi_I)\right) \gamma _A \tau _D^2 \tau _S \sigma _u^2 \left(\sigma _{\xi }^2 \left(\tau_A+\tau
   _D\right)+1\right){}^2 \times \\
   & (\lambda (c,k) \left(\pi _A V^2 \left((\tau_{\mu,A|s})\right)-\alpha ^2 \sigma _u^2
   \left(\sigma _{\xi }^2 \left((\tau_{\mu,A|s})\right)+1\right) \left(\left((-\pi_I)\right) \sigma _{\xi }^2 \left(\gamma
   _A \tau _S+\tau _D\right)-1\right)\right) \times \\
   & \left(\alpha ^2 \sigma _u^2 \left(\sigma _{\xi }^2 \left(\tau_A+\tau
   _D\right)+1\right) \left(\left((-\pi_I)\right) \tau _D^2 \sigma _{\xi }^2+\tau _D \left(-\left(\left((-\pi_I)\right) (\omega -3)
   \gamma _A \sigma _{\xi }^2 \tau _S\right)-1\right)+\tau_A \left(-\left(\left((-\pi_I)\right) (\omega -2) \gamma _A
   \sigma _{\xi }^2 \tau _S\right)-\pi _A \omega +\pi _A+\omega -2\right)\right)-\pi _A V^2 \left(\tau_A+\tau
   _D\right){}^2\right)+ \\
   & \gamma _A \left(\alpha ^2 \sigma _u^2 \left(\sigma _{\xi }^2 \left((\tau_{\mu,A|s})\right)+1\right)
   \left(\left((-\pi_I)\right) \sigma _{\xi }^2 \left((\tau_{\mu,A|s})\right)-1\right)-\pi _A V^2 \left(\gamma _A \tau
   _S+\tau _D\right)\right) \times \\
   & \left(\alpha ^2 \sigma _u^2 \left(\sigma _{\xi }^2 \left((\tau_{\mu,A|s})\right)+1\right)
   \left(\left((-\pi_I)\right) (2 \omega -1) \tau _D^2 \sigma _{\xi }^2+\tau _D \left(\left((-\pi_I)\right) (3 \omega -1) \gamma _A
   \sigma _{\xi }^2 \tau _S+2 \pi _A \omega -2 \pi _A-2 \omega +1\right)+\tau_A \left(\left((-\pi_I)\right) \omega 
   \left(\gamma _A \sigma _{\xi }^2 \tau _S+1\right)-\pi _A\right)\right)-\pi _A V^2 \left(\tau_A+\tau
   _D\right){}^2\right)))/\\
   & (\left((\tau_{\mu,A|s})\right) \left(\alpha ^2 \sigma _u^2 \left(\sigma _{\xi }^2
   \left((\tau_{\mu,A|s})\right)+1\right) \left(\left((-\pi_I)\right) \tau _D^2 \sigma _{\xi }^2+\tau _D \left(\left(\pi
   _A-1\right) (\omega +1) \gamma _A \sigma _{\xi }^2 \tau _S-1\right)+\tau_A \left(\left((-\pi_I)\right) \omega 
   \left(\gamma _A \sigma _{\xi }^2 \tau _S+1\right)-\pi _A\right)\right)-\pi _A V^2 \left(\tau_A+\tau
   _D\right){}^2\right){}^3)\\
   < 0.
\end{align}\end{tiny}

\section{Theorem \ref{theorem:market_efficiency}}

In the case of inattentive agents using LLMs, the price is given by:

\begin{tiny}\begin{align}
    p_{1|I} = \frac{-\left((-\pi_I)\right) t \left(\bar{d} \tau _D+s \tau _S \lambda (c,k)\right)+\left((-\pi_I)\right) (t-1) \left(\bar{d} \tau
   _D+s \omega  \tau_A\right)+\frac{\pi _A \left(\bar{d} \tau _D+s \tau_A\right) \left(\alpha ^2 \sigma _u^2
   \left(\sigma _{\xi }^2 \left((\tau_{\mu,A|s})\right)+1\right)+V^2 \left((\tau_{\mu,A|s})\right)\right)}{\alpha
   ^2 \sigma _u^2 \left(\sigma _{\xi }^2 \left((\tau_{\mu,A|s})\right)+1\right){}^2}}{-\left((-\pi_I)\right) t \left(\tau _S
   \lambda (c,k)+\tau _D\right)+\left((-\pi_I)\right) (t-1) \left(\omega  (\tau_{\mu,A|s})\right)+\frac{\pi _A
   \left(\frac{1}{(\tau_{\mu,A|s})}+\sigma _{\xi }^2+\frac{V^2}{\alpha ^2 \sigma _u^2}\right)}{\left(\frac{1}{\gamma _A \tau
   _S+\tau _D}+\sigma _{\xi }^2\right){}^2}}.
\end{align}\end{tiny}

Hence, mispricing in this case is:

\begin{tiny}\begin{align}
  E[\alpha_{M|I}^2] =&\\
  &(\tau _D^2 \tau _S (-\alpha ^2 \left((-\pi_I)\right) t \sigma _u^2 \lambda (c,k) \left(\sigma _{\xi }^2 \left(\gamma _A
   \tau _S+\tau _D\right)+1\right){}^2+\\
   & \alpha ^2 \sigma _u^2 \left(\sigma _{\xi }^2 \left((\tau_{\mu,A|s})\right)+1\right)
   \left(\left((-\pi_I)\right) \tau _D \sigma _{\xi }^2+\gamma _A \left(\left((-\pi_I)\right) \left((t-1) \omega  \left(\tau _D
   \sigma _{\xi }^2+1\right)+\sigma _{\xi }^2 \tau _S\right)+\pi _A\right)+\left((-\pi_I)\right) (t-1) \omega  \gamma _A^2 \sigma
   _{\xi }^2 \tau _S-1\right) \\
   & +\pi _A V^2 \left(\gamma _A-1\right) \left((\tau_{\mu,A|s})\right)){}^2)/\\
   &(\left(\tau
   _D+\tau _S\right){}^2 (\alpha ^2 \left((-\pi_I)\right) t \tau _S \sigma _u^2 \lambda (c,k) \left(\sigma _{\xi }^2
   \left((\tau_{\mu,A|s})\right)+1\right){}^2+\\
   & \alpha ^2 \sigma _u^2 \left(\sigma _{\xi }^2 \left(\tau_A+\tau
   _D\right)+1\right) \left(\left((-\pi_I)\right) \tau _D^2 \sigma _{\xi }^2+\tau _D \left(-\left(\left((-\pi_I)\right) \gamma _A
   \sigma _{\xi }^2 \tau _S ((t-1) \omega -1)\right)-1\right)-\tau_A \left(\left((-\pi_I)\right) (t-1) \omega 
   \left(\gamma _A \sigma _{\xi }^2 \tau _S+1\right)+\pi _A\right)\right)-\pi _A V^2 \left(\tau_A+\tau
   _D\right){}^2){}^2).
\end{align}\end{tiny}

The derivative of mispricing w.r.t. the proportion of inattentive agents using LLMs, $\theta$, is negative:

\begin{tiny}\begin{align}
   \frac{\partial E[\alpha_{M|I}^2]}{\partial \theta}=&\\
   &(2 \alpha ^2 \left((-\pi_I)\right) \tau _D^2 \tau _S \sigma _u^2 \left(\omega  \gamma _A-\lambda (c,k)\right) \times \\
   & \left(\sigma _{\xi
   }^2 \left((\tau_{\mu,A|s})\right)+1\right){}^2 \times
   \left(\alpha ^2 \sigma _u^2 \left(\sigma _{\xi }^2 \left(\gamma _A \tau
   _S+\tau _D\right)+1\right) \left(\left((-\pi_I)\right) \sigma _{\xi }^2 \left((\tau_{\mu,A|s})\right)-1\right)-\pi _A V^2
   \left((\tau_{\mu,A|s})\right)\right) \times \\
   & (-\alpha ^2 \left((-\pi_I)\right) t \sigma _u^2 \lambda (c,k) \left(\sigma
   _{\xi }^2 \left((\tau_{\mu,A|s})\right)+1\right){}^2+\\
   & \alpha ^2 \sigma _u^2 \left(\sigma _{\xi }^2 \left(\gamma _A \tau
   _S+\tau _D\right)+1\right) \left(\left((-\pi_I)\right) \tau _D \sigma _{\xi }^2+\gamma _A \left(\left((-\pi_I)\right) \left((t-1)
   \omega  \left(\tau _D \sigma _{\xi }^2+1\right)+\sigma _{\xi }^2 \tau _S\right)+\pi _A\right)+\left((-\pi_I)\right) (t-1) \omega 
   \gamma _A^2 \sigma _{\xi }^2 \tau _S-1\right)+\pi _A V^2 \left(\gamma _A-1\right) \left(\tau_A+\tau
   _D\right)))/\\
   &\left(\tau _D+\tau _S\right) (\alpha ^2 \left((-\pi_I)\right) t \tau _S \sigma _u^2 \lambda (c,k)
   \left(\sigma _{\xi }^2 \left((\tau_{\mu,A|s})\right)+1\right){}^2+\\
   & \alpha ^2 \sigma _u^2 \left(\sigma _{\xi }^2
   \left((\tau_{\mu,A|s})\right)+1\right) \left(\left((-\pi_I)\right) \tau _D^2 \sigma _{\xi }^2+\tau _D
   \left(-\left(\left((-\pi_I)\right) \gamma _A \sigma _{\xi }^2 \tau _S ((t-1) \omega -1)\right)-1\right)-\tau_A
   \left(\left((-\pi_I)\right) (t-1) \omega  \left(\gamma _A \sigma _{\xi }^2 \tau _S+1\right)+\pi _A\right)\right)-\pi _A V^2
   \left((\tau_{\mu,A|s})\right){}^2){}^3)\\
   < 0.
\end{align}\end{tiny}

The derivative  of mispricing w.r.t. the LLM's model size, $k$, is also negative:

\begin{tiny}\begin{align}
   \frac{\partial E[\alpha_{M|I}^2]}{\partial k}=&\\
   &-(2 \alpha ^2 \left((-\pi_I)\right) t \tau _D^2 \tau _S \sigma _u^2 \frac{\partial \lambda}{\partial k} \left(\sigma _{\xi }^2 \left(\gamma _A
   \tau _S+\tau _D\right)+1\right){}^2 \times \\
   & \left(\alpha ^2 \sigma _u^2 \left(\sigma _{\xi }^2 \left(\tau_A+\tau
   _D\right)+1\right) \left(\left((-\pi_I)\right) \sigma _{\xi }^2 \left((\tau_{\mu,A|s})\right)-1\right)-\pi _A V^2
   \left((\tau_{\mu,A|s})\right)\right) \times \\
   & (-\alpha ^2 \left((-\pi_I)\right) t \sigma _u^2 \lambda (c,k) \left(\sigma
   _{\xi }^2 \left((\tau_{\mu,A|s})\right)+1\right){}^2+\\
   & \alpha ^2 \sigma _u^2 \left(\sigma _{\xi }^2 \left(\gamma _A \tau
   _S+\tau _D\right)+1\right) \left(\left((-\pi_I)\right) \tau _D \sigma _{\xi }^2+\gamma _A \left(\left((-\pi_I)\right) \left((t-1)
   \omega  \left(\tau _D \sigma _{\xi }^2+1\right)+\sigma _{\xi }^2 \tau _S\right)+\pi _A\right)+\left((-\pi_I)\right) (t-1) \omega 
   \gamma _A^2 \sigma _{\xi }^2 \tau _S-1\right)+\pi _A V^2 \left(\gamma _A-1\right) \left(\tau_A+\tau
   _D\right)))/\\
   &(\left(\tau _D+\tau _S\right)(\alpha ^2 \left((-\pi_I)\right) t \tau _S \sigma _u^2 \lambda (c,k)
   \left(\sigma _{\xi }^2 \left((\tau_{\mu,A|s})\right)+1\right){}^2+\\
   & \alpha ^2 \sigma _u^2 \left(\sigma _{\xi }^2
   \left((\tau_{\mu,A|s})\right)+1\right) \left(\left((-\pi_I)\right) \tau _D^2 \sigma _{\xi }^2+\tau _D
   \left(-\left(\left((-\pi_I)\right) \gamma _A \sigma _{\xi }^2 \tau _S ((t-1) \omega -1)\right)-1\right)-\tau_A
   \left(\left((-\pi_I)\right) (t-1) \omega  \left(\gamma _A \sigma _{\xi }^2 \tau _S+1\right)+\pi _A\right)\right)-\pi _A V^2
   \left((\tau_{\mu,A|s})\right){}^2){}^3)\\
   < 0.
\end{align}\end{tiny}

\subsection{Proposition \ref{prop:inattentive_llm}}

The profitability of LLM strategies when inattentive agents  is given by

\begin{tiny}\begin{align}
    \text{Profitability}_{LLM|I} \equiv \\
    & (\alpha  \left((-\pi_I)\right) \tau _D^2 \tau _S \sigma _u^2 \left(\sigma _{\xi }^2 \left(\tau_A+\tau
   _D\right)+1\right){}^2 \times \\
   & \left(-t \lambda (c,k)+\gamma _A+(t-1) \omega  \gamma _A\right) (\gamma _A \left(\alpha ^2 \sigma _u^2
   \left(\sigma _{\xi }^2 \left((\tau_{\mu,A|s})\right)+1\right) \left(\left((-\pi_I)\right) (t-1) \omega  \left(\sigma
   _{\xi }^2 \left((\tau_{\mu,A|s})\right)+1\right)+\pi _A\right)+\pi _A V^2 \left(\tau_A+\tau
   _D\right)\right)-\\
   & \lambda (c,k) \left(\alpha ^2 \sigma _u^2 \left(\sigma _{\xi }^2 \left((\tau_{\mu,A|s})\right)+1\right)
   \left(-\left(\left((-\pi_I)\right) \sigma _{\xi }^2 \left((\tau_{\mu,A|s})\right)\right)+\left((-\pi_I)\right) t
   \left(\sigma _{\xi }^2 \left((\tau_{\mu,A|s})\right)+1\right)+1\right)+\pi _A V^2 \left(\tau_A+\tau
   _D\right)\right)))/\\
   & (\left((\tau_{\mu,A|s})\right) (\alpha ^2 \left((-\pi_I)\right) t \tau _S \sigma _u^2
   \lambda (c,k) \left(\sigma _{\xi }^2  \left((\tau_{\mu,A|s})\right)+1\right){}^2+\\
   & \alpha ^2 \sigma _u^2 \left(\sigma _{\xi
   }^2 \left((\tau_{\mu,A|s})\right)+1\right) \times \\
   & \left(\left((-\pi_I)\right) \tau _D^2 \sigma _{\xi }^2+\tau _D
   \left(-\left(\left((-\pi_I)\right) \gamma _A \sigma _{\xi }^2 \tau _S ((t-1) \omega -1)\right)-1\right)-\tau_A
   \left(\left((-\pi_I)\right) (t-1) \omega  \left(\gamma _A \sigma _{\xi }^2 \tau _S+1\right)+\pi _A\right)\right)-\pi _A V^2
   \left((\tau_{\mu,A|s})\right){}^2){}^2).
\end{align}\end{tiny}

And the proposition follows immediately by substituting $\lambda(c, k') = \gamma$ for large enough $k'$, and $\theta = 1$.

\section{Proposition \ref{prop:attentive_efficiency}}

This proposition follows immediately from Proposition \ref{prop:mispricing} because the parameter $\gamma$ is substituted by $\lambda(c, k)$.

\end{landscape}
\newpage

\section{Proof for Partial Adoption by Attentive Agents}
% This file is part of the appendix and is NOT meant to be compiled standalone.
% It should be included in the main document which provides the preamble, bibliography, and other dependencies.
% Citations, theorem counters, and cross-references are resolved in the parent document.

This section provides the complete proof for the case where a fraction of attentive agents adopt superior LLM technology while the remaining attentive agents learn from the equilibrium price. This case involves a rational expectations equilibrium with information feedback through prices.

\textbf{Note on the Modeling Framework:} This section adapts the main model's framework to specifically analyze the case of learning from prices. In the main model (Section 3 and the preceding section of this appendix), predictability arises from different agents processing a single \textit{public} signal with varying capacities—attentive agents extract precision $\tau_A = \gamma_A \tau_S$ while LLMs extract $\tau_L = \lambda(c,k) \tau_S$ from the same signal $s$. To formally model a scenario where some agents can learn from the equilibrium price, it is necessary to introduce \textit{private} information. A price based only on differential processing of public information would not convey any new information about the fundamental to agents who already observe that public signal—they could perfectly infer the noise component and learn nothing new.

Here, we operationalize the LLM's superior capability as providing access to a private signal $\eta$ that represents information processed more quickly by LLMs and revealed publicly with a lag. This approach is conceptually consistent with the main model: an agent with a private signal has higher total information precision (i.e., $\tau_d + \tau_A + \tau_\eta$ vs. $\tau_d + \tau_A$), which is economically equivalent to superior processing capacity. The private signal $\eta$ can be interpreted as complex information that sophisticated AI systems extract from news before it becomes obvious to all market participants—for example, subtle patterns in earnings reports or connections between disparate pieces of information. This modification allows for a tractable analysis of prices as a channel of information aggregation and learning, which is the central question for this extension.\footnote{An alternative interpretation: the private signal represents the output of a proprietary LLM model that some investors have exclusive access to in the short run, before the same analysis becomes widely available. This maps to our empirical setting where sophisticated investors may have deployed GPT-4 or similar models before they became publicly accessible.}

\subsection{Model Setup and Assumptions}

\subsubsection{Primitive Assumptions}

\begin{assumption}[Market Structure]\label{ass:market}
\begin{enumerate}[label=(\roman*)]
\item The economy consists of a continuum of agents with total mass $V > 0$, partitioned into attentive agents (measure $\pi_A V$) and inattentive agents (measure $\pi_I V = (1-\pi_A)V$).
\item All agents have CARA utility with risk aversion parameter $\alpha > 0$: $U(w) = -\exp(-\alpha w)$.
\item There exists a single risky asset in zero net supply with payoff $\tilde{d}$.
\item Markets are competitive (agents are price-takers).
\end{enumerate}
\end{assumption}

\begin{assumption}[Information Structure]\label{ass:info}
\begin{enumerate}[label=(\roman*)]
\item The asset payoff is $\tilde{d} = \mu_d + \sigma_{\xi} \tilde{\xi}$, where $\tilde{\xi} \sim N(0,1)$ is independent of all other random variables.
\item The fundamental $\mu_d \sim N(\bar{d}, \sigma_d^2)$ with precision $\tau_d = \sigma_d^{-2}$.
\item A public signal $s = \mu_d + \varepsilon_s$ is revealed, where $\varepsilon_s \sim N(0, \sigma_S^2)$ with precision $\tau_S = \sigma_S^{-2}$. This signal is observed by all agents in Period 1.
\item In Period 1, LLM adopters gain exclusive access to an additional private signal $\eta = \mu_d + \varepsilon_\eta$, where $\varepsilon_\eta \sim N(0, \sigma_\eta^2)$ with precision $\tau_\eta = \sigma_\eta^{-2}$. This signal represents information that is processed more quickly by LLMs and is publicly revealed in Period 2 (see Assumption \ref{ass:timing}).
\item Independence: The noise terms $\varepsilon_s$ and $\varepsilon_\eta$ are independent of each other and of $\mu_d$.
\item Information processing capacities for the public signal are:
\begin{align}
\tau_A &= \gamma_A \tau_S, \quad \gamma_A \in (0,1) \quad \text{(Attentive processing of public signal)} \\
\tau_I &= \omega \tau_A, \quad \omega \in (0,1) \quad \text{(Inattentive processing of public signal)}
\end{align}
\item The LLM's superiority is captured by the precision of its private signal:
\begin{equation}
\tau_\eta = \lambda(c,k) \tau_S
\end{equation}
where $\lambda: \mathbb{R}_+^2 \to [0,1]$ satisfies the regularity conditions in Assumption \ref{ass:llm}. For LLM adopters with technology quality $k > 0$, the boundary conditions and monotonicity imply $\lambda(c,k) > 0$, hence $\tau_\eta > 0$. The restriction $\lambda \leq 1$ ensures $\tau_\eta \leq \tau_S$ (the private signal is no more precise than the raw public signal), a natural modeling choice though not essential for the equilibrium characterization. We assume attentive agents who adopt LLMs still process the public signal with precision $\tau_A$.
\end{enumerate}
\end{assumption}

\begin{assumption}[LLM Technology]\label{ass:llm}
The function $\lambda(c,k)$, where $k$ represents the quality (e.g., model size) of the LLM technology and $c$ represents the complexity of the information being processed, is twice continuously differentiable with:
\begin{enumerate}[label=(\roman*)]
\item $\frac{\partial \lambda}{\partial k} > 0$, $\frac{\partial \lambda}{\partial c} \leq 0$ (monotonicity)
\item $\frac{\partial^2 \lambda}{\partial k^2} \leq 0$, $\frac{\partial^2 \lambda}{\partial c^2} \leq 0$ (concavity)
\item $\frac{\partial^2 \lambda}{\partial k \partial c} > 0$ (complementarity)
\item $\lambda(c,0) = 0$ for all $c \geq 0$, and $\lim_{k \to \infty} \lambda(c,k) = 1$ for all $c \geq 0$
\end{enumerate}
\end{assumption}

\begin{assumption}[Timing and Trading]\label{ass:timing}
\begin{enumerate}[label=(\roman*)]
\item Period 0: No news, initial price $p_0 = \bar{d}$.
\item Period 1: Signal $s$ revealed, agents trade based on partial information processing, noise traders enter with aggregate demand $u_1 \sim N(0, \sigma_{u1}^2)$ (representing a continuum of small noise traders).
\item Period 2: Full information processing. The private signal $\eta$ is publicly revealed as the raw signal $\mu_d + \varepsilon_\eta$, where $\varepsilon_\eta \sim N(0, \sigma_\eta^2)$. By Period 2, information processing frictions lapse: all agents process the public signal $s$ at attentive precision $\tau_A$ and observe $\eta$. This yields a common posterior $\mu_{L|s,\eta}$ with payoff variance $\sigma_{d,L}^2 = T_L^{-1} + \sigma_\xi^2$, where $T_L = \tau_d + \tau_A + \tau_\eta$. Agents retrade based on this unified information set, and noise traders enter with additional demand $u_2 \sim N(0, \sigma_{u2}^2)$.
\item Period 3: Asset pays $\tilde{d}$.
\item Independence: Noise demands $u_1, u_2$ are independent of $(\mu_d, s, \eta)$ and of each other.
\end{enumerate}
\end{assumption}

\begin{assumption}[Inattentive Agent Behavior]\label{ass:inattentive}
Inattentive agents:
\begin{enumerate}[label=(\roman*)]
\item Trade only on their noisy processing of the public signal $s$ (with precision $\tau_I$) and do not attempt to extract information from the equilibrium price $p_1$.
\item Do not engage in speculative trading. They trade only based on the asset's fundamental payoff $\tilde{d}$ and do not anticipate retrading at price $p_2$.
\end{enumerate}
This assumption reflects limited sophistication: inattentive agents lack both the information processing capacity to learn from prices and the strategic sophistication to engage in speculative arbitrage across periods. This behavior is consistent with models of noise traders or unsophisticated retail investors who trade on simple heuristics.
\end{assumption}

\textbf{Note on Private Information and Noise Trading:} A key feature of this equilibrium is that learning from prices requires two components: private information and noise. The private signal, $\eta$, is essential because it introduces information into the market that is not common knowledge. Without it, the price would only be a function of the public signal $s$ and noise $u_1$. Any agent observing $s$ could perfectly infer the noise component, learning nothing new about the asset's fundamental value.

The noise trading, $u_1$, is equally essential. It garbles the price signal, preventing the private information $\eta$ from being perfectly revealed. Without noise, the price would become a deterministic function of the signals, allowing non-adopters to deduce $\eta$ exactly (and thus match adopters' information), eliminating the LLM adopters' informational advantage. Together, private information and noise create a meaningful signal-extraction problem for non-adopting agents.

\textit{Units:} The noise demands $u_1, u_2$ are measured in shares, so the coefficient $\chi$ in $p_1 = a + bs + g\eta + \chi u_1$ has units (price/share), ensuring $(\chi/g)u_1$ has the same units as $\eta$.

\subsection{Equilibrium Characterization}

\subsubsection{Bayesian Updating}

Given their information, agent types form different posterior beliefs.

\begin{lemma}[Posterior Beliefs]\label{lem:posterior}
\begin{enumerate}[label=(\roman*)]
\item For an LLM-adopting agent (type $L$) observing public signal $s$ (processed with precision $\tau_A$) and private signal $\eta$ (with precision $\tau_\eta$):
\begin{align}
\mu_{L|s,\eta} &\equiv E[\tilde{d}|s, \eta] = \frac{\bar{d} \tau_d + s \tau_A + \eta \tau_\eta}{\tau_d + \tau_A + \tau_\eta} \\
\sigma_{\mu,L|s,\eta}^2 &\equiv Var[\mu_d|s, \eta] = \frac{1}{\tau_d + \tau_A + \tau_\eta}
\end{align}
\item For an attentive, non-adopting agent (type $A$) processing only the public signal $s$ with precision $\tau_A$:
\begin{align}
\mu_{A|s} &\equiv E[\tilde{d}|s, \tau_A] = \frac{\bar{d} \tau_d + s \tau_A}{\tau_d + \tau_A} \\
\sigma_{\mu,A|s}^2 &\equiv Var[\mu_d|s, \tau_A] = \frac{1}{\tau_d + \tau_A}
\end{align}
\item For an inattentive agent (type $I$) processing only the public signal $s$ with precision $\tau_I$:
\begin{align}
\mu_{I|s} &\equiv E[\tilde{d}|s, \tau_I] = \frac{\bar{d} \tau_d + s \tau_I}{\tau_d + \tau_I} \\
\sigma_{\mu,I|s}^2 &\equiv Var[\mu_d|s, \tau_I] = \frac{1}{\tau_d + \tau_I}
\end{align}
\item For any agent $j$ with information set $\mathcal{I}_j$, the posterior variance of the fundamental is $\sigma_{\mu,j|\mathcal{I}_j}^2 \equiv Var[\mu_d|\mathcal{I}_j]$. The total conditional variance of the payoff is $\sigma_{d,j|\mathcal{I}_j}^2 \equiv Var[\tilde{d}|\mathcal{I}_j] = \sigma_{\mu,j|\mathcal{I}_j}^2 + \sigma_{\xi}^2$.
\end{enumerate}
\end{lemma}

\begin{proof}
Standard Bayesian updating with normal distributions. The posterior precision is the sum of the prior and signal precisions. The total variance includes both fundamental uncertainty and payoff noise.
\end{proof}

\subsubsection{Key Definitions for Equilibrium Analysis}

\begin{definition}[Precision Terms]\label{def:precision}
The following precision terms are used in the equilibrium characterization:
\begin{enumerate}[label=(\roman*)]
\item \textbf{Fundamental precision after signals:}
\begin{align}
T_L &\equiv \tau_d + \tau_A + \tau_\eta \quad \text{(LLM adopters)} \\
T_{A^*} &\equiv \tau_d + \tau_A + \tau_{\text{learn}} \quad \text{(Attentive non-adopters after price learning)} \\
T_I &\equiv \tau_d + \tau_I \quad \text{(Inattentive agents)}
\end{align}
where $\tau_{\text{learn}}$ is the precision of information about $\mu_d$ extracted from the price (derived in the proof).

\item \textbf{Type-specific speculative precisions:} Following the logic of the main model, speculative precision arises from the risk of trading against future noise. As established in Assumption \ref{ass:timing}, by period 2 all agents trade on the unified information set $(s, \eta)$. The consensus expectation is therefore $\mu_{L|s,\eta}$ with total payoff variance $\sigma_{d,L}^2 = T_L^{-1} + \sigma_\xi^2$. The period 2 price is thus $p_2 = \mu_{L|s,\eta} + (\alpha/V)\sigma_{d,L}^2 u_2$.

\textit{For LLM adopters:} Since they observe $(s, \eta)$, the uncertainty in the period 2 price arises solely from the future noise term $u_2$. The variance of the future price is $Var_1(p_2|s,\eta) = ((\alpha/V)\sigma_{d,L}^2)^2 \sigma_{u2}^2$. Their speculative precision is:
\begin{equation}
\tau_{p,L} = \frac{V^2}{\alpha^2 \sigma_{u2}^2 (\sigma_{d,L}^2)^2}
\end{equation}

\textit{For non-adopting attentive agents:} They observe $(s, p_1)$ but not $\eta$ directly. Their uncertainty about $p_2$ has two components: (a) uncertainty about $\mu_{L|s,\eta}$ given their information, and (b) the noise $u_2$. Define the incremental uncertainty from the law of total variance:
\begin{equation}
\Delta \equiv Var(\mu_{L|s,\eta}|s,p_1) = Var(\mu_d|s,p_1) - Var(\mu_d|s,\eta) = \frac{1}{T_{A^*}} - \frac{1}{T_L} \geq 0
\end{equation}

\item \textbf{Effective trading precisions:} From Lemma \ref{lem:demand}, when agents can retrade and the price expectation is unbiased ($E_1[p_2] = E_1[\tilde{d}]$), the optimal demand can be expressed using effective precision terms:

\textit{LLM adopters:} Their demand precision exhibits additive structure:
\begin{equation}
\tau_{\text{eff},L} = \sigma_{d,L}^{-2} + \tau_{p,L} = (T_L^{-1} + \sigma_\xi^2)^{-1} + \tau_{p,L}
\end{equation}

\textit{Non-adopters:} Their demand precision incorporates the parallel variance operator $(a \,|\, b) \equiv (a^{-1} + b^{-1})^{-1}$:
\begin{equation}
\tau_{\text{eff},A^*} = \left[\Delta + (\Sigma_p \,|\, \sigma_{d,L}^2)\right]^{-1} = \left[\Delta + \frac{\Sigma_p \sigma_{d,L}^2}{\Sigma_p + \sigma_{d,L}^2}\right]^{-1}
\end{equation}
where $\Sigma_p = \tau_{p,L}^{-1}$ is the variance of future price risk. This is derived in Lemma \ref{lem:demand}.

\textit{Inattentive agents:} By Assumption \ref{ass:inattentive}, they neither learn from prices nor engage in speculative trading:
\begin{equation}
\tau_{\text{eff},I} = \sigma_{d,I}^{-2} = (T_I^{-1} + \sigma_\xi^2)^{-1}
\end{equation}
\end{enumerate}
\end{definition}

\subsection{The Case of Partial Adoption by Attentive Agents}

Consider the case where fraction $\phi \in [0,1]$ of attentive agents adopt LLM technology, gaining access to a private signal $\eta$ with precision $\tau_\eta > 0$. This gives them superior total information precision: $T_L = \tau_d + \tau_A + \tau_\eta > \tau_d + \tau_A$. The remaining $(1-\phi)$ fraction of attentive agents do not have this signal but may infer it from the equilibrium price.

\textbf{Economic Interpretation:} This scenario captures a sophisticated market where some investors deploy advanced (e.g., proprietary LLM) technology, while others engage in price discovery to infer the conclusions of those advanced models. The learning-from-prices mechanism creates strategic complementarity: as more agents adopt LLMs, prices become more informative about their private signals, making non-adoption a potentially viable strategy and leading to a stable partial adoption equilibrium.

\begin{definition}[Rational Expectations Equilibrium with Learning]
An equilibrium consists of:
\begin{enumerate}[label=(\roman*)]
\item Price function $P: \mathbb{R}^3 \to \mathbb{R}$ mapping signals $(s, \eta)$ and noise $u_1$ to price $p_1$
\item Belief functions $\mu_j: \mathbb{R}^2 \to \mathbb{R}$ for each agent type $j$
\item Demand functions $x_j: \mathbb{R}^2 \to \mathbb{R}$ for each agent type $j$
\end{enumerate}
such that markets clear and non-adopting attentive agents optimally extract information from prices.
\end{definition}

\begin{theorem}[Partial Adoption by Attentive with Learning]\label{thm:partial_attentive}
Assume a fraction $\phi \in (0,1]$ of attentive agents adopt LLM technology, giving them a private signal $\eta$ with precision $\tau_\eta > 0$. In a linear Rational Expectations Equilibrium:

\begin{enumerate}[label=(\roman*)]
\item The price function is $p_1 = a + bs + g\eta + \chi u_1$ where the coefficients $(a,b,g,\chi)$ are determined by market clearing.

\item Non-adopting attentive agents observe $(s, p_1)$ and construct an effective signal about the fundamental $\mu_d$. Their updated fundamental precision is:
\begin{equation}
T_{A^*} = \tau_d + \tau_A + \tau_{\text{learn}}
\end{equation}
where $\tau_{\text{learn}} = (\sigma_\eta^2 + (\chi/g)^2 \sigma_{u1}^2)^{-1}$ is the precision of the information about $\mu_d$ extracted from the price. The noise-to-signal ratio has a closed-form expression: $(\chi/g) = \alpha T_L/(V m_L \tau_{\text{eff},L} \tau_\eta)$, where $T_L = \tau_d + \tau_A + \tau_\eta$, $\tau_{\text{eff},L} = \sigma_{d,L}^{-2} + \tau_{p,L}$, and $m_L = \phi \pi_A$, depending only on adopter-side objects and primitives.

\item Market efficiency is increasing in the fraction of LLM adopters, $\phi$, and the quality of the LLM technology, $\tau_\eta$. We measure efficiency by the forecast variance $\mathrm{Var}(\mu_d\mid s,p_1)=1/T_{A^*}$; thus efficiency increases if and only if $T_{A^*}$ increases.
\end{enumerate}
\end{theorem}

\begin{proof}
\textbf{Step 1: Price conjecture.} Conjecture a linear price function of all random variables: $p_1 = a + bs + g\eta + \chi u_1$.

\textbf{Step 2: Information extraction by Non-Adopters.} Non-adopting attentive agents observe the public signal $s$ and the price $p_1$. They know the conjectured form of the price function. They can then construct a new signal, $z$, by isolating the parts of the price they do not already know:
\begin{equation}
z \equiv \frac{p_1 - a - bs}{g} = \eta + \frac{\chi}{g} u_1
\end{equation}
This signal $z$ is a noisy measure of the LLM adopters' private signal $\eta$. Since $\eta = \mu_d + \varepsilon_\eta$, the signal $z$ can be rewritten as a noisy signal about the fundamental $\mu_d$:
\begin{equation}
z = \mu_d + \varepsilon_\eta + \frac{\chi}{g} u_1
\end{equation}
The total noise in this signal is $\varepsilon_z = \varepsilon_\eta + \frac{\chi}{g} u_1$, which has a variance of $Var(\varepsilon_z) = \sigma_\eta^2 + (\frac{\chi}{g})^2\sigma_{u1}^2$. The precision of the information about $\mu_d$ that non-adopters can learn from the price is:
\begin{equation}
\tau_{\text{learn}} = \frac{1}{\sigma_\eta^2 + (\chi/g)^2 \sigma_{u1}^2}
\end{equation}
where $\chi/g$ represents the noise-to-signal ratio in the price function, capturing how much noise trading affects the informativeness of prices relative to the private signal component.

\begin{lemma}[Bound on Learning Precision]\label{lem:learn_bound}
$\tau_{\text{learn}} \leq \tau_\eta$, with equality if and only if $\sigma_{u1}^2 = 0$ (no noise trading in Period 1).
\end{lemma}
\begin{proof}
Immediate from the definition: $\tau_{\text{learn}} = [\sigma_\eta^2 + (\chi/g)^2 \sigma_{u1}^2]^{-1} \leq \sigma_\eta^{-2} = \tau_\eta$.
\end{proof}

The non-adopter now has an information set effectively equivalent to observing $(s, z)$. Their posterior belief about $\mu_d$ is formed by combining the prior, the public signal $s$ (processed with precision $\tau_A$), and their inferred signal $z$ (which provides information about $\mu_d$ with precision $\tau_{\text{learn}}$).

The non-adopter's updated posterior mean and fundamental precision are:
\begin{align}
\mu_{A^*|s,p} &= \frac{\bar{d}\tau_d + s\tau_A + z\tau_{\text{learn}}}{T_{A^*}} \\
T_{A^*} &= \tau_d + \tau_A + \tau_{\text{learn}}
\end{align}

\textit{Key insight for closed-form solution:} As shown in Step 5 below, the noise-to-signal ratio can be expressed as:
\begin{equation}
\frac{\chi}{g} = \frac{\alpha T_L}{V m_L \tau_{\text{eff},L} \tau_\eta}
\end{equation}
where $m_L = \phi \pi_A$ is the mass of LLM adopters and $\tau_{\text{eff},L} = \sigma_{d,L}^{-2} + \tau_{p,L}$. Crucially, this ratio depends only on adopter-side objects and primitives, breaking what would otherwise be a circular dependence on non-adopter precisions. This allows us to compute $\tau_{\text{learn}}$ in closed form before determining non-adopter demands.

\begin{lemma}[Gaussian Linear-Quadratic MGF]\label{lem:mgf_quadratic_r2}
If $U \sim N(0, \sigma^2)$, then for any $t \in \mathbb{R}$ and $a > -1/\sigma^2$:
\begin{equation}
E\left[\exp\left(tU - \frac{a}{2}U^2\right)\right] = \frac{1}{\sqrt{1 + a\sigma^2}} \exp\left(\frac{t^2 \sigma^2}{2(1 + a\sigma^2)}\right)
\end{equation}
\end{lemma}
\begin{proof}
Complete the square in the exponent and integrate the resulting Gaussian density.
\end{proof}

\begin{lemma}[Optimal Demand with Retrading]\label{lem:demand}
Consider a CARA agent with risk aversion $\alpha$ who can trade at price $p_1$ in Period 1 and retrade at random price $\tilde{p}_2$ in Period 2 before the asset pays dividend $\tilde{d}$ in Period 3.

\textbf{(i) When the agent observes all information in Period 1:} If the agent's Period-1 conditional expectation satisfies $E_1[p_2] = E_1[\tilde{d}] = \mu$ and the agent knows the Period-2 consensus, the optimal Period-1 demand exhibits additive precision structure:
\begin{equation}
x_1 = \left( \tau_{\text{fundamental}} + \tau_p \right) \frac{\mu - p_1}{\alpha}
\end{equation}
where $\tau_{\text{fundamental}} = 1/Var_1(\tilde{d})$ and $\tau_p = 1/Var_1(p_2)$.

\textbf{(ii) When the agent learns the consensus in Period 2:} If the Period-2 consensus $\mu_{L|s,\eta}$ is uncertain from the agent's Period-1 perspective (with variance $\Delta > 0$), the optimal demand is:
\begin{equation}
x_1 = \frac{\mu_{A^*|s,p} - p_1}{\alpha \left[\Delta + (\Sigma_p \,|\, \sigma_{d,L}^2)\right]}
\end{equation}
where $(a \,|\, b) \equiv (a^{-1} + b^{-1})^{-1}$ is the parallel variance operator, $\Sigma_p = \tau_p^{-1}$ is the variance of future price risk, and $\sigma_{d,L}^2$ is the Period-2 payoff variance.

\end{lemma}
\begin{proof}
We solve by backward induction. The key challenge is that agents have heterogeneous information sets in Period 1, leading to different posterior beliefs and trading demands.

\vspace{0.3cm}
\noindent\textbf{Case 1: LLM Adopters}

For LLM adopters who observe $(s,\eta)$ in Period 1, the derivation follows standard backward induction with CARA-normal preferences:

\begin{enumerate}
\item \textbf{Period 2 value function}: Standard CARA-normal gives indirect value $V_2(w_2) = -\exp(-\alpha w_2 - \frac{1}{2}\frac{(\mu_{L|s,\eta} - p_2)^2}{\sigma_{d,L}^2})$

\item \textbf{Period 1 optimization}: The agent maximizes $E_1[V_2(\tilde{w}_2)|s,\eta]$ where $\tilde{w}_2 = w_1 + x_1(\tilde{p}_2 - p_1)$

\item \textbf{Quadratic expansion}: The exponent becomes $a(x_1) + b(x_1)\tilde{p}_2 + c\tilde{p}_2^2$ with:
\begin{align*}
a(x_1) &= \alpha p_1 x_1 - \frac{\mu_{L|s,\eta}^2}{2\sigma_{d,L}^2} - \alpha w_1 \\
b(x_1) &= \frac{\mu_{L|s,\eta}}{\sigma_{d,L}^2} - \alpha x_1 \\
c &= -\frac{1}{2\sigma_{d,L}^2}
\end{align*}

\item \textbf{Applying Lemma \ref{lem:mgf_quadratic_r2}}: With $\tilde{p}_2 \sim N(\mu_{L|s,\eta}, \sigma_{p,L}^2)$ where $\sigma_{p,L}^2 = 1/\tau_{p,L}$, taking the FOC and solving yields:
\begin{equation}
x_{1,L} = \frac{(\mu_{L|s,\eta} - p_1)(\sigma_{d,L}^2 + \sigma_{p,L}^2)}{\alpha \sigma_{p,L}^2 \sigma_{d,L}^2} = \underbrace{\frac{\mu_{L|s,\eta} - p_1}{\alpha \sigma_{d,L}^2}}_{\text{Fundamental}} + \underbrace{\frac{\mu_{L|s,\eta} - p_1}{\alpha \sigma_{p,L}^2}}_{\text{Speculative}}
\end{equation}
\end{enumerate}

\vspace{0.3cm}
\noindent\textbf{Case 2: Non-Adopting Attentive Agents}

For non-adopters who observe $(s,p_1)$ in Period 1 but learn $\eta$ in Period 2, the Period-2 consensus belief $E_2[\tilde{d}] = \mu_{L|s,\eta}$ is random from their Period-1 perspective. The key to solving this problem is to substitute the market-clearing law of motion for $p_2$ before taking expectations.

\textbf{Step 1: Market clearing in Period 2.} From Assumption \ref{ass:timing}, in Period 2 all agents observe $(s,\eta)$, yielding consensus $\mu_{L|s,\eta}$ and variance $\sigma_{d,L}^2 = T_L^{-1} + \sigma_\xi^2$. Market clearing gives:
\begin{equation}
p_2 = \mu_{L|s,\eta} + \kappa u_2, \qquad \kappa \equiv \frac{\alpha}{V}\sigma_{d,L}^2, \quad u_2 \sim N(0,\sigma_{u2}^2), \quad u_2 \perp (\mu_d, s, \eta)
\end{equation}

\textit{Why this substitution is valid:} The Period-2 market-clearing condition is an equilibrium outcome, not a choice variable. Since $\kappa$ and the distribution of $u_2$ are primitives (independent of agent $A^*$'s Period-1 choice), we can substitute this law of motion into the agent's Period-1 optimization without introducing circularity. The agent takes the stochastic relationship $p_2 = \mu_{L|s,\eta} + \kappa u_2$ as given when choosing $x_1$.

\textbf{Step 2: The key cancellation.} Let $\mu \equiv \mu_{L|s,\eta}$ denote the Period-2 consensus (random from Period-1 perspective). Substituting the market-clearing condition:
\begin{equation}
(\mu - p_2)^2 = (\mu - (\mu + \kappa u_2))^2 = \kappa^2 u_2^2
\end{equation}

This eliminates both the $\mu^2$ term and the $\mu \cdot p_2$ cross term that would otherwise create a bivariate problem!

\textbf{Step 3: Factorization.} The Period-1 objective becomes:
\begin{align}
\max_{x_1} E_1\left[-\exp\left(-\alpha(w_1 + x_1(\mu + \kappa u_2 - p_1)) - \frac{\kappa^2 u_2^2}{2\sigma_{d,L}^2}\right) \bigg| s,p_1\right]
\end{align}

The exponent is linear in $\mu$ and quadratic in $u_2$. Since $\mu \perp u_2$, the expectation factorizes:
\begin{equation}
e^{-\alpha(w_1 - x_1 p_1)} \cdot \underbrace{E_1[e^{-\alpha x_1 \mu}|s,p_1]}_{\mu\text{-part}} \cdot \underbrace{E[e^{-\alpha x_1 \kappa u_2 - \kappa^2 u_2^2/(2\sigma_{d,L}^2)}]}_{u_2\text{-part}}
\end{equation}

\textbf{Step 4: Computing the expectations.} From the non-adopter's Period-1 perspective:
\begin{itemize}
\item $\mu \mid (s,p_1) \sim N(\mu_p, \Delta)$ where $\mu_p = E_1[\mu_{L|s,\eta}|s,p_1] = \mu_{A^*|s,p}$ and $\Delta = Var_1(\mu_{L|s,\eta}|s,p_1) = T_{A^*}^{-1} - T_L^{-1} \geq 0$
\item $u_2 \sim N(0, \sigma_{u2}^2)$ independent of all Period-1 information
\end{itemize}

Using standard MGFs for univariate normals:
\begin{align}
E_1[e^{-\alpha x_1 \mu}|s,p_1] &= \exp\left(-\alpha x_1 \mu_p + \frac{\alpha^2 x_1^2 \Delta}{2}\right)\\
E[e^{-\alpha x_1 \kappa u_2 - \kappa^2 u_2^2/(2\sigma_{d,L}^2)}] &= \frac{1}{\sqrt{1 + \Sigma_p/\sigma_{d,L}^2}} \exp\left(\frac{\alpha^2 x_1^2 \Sigma_p}{2(1 + \Sigma_p/\sigma_{d,L}^2)}\right)
\end{align}
where $\Sigma_p \equiv \kappa^2 \sigma_{u2}^2 = \left(\frac{\alpha}{V}\sigma_{d,L}^2\right)^2 \sigma_{u2}^2 = 1 / \tau_{p,L}$.

\textbf{Step 5: FOC and solution.} Ignoring constants in $x_1$, the log-objective is:
\begin{equation}
\mathcal{L}(x_1) = \alpha x_1(p_1 - \mu_p) + \frac{\alpha^2 x_1^2}{2}\left[\Delta + \frac{\Sigma_p}{1 + \Sigma_p/\sigma_{d,L}^2}\right]
\end{equation}

Define the \textbf{parallel variance} operator $(a \,|\, b) \equiv (a^{-1} + b^{-1})^{-1}$ (the variance of two independent Gaussian risks borne in parallel) and the effective variance:
\begin{equation}
\sigma_{\text{eff}}^2 \equiv \Delta + (\Sigma_p \,|\, \sigma_{d,L}^2) = \Delta + \frac{\Sigma_p \sigma_{d,L}^2}{\Sigma_p + \sigma_{d,L}^2}
\end{equation}

The independence $u_2 \perp (\mu_d, s, \eta)$ allows the expectation to factorize, yielding a closed-form solution.

Taking the FOC: $\frac{d\mathcal{L}}{dx_1} = \alpha(p_1 - \mu_p) + \alpha^2 x_1 \sigma_{\text{eff}}^2 = 0$, yielding:
\begin{equation}
\boxed{x_{1,A^*} = \frac{\mu_p - p_1}{\alpha \sigma_{\text{eff}}^2} = \frac{\mu_{A^*|s,p} - p_1}{\alpha \left[\Delta + (\Sigma_p \,|\, \sigma_{d,L}^2)\right]}}
\end{equation}

\textbf{Sanity checks:}
\begin{itemize}
\item If $\Delta = 0$ (non-adopters already know $\eta$), then $\sigma_{\text{eff}}^2 = \Sigma_p \,|\, \sigma_{d,L}^2$ and $x_{1,A^*} = (\sigma_{d,L}^{-2} + \tau_{p,L}) \frac{\mu_{L|s,\eta} - p_1}{\alpha}$ (identical to $L$).
\item If $\sigma_{d,L}^2 \gg \Sigma_p$ (future-price risk dominates payoff risk), then $\Sigma_p \,|\, \sigma_{d,L}^2 \approx \Sigma_p$ and $x_{1,A^*} \approx \frac{\mu_{A^*|s,p} - p_1}{\alpha(\Delta + \Sigma_p)}$.
\end{itemize}

\vspace{0.3cm}
\noindent\textbf{General Result}

The demand formulas for all agent types are:
\begin{itemize}
\item \textbf{LLM adopters} ($\Delta = 0$):
\begin{equation}
x_{1,L} = \left(\sigma_{d,L}^{-2} + \tau_{p,L}\right) \frac{\mu_{L|s,\eta} - p_1}{\alpha}
\end{equation}

\item \textbf{Non-adopters} ($\Delta = T_{A^*}^{-1} - T_L^{-1} \geq 0$):
\begin{equation}
x_{1,A^*} = \frac{\mu_{A^*|s,p} - p_1}{\alpha \left[\Delta + (\Sigma_p \,|\, \sigma_{d,L}^2)\right]}
\end{equation}

\item \textbf{Inattentive agents}: Only fundamental component (no speculation), $x_{1,I} = \frac{\mu_{I|s} - p_1}{\alpha \sigma_{d,I}^2}$
\end{itemize}

where $\Sigma_p = \tau_{p,L}^{-1} = (\alpha/V)^2 (\sigma_{d,L}^2)^2 \sigma_{u2}^2$ is the variance of future price risk and $(a \,|\, b) = (a^{-1} + b^{-1})^{-1}$ is the parallel variance operator.
\end{proof}

\textbf{Step 3: Demands.} All agents form their demands based on their information, as derived in Lemma \ref{lem:demand}.

For LLM adopters, since $E_1[\tilde{d}|s,\eta] = E_1[p_2|s,\eta] = \mu_{L|s,\eta}$:
\begin{equation}
x_{L} = \left(\sigma_{d,L}^{-2} + \tau_{p,L}\right) \frac{\mu_{L|s,\eta} - p_1}{\alpha}
\end{equation}
where $\sigma_{d,L}^2 = T_L^{-1} + \sigma_\xi^2$ and $\tau_{p,L} = V^2/(\alpha^2 \sigma_{u2}^2 (\sigma_{d,L}^2)^2)$.

For non-adopters, their demand incorporates the additional uncertainty $\Delta = T_{A^*}^{-1} - T_L^{-1}$ about the Period-2 consensus:
\begin{equation}
x_{A^*} = \frac{\mu_{A^*|s,p} - p_1}{\alpha \left[\Delta + (\Sigma_p \,|\, \sigma_{d,L}^2)\right]}
\end{equation}
where $\Sigma_p = \tau_{p,L}^{-1}$ and $(a \,|\, b) = (a^{-1} + b^{-1})^{-1}$ is the parallel variance operator.

For inattentive agents, by Assumption \ref{ass:inattentive}, they neither learn from prices nor engage in speculative trading:
\begin{equation}
x_{I} = \sigma_{d,I}^{-2} \frac{\mu_{I|s} - p_1}{\alpha}
\end{equation}
where $\sigma_{d,I}^2 = T_I^{-1} + \sigma_\xi^2$.

\textbf{Step 4: Market clearing.} Define group masses: $m_L = \phi \pi_A$, $m_A = (1-\phi)\pi_A$, $m_I = \pi_I$. The aggregate demand must clear the market:
\begin{equation}
m_L V \cdot x_{L} + m_A V \cdot x_{A^*} + m_I V \cdot x_{I} = -u_1
\end{equation}

Substituting the demand functions from Step 3 and solving for $p_1$ yields a linear price function. Define effective precision weights:
\begin{align}
\tau_{\text{eff},L} &\equiv \sigma_{d,L}^{-2} + \tau_{p,L} \\
\tau_{\text{eff},A^*} &\equiv \left[\Delta + (\Sigma_p \,|\, \sigma_{d,L}^2)\right]^{-1} \\
\tau_{\text{eff},I} &\equiv \sigma_{d,I}^{-2}
\end{align}

Define the aggregate effective precision mass:
\begin{equation}
S \equiv m_L \tau_{\text{eff},L} + m_A \tau_{\text{eff},A^*} + m_I \tau_{\text{eff},I}
\end{equation}
and precision-weighted masses $w_j = m_j \tau_{\text{eff},j}/S$ for $j \in \{L, A^*, I\}$. Then:
\begin{equation}
p_1 = \sum_{j \in \{L,A^*,I\}} w_j \mu_j + \frac{\alpha}{VS} u_1
\end{equation}
where $\mu_L = \mu_{L|s,\eta}$, $\mu_{A^*} = \mu_{A^*|s,p}$, and $\mu_I = \mu_{I|s}$.

\begin{remark}[Existence and uniqueness of linear REE]
For $\phi>0$, $\tau_\eta>0$, and finite $\sigma_{u1}^2,\sigma_{u2}^2$, a linear REE exists and is unique.

\textit{Existence} follows from the constructive sequence: adopters' objects $\Rightarrow$ the ratio
\[
\frac{\chi}{g}=\frac{\alpha T_L}{V m_L \tau_{\text{eff},L}\tau_\eta}
\]
(depends only on adopter-side primitives) $\Rightarrow$ $\tau_{\text{learn}}$, $T_{A^*}$, $S$, $w_j$ $\Rightarrow$ $(a,b,g,\chi)$. Well-posedness requires $S > 0$ (satisfied because each $\tau_{\text{eff},j} > 0$ under finite variances) and $g > 0$ (satisfied because $\kappa_\eta > 0$ when $m_L > 0$ and $\tau_\eta > 0$).

\textit{Uniqueness} holds because any linear REE must satisfy the same coefficient-matching, which forces the same unique $\chi/g$, hence the same $\tau_{\text{learn}}$, $T_{A^*}$, $S$, $w_j$, and therefore the same $(a,b,g,\chi)$.

\textit{Economic intuition:} Adopters' trading against noise pins the price's informativeness ($\chi/g$). Once that ratio is fixed, what non-adopters can learn, their risk, their demand, and therefore the entire price—all become determinate. This mirrors the standard CARA–Gaussian linear-REE uniqueness logic \citep{Hellwig1980OnMarkets, grossman1980impossibility}, with the refinement that the key fixed point collapses to the single ratio $\chi/g$, which we solve in closed form.
\end{remark}

\textbf{Step 5: Solving for price coefficients.} Substituting the expressions for posterior means:
\begin{align}
\mu_L &= \frac{\tau_d \bar{d} + \tau_A s + \tau_\eta \eta}{T_L} \\
\mu_{A^*} &= \frac{\tau_d \bar{d} + \tau_A s + \tau_{\text{learn}} z}{T_{A^*}}, \quad z = \frac{p_1 - a - bs}{g} \\
\mu_I &= \frac{\tau_d \bar{d} + \tau_I s}{T_I}
\end{align}

Matching coefficients with $p_1 = a + bs + g\eta + \chi u_1$ gives:
\begin{align}
a &= \left(\frac{w_L \tau_d}{T_L} + \frac{w_{A^*} \tau_d}{T_{A^*}} + \frac{w_I \tau_d}{T_I}\right) \bar{d} \\
b &= \frac{w_L \tau_A}{T_L} + \frac{w_{A^*} \tau_A}{T_{A^*}} + \frac{w_I \tau_I}{T_I}
\end{align}

For the coefficients on $\eta$ and $u_1$, define:
\begin{align}
\kappa_\eta &\equiv \frac{w_L \tau_\eta}{T_L}, \quad \kappa_A \equiv \frac{w_{A^*} \tau_{\text{learn}}}{T_{A^*}}, \quad \kappa_u^0 \equiv \frac{\alpha}{VS}
\end{align}

Collecting terms yields:
\begin{align}
g &= \kappa_\eta + \kappa_A \quad \text{($\eta$-coefficient: adopters and non-adopters)}\\
\chi &= \kappa_A \left(\frac{\chi}{g}\right) + \kappa_u^0 \quad \text{($u_1$-coefficient: direct and learned channels)}
\end{align}

\textbf{Key algebraic simplification:} Dividing the $\eta$-coefficient by the $u_1$-coefficient immediately yields the noise-to-signal ratio:
\begin{equation}
\boxed{\frac{\chi}{g} = \frac{\kappa_u^0}{\kappa_\eta} = \frac{\alpha T_L}{V m_L \tau_{\text{eff},L} \tau_\eta}}
\end{equation}

\textit{Why this works:} From the second equation, $\chi = \kappa_A (\chi/g) + \kappa_u^0$, rearranging gives $\chi(1 - \kappa_A/g) = \kappa_u^0$. Since $g = \kappa_\eta + \kappa_A$, we have $1 - \kappa_A/g = \kappa_\eta/g$, so $\chi/g = \kappa_u^0/\kappa_\eta$. The $\kappa_A$ terms—reflecting non-adopter behavior—cancel perfectly in the ratio. The aggregate precision $S$ also cancels. This holds regardless of the specific non-adopter demand formula.

\textbf{Economic interpretation:} The noise-to-signal ratio—the measure of price informativeness—is determined entirely by LLM adopters' behavior. Non-adopters are "price-takers" in two senses: they take $p_1$ as given, and they take the informativeness of $p_1$ (captured by $\chi/g$) as given. The informativeness is created by LLM adopters trading on $\eta$ against noise $u_1$; non-adopters' trades merely scale both $g$ and $\chi$ proportionally, leaving $\chi/g$ unchanged.

Therefore, $\chi/g$ depends only on:
\begin{itemize}
\item Primitives: $\alpha, V, \tau_d, \tau_A, \tau_\eta, \phi, \pi_A, \sigma_\xi^2, \sigma_{u2}^2$
\item Adopter-side objects: $T_L = \tau_d + \tau_A + \tau_\eta$ and $\tau_{\text{eff},L} = \sigma_{d,L}^{-2} + \tau_{p,L}$
\end{itemize}

This breaks the circularity and allows closed-form solution without iteration.

The individual coefficients can be computed from the weights, but are not needed for the main result. The key is that the ratio $\chi/g$ depends only on adopter-side primitives.

\textbf{Closed-form solution procedure:}
\begin{enumerate}
\item Compute adopter objects:
\begin{align*}
T_L &= \tau_d + \tau_A + \tau_\eta \\
\sigma_{d,L}^2 &= T_L^{-1} + \sigma_\xi^2 \\
\tau_{p,L} &= \frac{V^2}{\alpha^2 \sigma_{u2}^2 (\sigma_{d,L}^2)^2} \\
\tau_{\text{eff},L} &= \sigma_{d,L}^{-2} + \tau_{p,L}
\end{align*}

\item Compute learning precision:
\begin{equation}
\tau_{\text{learn}} = \left[\sigma_\eta^2 + \left(\frac{\alpha T_L}{V m_L \tau_{\text{eff},L} \tau_\eta}\right)^2 \sigma_{u1}^2\right]^{-1}
\end{equation}

\item Compute non-adopter objects:
\begin{align*}
T_{A^*} &= \tau_d + \tau_A + \tau_{\text{learn}} \\
\Delta &= T_{A^*}^{-1} - T_L^{-1} \\
\Sigma_p &= \tau_{p,L}^{-1} \\
\tau_{\text{eff},A^*} &= \left[\Delta + (\Sigma_p \,|\, \sigma_{d,L}^2)\right]^{-1} = \left[\Delta + \frac{\Sigma_p \sigma_{d,L}^2}{\Sigma_p + \sigma_{d,L}^2}\right]^{-1}
\end{align*}

\item Compute inattentive objects:
\begin{align*}
\sigma_{d,I}^2 &= T_I^{-1} + \sigma_\xi^2 \\
\tau_{\text{eff},I} &= \sigma_{d,I}^{-2}
\end{align*}

\item Compute aggregates: $S = m_L \tau_{\text{eff},L} + m_A \tau_{\text{eff},A^*} + m_I \tau_{\text{eff},I}$, $w_j = m_j \tau_{\text{eff},j}/S$

\item Read off coefficients $(a,b,g,\chi)$ from the formulas above
\end{enumerate}

This sequential procedure yields an equilibrium characterization without iterative fixed-point computation. The rational expectations consistency condition is satisfied because beliefs are formed correctly given the price function, and the price function correctly aggregates demands given beliefs.

\textbf{Step 6: Efficiency and comparative statics.} An increase in the proportion of adopters ($\phi$) or the quality of their information ($\tau_\eta$) makes the price more informative about the private signal $\eta$. The mechanism operates through the noise-to-signal ratio:
\begin{equation}
\frac{\chi}{g} = \frac{\alpha T_L}{V m_L \tau_{\text{eff},L} \tau_\eta}
\end{equation}

\textit{Effect of adoption rate ($\phi$):} As $\phi$ increases, $m_L = \phi \pi_A$ increases, which directly reduces $(\chi/g)$. Intuitively, more LLM adopters trading on their private signal $\eta$ makes the price more informative about $\eta$ relative to the noise.

\textit{Effect of LLM quality ($\tau_\eta$):} As $\tau_\eta$ rises, $T_L$ increases linearly in the numerator, while the denominator $\tau_{\text{eff},L} \cdot \tau_\eta$ grows faster, ensuring $\chi/g$ falls (making prices more informative). The strength of this effect depends critically on the relative magnitude of payoff noise $\sigma_\xi^2$ versus fundamental uncertainty $T_L^{-1}$. When payoff noise is small ($\sigma_\xi^2 \ll T_L^{-1}$), the effect is dramatic: $\sigma_{d,L}^2 \approx T_L^{-1}$ shrinks rapidly with $\tau_\eta$, creating a quadratic relationship $\tau_{p,L} \propto T_L^2$ that causes adopters to trade very aggressively on their superior signal. The denominator $\tau_{\text{eff},L} \cdot \tau_\eta$ grows approximately like $\tau_\eta^3$, so $\chi/g \propto \tau_\eta^{-2}$ falls rapidly. Conversely, when payoff noise is large ($\sigma_\xi^2 \gg T_L^{-1}$), the effect saturates: $\sigma_{d,L}^2 \approx \sigma_\xi^2$ remains nearly constant, capping both fundamental precision and speculative trading intensity. In this regime, $\chi/g$ remains a decreasing function of $\tau_\eta$, but the effect saturates as $\tau_\eta \to \infty$: $\chi/g$ approaches a positive constant. Intuitively, large irreducible payoff noise limits how aggressively agents can trade on any signal, thereby limiting the informativeness of prices.

Both channels increase the learning precision:
\begin{equation}
\tau_{\text{learn}} = \left[\sigma_\eta^2 + \left(\frac{\alpha T_L}{V m_L \tau_{\text{eff},L} \tau_\eta}\right)^2 \sigma_{u1}^2\right]^{-1}
\end{equation}

Since market efficiency is increasing in the precision with which agents can forecast the fundamental, and non-adopters' forecast precision is $T_{A^*} = \tau_d + \tau_A + \tau_{\text{learn}}$, market efficiency improves as $\phi$ or $\tau_\eta$ increase. This creates strategic complementarity: as more agents adopt superior LLMs, prices become more informative, benefiting even non-adopters through the information aggregation mechanism. This feedback effect can support partial adoption equilibria where some sophisticated investors rely on price discovery rather than directly acquiring the technology.
\end{proof}

\newpage

\newpage
\begin{center}
    \section*{Online Appendix E: Interpretability Methodology} \label{appendix:topic}

\end{center}

\onehalfspacing

To analyze heterogeneity in information processing across different types of news, we develop a technique involving topic modeling and linear regression. For topic modeling, we employ BERTopic \citep{grootendorst2022bertopic}, a topic modeling approach that combines modern embedding techniques with clustering algorithms. This method enables us to categorize headlines into coherent semantic categories without imposing predefined topic structures.
\\

\subsection*{Sample and Data Preparation}

Our analysis focuses on overnight news headlines with directional GPT-4 predictions (excluding neutral scores of 0), yielding 81,700 observations. We preprocess the headlines by removing the prediction labels (`YES,' `NO', or `UNKNOWN') from GPT-4's explanations to avoid mechanical clustering based on these markers rather than underlying news content.

\subsection*{Technical Implementation}

The BERTopic pipeline consists of five steps:

\paragraph{1. Sentence Embeddings.} We transform each headline into a dense vector representation using the SentenceTransformer model \citep{reimers2019sentence}. Specifically, we employ the \texttt{all-MiniLM-L6-v2} model from Hugging Face, which maps text to a 384-dimensional embedding space where semantically similar headlines cluster together.

\paragraph{2. Dimensionality Reduction.} We apply UMAP (Uniform Manifold Approximation and Projection; \citealt{McInnes2018UMAP:Projection}) to reduce the embedding dimensionality from 384 to 5 dimensions while preserving both local and global structure. We use standard UMAP parameters: 15 neighbors and minimum distance of 0. This step concentrates semantic similarities in a lower-dimensional space to enhance clustering performance.

\paragraph{3. Initial Clustering.} We apply K-Means clustering to identify initial groupings of headlines in the reduced embedding space. We set K=50 clusters as the initial partition, which provides sufficient granularity to capture fine-grained distinctions in news content while maintaining adequate cluster sizes for analysis.

\paragraph{4. Topic Reduction.} To facilitate interpretation and statistical power, we reduce the 50 initial clusters to 20 final topics using BERTopic's hierarchical topic merging algorithm. This algorithm iteratively merges the most similar topics based on their c-TF-IDF representations until reaching the target number. This reduction step consolidates semantically related clusters while preserving the major thematic categories in the data. The final 15 topics range from large categories like ``Insider Stock Transactions'' (16,790 headlines) to smaller but distinct categories like ``Retail Store Openings'' (175 headlines).

\paragraph{5. Topic Labeling.} After identifying the final topics, we generate human-interpretable topic labels using GPT-4o. For each topic, we extract representative headlines and keywords, then prompt GPT-4o to create a concise, descriptive label. The prompt instructs the model to avoid firm-specific names and generic terms, focusing instead on the underlying business action or market event common across headlines in each topic. The specific prompt is as follows:

\begin{quote}
    
You are tasked with extracting a short but highly descriptive topic label from a topic model and a set of example documents from that topic.

The documents contain [DOCUMENT TYPE].

Follow these instructions carefully:

1. Review the following documents:
[DOCUMENTS]

2. Consider the following keywords that describe the topic:
[KEYWORDS]

3. Based on the information provided in the documents and keywords, extract a short but highly descriptive topic label. 

a. Keep the label at most 4 words long.

b. Do NOT include specific names, firm names, or stock tickers in the label, as the examples may come from only a few companies.

c. Interpret the concise topic rather than focusing on specific entities.

d. Make the label MORE specific than general terms like `Financial Results', `Earnings Announcement', `Financial Announcements', `Financial Updates', or `Corporate Finance Announcements.'

e. Avoid using firm names in whole or in part, tickers, and the words `Quarterly' or `Financial' in your label.

f. Focus on the specific theme that best describes the collection of explanations and keywords.

Keep your topic label concise, specific, and descriptive. 

Provide your final topic label in the following format:
topic: $<$topic label$>$
\end{quote}

\subsection*{Methodological Advantages}

This approach offers several advantages over traditional topic modeling methods (e.g., Latent Dirichlet Allocation) for our purposes. First, the embedding-based approach captures semantic similarity better than bag-of-words representations, which is particularly important given the brevity and density of news headlines. Second, K-Means clustering ensures all headlines are assigned to topics (no outliers), which maximizes sample utilization for our statistical analysis. Third, the two-stage approach (initial K-Means followed by hierarchical merging) balances granularity and interpretability: we capture fine-grained semantic distinctions in the initial clustering, then consolidate them into interpretable categories. Fourth, reducing to 20 final topics provides sufficient statistical power per category while maintaining conceptual coherence. Finally, using GPT-4o for labeling ensures consistency with our broader analytical framework while maintaining human interpretability of the resulting categories.

The resulting topic assignments enable us to decompose GPT-4's performance and market reactions across 15 distinct categories of corporate news, as presented in Table~\ref{tab:decomposition_analysis} of the main text. Table~\ref{table:news_topics} below provides the complete list of topics with their representative keywords and example headlines. In the regression analysis underlying the decomposition table, we do not include firm or date fixed effects to preserve topic interpretability, but standard errors remain double-clustered by firm and date to account for potential correlation in residuals.

\begin{comment}
\begin{equation}
    \text{Performance} = \text{GPT Score} \times \text{Return Next Period}.
\end{equation}
\end{comment}

\newpage
\begin{center}
\section*{Online Appendix F: Transaction Cost Analysis}
\end{center}
%\subsection*{Transaction Cost Analysis}

This section presents a comprehensive analysis of transaction costs and portfolio turnover for drift-based trading strategies. We examine strategies that enter positions after news announcements and exit after one trading session, focusing on the post-announcement drift component of returns.
\\

\subsubsection*{1. Methodology}

We analyze drift returns (\texttt{ret\_o}) which measure the post-announcement price adjustment:
\begin{itemize}
    \item \textbf{Overnight News:} For news released before 9 a.m. or after 4 p.m., we enter at market open and exit at market close, capturing the open-to-close drift return.
    \item \textbf{Intraday News:} For news released during trading hours (9:30 a.m. to 4 p.m.), we enter at market close and exit at the next day's close, capturing the close-to-close drift return.
\end{itemize}

To address concerns about high turnover, we implement partial rebalancing strategies where the portfolio weight at time $t$ is:
\[
w_t = x \cdot w^*_t + (1-x) \cdot w_{t-1}
\]
where $w^*_t$ is the target weight, $w_{t-1}$ is the previous weight, and $x \in \{1.0, 0.5, 0.25\}$ is the rebalancing fraction.

Importantly, partial rebalancing reduces \textit{both} turnover and portfolio exposure. When targets change rapidly, positions converge only partway toward each new target before it changes again. For example, if the target flips from +100\% long to -100\% short, full rebalancing ($x=1.0$) immediately shifts to -100\%, while 50\% rebalancing ($x=0.5$) moves to 0\% on day 1, then -50\% on day 2. If targets continue changing, positions perpetually lag behind. In steady state, 50\% rebalancing achieves approximately 100\% gross exposure (50\% long + 50\% short) versus 200\% for full rebalancing (100\% long + 100\% short). 

Transaction costs are modeled as a fixed cost per round-trip trade, measured in basis points. We test cost levels of 0, 5, 10, and 20 basis points, which span the range from highly sophisticated investors to retail traders. We use fixed round-trip costs rather than bid-ask spreads because our baseline trades execute at the opening and closing auctions, which are call auctions that establish a single clearing price with no bid-ask spread. For overnight news released before 9 a.m., we enter at the opening auction and exit at the closing auction of the same day. For news released after market close, we enter at the next day's opening auction and exit at that day's closing auction. Because these auction mechanisms involve no continuous bid-ask spread, fixed costs parsimoniously capture commissions and price impact.

\subsubsection*{2. Results}

Tables \ref{tab:tcosts_overnight} and \ref{tab:tcosts_intraday} present a comprehensive analysis of overnight and intraday news strategies, respectively. Each table reports: Sharpe ratios (Panel A), mean daily returns (Panel B), turnover and gross exposure (Panel C), and daily return volatility (Panel D) for both equal-weighted (EW) and value-weighted (VW) portfolios under different rebalancing intensities and transaction cost assumptions.

Panel C shows both average daily one-way turnover and average gross exposure (sum of absolute long and short positions). This dual reporting clarifies the fundamental trade-off: reducing rebalancing from 100\% to 50\% cuts turnover in half (from ~190\% to ~94\%) but also cuts gross exposure in half (from ~196\% to ~101\%).

The reduction in gross exposure from partial rebalancing has proportional effects on both mean returns and volatility, leaving Sharpe ratios relatively unchanged. For overnight news with equal-weighting at zero transaction costs, reducing rebalancing from 100\% to 50\% cuts mean daily returns approximately in half (from 0.34\% to 0.17\%) while also cutting volatility approximately in half (from 1.80\% to 0.91\%, Panel D). Because both mean and volatility scale proportionally with exposure, the Sharpe ratio remains nearly constant (2.97 for 100\% rebalancing vs 2.95 for 50\% rebalancing). This explains why Panel A shows minimal Sharpe ratio differences across rebalancing intensities at zero costs, despite the substantial differences in mean returns (Panel B) and volatility (Panel D): partial rebalancing reduces signal exposure but maintains similar risk-adjusted returns per unit of exposure.

\subsubsection*{3. Turnover and Transaction Costs}

Panel C shows that the 100\% rebalancing strategy exhibits extremely high turnover of approximately 190\% per day for overnight news and 174\% per day for intraday news.\footnote{In Figure OA3, we show the daily number of stocks held in the long and short legs of the overnight news strategy over time. The strategy holds an average of 74 stocks in the long leg and 21 stocks in the short leg per day, with substantial day-to-day variation reflecting fluctuations in news availability and sentiment. Additionally, Figure OA4 shows portfolio turnover for that strategy over time.} Reducing the rebalancing fraction to 25\% decreases turnover to approximately 45\% per day for both strategies—a reduction of approximately 76\%.

Panel A demonstrates that high turnover makes the 100\% rebalancing strategy highly sensitive to transaction costs. For overnight news, the equal-weighted Sharpe ratio falls from 2.97 (zero costs) to -0.39 with 20 basis points costs. In contrast, the 25\% rebalancing strategy is more robust, with its Sharpe ratio declining from 2.89 to -0.21 under the same cost assumptions.

Panel B shows mean daily returns under different cost levels. The 100\% rebalancing strategy generates mean daily returns of 0.34\% for overnight news at zero costs, but becomes unprofitable (-0.04\%) at 20 basis points costs. The 25\% rebalancing strategy shows more gradual performance degradation, maintaining positive returns up to 10 basis points costs.

\subsubsection*{4. Equal-Weighted vs Value-Weighted Portfolios}

To assess whether the predictability is concentrated in smaller or larger stocks, we compare equal-weighted (EW) and value-weighted (VW) portfolios. In VW portfolios, stocks are weighted by lagged market capitalization (\texttt{me\_lag}), giving more weight to larger firms.

The results show dramatically lower performance for VW portfolios:

\begin{itemize}
    \item \textbf{Overnight news (100\% rebalancing, 0 bps costs):} EW Sharpe = 2.97 vs VW Sharpe = 0.56
    \item \textbf{Intraday news (100\% rebalancing, 0 bps costs):} EW Sharpe = 2.63 vs VW Sharpe = 1.28
\end{itemize}

The substantially lower Sharpe ratios for VW portfolios confirm that the predictability is heavily concentrated in smaller stocks. This finding is consistent with limits-to-arbitrage theories: smaller stocks face higher transaction costs and less attention from institutional investors, allowing mispricings to persist longer.

Notably, VW portfolios also show higher sensitivity to transaction costs. The overnight VW strategy becomes unprofitable even at 5 basis points costs (Sharpe = -0.40), while the EW strategy remains highly profitable (Sharpe = 2.13) at the same cost level. This occurs because the VW strategy concentrates capital in larger positions that face higher price impact costs.

\subsubsection*{5. Implications for Tradability}

This analysis demonstrates that while the GPT-4 signal generates significant predictability, exploiting it requires careful attention to implementation costs. The 25\% rebalancing strategy provides a more realistic alternative that:
\begin{enumerate}
    \item Reduces daily turnover by approximately 76\% (from ~190\% to ~46\% for overnight news)
    \item Achieves approximately 28\% of full signal exposure (~54\% gross exposure vs ~196\%)
    \item Maintains similar risk-adjusted returns (Sharpe ratio) per unit of exposure
    \item Shows more gradual performance degradation as costs increase
    \item Remains profitable at moderate cost levels (5-10 basis points)
\end{enumerate}

The comparison between EW and VW portfolios reveals that predictability is concentrated in smaller stocks, consistent with limits-to-arbitrage theories. While this concentration raises concerns about practical tradability for large institutional investors, it also validates the economic mechanism: markets are less efficient precisely where arbitrage is most costly.

\subsubsection*{6. Understanding the Size Effect: Reaction vs Drift}

To understand \textit{why} predictability concentrates in smaller stocks, we decompose returns into initial reaction (ret\_n) and subsequent drift (ret\_o) across market capitalization terciles (i.e., small, medium, and large caps). Tables \ref{tab:size_reaction_drift} and \ref{tab:size_portfolio} present this analysis.

The key finding is a striking asymmetry in drift patterns following negative news. Table \ref{tab:size_reaction_drift} Panel B shows that after stocks receive negative GPT-4 signals:

\begin{itemize}
    \item \textbf{Small caps:} Initial reaction of -2.33\% followed by drift of -0.70\% (continued decline)
    \item \textbf{Medium caps:} Initial reaction of -1.34\% followed by drift of -0.18\% (modest continuation)
    \item \textbf{Large caps:} Initial reaction of -1.12\% followed by drift of +0.06\% (mean reversion)
\end{itemize}

This pattern explains the value-weighting underperformance. The short side of the strategy profits from drift continuation after negative news, which occurs for small and medium caps but not for large caps. Table \ref{tab:size_portfolio} confirms this at the portfolio level: short-only strategies achieve Sharpe ratios of 3.17 for small caps versus -0.61 for large caps.

In contrast, positive signals (Panel A) show positive drift across all size groups, though with declining magnitude for larger stocks. This explains why long-only portfolios perform reasonably well across all sizes (Sharpe ratios ranging from 0.30 to 1.09), while short-only portfolios exhibit dramatic size dependence.

The asymmetric drift patterns suggest different market mechanisms operate for positive versus negative news, and these mechanisms vary systematically with firm size. Large cap stocks experiencing negative news appear to benefit from support mechanisms---perhaps informed buying, liquidity provision, or analyst attention---that cause prices to partially recover. Small-cap stocks lack these support mechanisms, allowing negative drift to continue as uninformed investors gradually process the information.

This size-stratified analysis provides direct evidence for the mechanism behind the EW versus VW performance gap. It is not merely that smaller stocks are more predictable overall; rather, the predictability operates asymmetrically, with the short side working only for smaller stocks that exhibit drift continuation rather than mean reversion following negative news.

\end{document}